
\input harvmac
\noblackbox
\let\chapfont\titlefont
\font\secfont=cmbx10 scaled\magstep1

\catcode`\@=11 
\ifx\hyperdef\UNd@FiNeD\def\hyperdef#1#2#3#4{#4}\def\hyperref#1#2#3#4{#4}\fi
\def\section#1{\global\advance\secno by1 \xdef\secn@m{\ifcase\secno
\or A\or B\or C\or D\or E\or F\or G\or H\or I\or J\or K\or L\or M%
\or N\or O\or P\or Q\or R\or S\or T\or U\or V\or W\or X\or Y\or Z\fi
}\message{(\secn@m. #1)}\bigbreak\bigskip
\global\subsecno=0\xdef\s@csym{\secn@m.}
\noindent{\bf\hyperdef\hypernoname{section}{\secn@m}{\secn@m.} #1}%
\writetoca{{\string\hyperref{}{section}{\secn@m}{\secn@m.}} {#1}}%
\par\nobreak\medskip\nobreak}
\catcode`\@=12 

\def\xvec{ {\vec  x}}

\def\bq{{\bf q}}

\def\Intq{\int {d^d \bq \over (2\pi)^d}}

\def\bt{\par\noindent $\bullet\quad$}
\def\nextp{\smallskip\noindent}
\def\tr{\mathop{\rm tr}}
\def\g2#1{(\nabla #1)^2}
 
\def\frac#1#2{{#1 \over #2}}
\def\xvec{{\bf x}}
\def\xxSqrt{$\overline{\mid <\!\!\xvec\!\!> \mid}$}
\def\xx{$\overline{<\!\!\xvec\!\!>^{2}}$}
\def\xSq{$\overline{<\!\!\xvec^{2}\!\!>}$}
\def\xSqxx{$\overline{<\!\!\xvec^{2}\!\!> - <\!\!\xvec\!\!>^{2}}$}
\def\PsiPlus{W\!_{+}}     \def\PsiMinus{W\!_{-}}
\nopagenumbers\null\pageno0
\rightline{cond-mat/9411022}
\vfill\abstractfont
\centerline{\chapfont Lectures on}
\medskip
\centerline{\chapfont Directed Paths in Random Media}
\vfill\centerline{by} \bigskip
\centerline{Mehran Kardar}
\centerline{Department of Physics}
\centerline{Massachusetts Institute of Technology}
\centerline{Cambridge, Massachusetts 02139, USA}
\vfill
\centerline{Presented at:}
\vfill
\centerline{\secfont Les Houches Summer School }
\centerline{on}
\centerline{\secfont Fluctuating Geometries in }
\centerline{\secfont Statistical Mechanics and Field Theory}

\centerline{\secfont August 1994}

\vfill\eject\tenpoint
\footline={\hss\tenrm\hyperdef\hypernoname{page}\folio\folio\hss}
\par\centerline{\secfont Table of Contents}\bigskip\bigskip

\noindent {\hyperref{}{section}{A}{\bf A.}} {\ Introduction}
\leaderfill{\hyperref{}{page}{2}{2}} \par
\noindent {\hyperref{}{section}{B}{\bf B.}} {\ High Temperature Expansions for
the Ising Model} \leaderfill{\hyperref{}{page}{4}{4}} \par
\noindent {\hyperref{}{section}{C}{\bf C.}} {\ Characteristic Functions and
Cumulants} \leaderfill{\hyperref{}{page}{6}{6}} \par
\noindent {\hyperref{}{section}{D}{\bf D.}} {\ The One Dimensional Chain}
\leaderfill{\hyperref{}{page}{9}{9}} \par
\noindent {\hyperref{}{section}{E}{\bf E.}} {\ Directed Paths and the Transfer
Matrix} \leaderfill{\hyperref{}{page}{13}{13}} \par
\noindent {\hyperref{}{section}{F}{\bf F.}} {\ Moments of the Correlation
Function} \leaderfill{\hyperref{}{page}{18}{18}} \par
\noindent {\hyperref{}{section}{G}{\bf G.}} {\ The Probability Distribution in
Two Dimensions} \leaderfill{\hyperref{}{page}{22}{22}} \par
\noindent {\hyperref{}{section}{H}{\bf H.}} {\ Higher Dimensions}
\leaderfill{\hyperref{}{page}{25}{25}} \par
\noindent {\hyperref{}{section}{I}{\bf I.}} {\ Random Signs}
\leaderfill{\hyperref{}{page}{28}{28}} \par
\noindent {\hyperref{}{section}{J}{\bf J.}} {\ Other Realizations of DPRM}
\leaderfill{\hyperref{}{page}{32}{32}} \par
\noindent {\hyperref{}{section}{K }{\bf K.}} {\ Quantum Interference of
Strongly Localized Electrons} \leaderfill{\hyperref{}{page}{34}{34}} \par
\noindent {\hyperref{}{section}{L }{\bf L.}} {\ The Locator Expansion and
Forward Scattering Paths} \leaderfill{\hyperref{}{page}{37}{37}} \par
\noindent {\hyperref{}{section}{M }{\bf M.}} {\ Magnetic Field Response}
\leaderfill{\hyperref{}{page}{39}{39}} \par
\noindent {\hyperref{}{section}{N }{\bf N.}} {\ Unitary Propagation}
\leaderfill{\hyperref{}{page}{44}{44}} \par
\noindent {\hyperref{}{section}{O }{\bf O.}} {\ Unitary Averages}
\leaderfill{\hyperref{}{page}{47}{47}} \par
\noindent {\hyperref{}{section}{P }{\bf P.}} {\ Summing all Paths in High
Dimensions} \leaderfill{\hyperref{}{page}{52}{52}} \par
\noindent {\hyperref{}{section}{Q }{\bf Q.}} {\ The Ising Model on a Square
Lattice} \leaderfill{\hyperref{}{page}{57}{57}} \par
\noindent {\hyperref{}{section}{R }{\bf R.}} {\ Singular Behavior}
\leaderfill{\hyperref{}{page}{62}{62}} \par
\noindent {\hyperref{}{section}{S }{\bf S.}} {\ The Two Dimensional Spin Glass}
\leaderfill{\hyperref{}{page}{64}{64}} \par
\noindent {\hyperref{}{section}{T }{\bf T.}} {\ Results for the Two Dimensional
Spin Glass} \leaderfill{\hyperref{}{page}{67}{67}} \par
\bigskip\noindent {\it References}\leaderfill {\hyperref{}{page}{70}{70}} \par

\vfill\eject
\section{Introduction}

Many physical problems involve calculating sums over paths. Each path could
represent one possible physical realization of an object such as a polymer, in
which case the weight of the path is the probability of that configuration. The
weights themselves could be imaginary as in the case of Feynman paths
describing
the amplitude for the propagation of a particle. Path integral calculations are
now
a standard tool of the theoretical physicist, with many excellent books devoted
to
the subject\nref
\rFeynman{R. P. Feynman and A. R. Hibbs, {\it Quantum Mechanics and Path
Integrals} (McGraw--Hill, New York, 1965).}\nref
\rWiegel{F. W. Wiegel, {\it Introduction to Path--Integral Methods in Physics
and
Polymer Science} (World Scientific, Singapore, 1986).}
\refs{\rFeynman,\rWiegel}.

	What happens to sums over paths in the presence of quenched disorder in the
medium?  Individual paths are no longer weighted simply by their length, but
are influenced by the impurities along their route.  The sum may be
dominated by ``optimal"  paths pinned to the impurities; the optimal paths
usually forming
complex hierarchical structures.  Physical examples are provided by
the interface of the random bond Ising model in two dimensions, and by
magnetic flux lines in superconductors. The actual value of the sum
naturally depends on the particular realization of randomness and varies
from sample to sample.  I shall initially describe the problem in
the context of the high temperature expansion for the random bond Ising
model. Introducing the sums over paths for such a lattice model
avoids the difficulties associated with short distance cutoffs. Furthermore,
the Ising model is sufficiently well understood to make the nature of various
approximations more evident.

The high temperature correlation functions of the Ising model are
dominated by the shortest paths connecting the spins.
Such configurations, that exclude loops and overhangs, are referred to as
{\it directed paths}.  They dominate the asymptotic behavior of the sum over
distances that are much longer than the correlation length.
Most of the lectures are devoted to describing the statistical properties
of sums over such directed paths.
As in all multiplicative noise processes, the probability distribution for the
sum
is broad. Hence Monte Carlo simulations may not be an appropriate tool
for numerical studies; failing to find typical members of the ensemble.
Instead, we shall present a transfer matrix method
that allows a numerical evaluation of the sum in polynomial time in the length
of the path.  The results indeed show that the sum has a broad probability
distribution that resembles (but is not quite) log--normal.

To obtain analytical information about this probability distribution we shall
introduce the replica method for examining the moments. A brief review
explains the relationship between the moments, the characteristic
function, and cumulants. It can be shown easily
that the one dimensional sum has a log--normal distribution.  The moments of
the sum over directed paths in two dimensions can be obtained by using a
simple Bethe Ansatz. The implications and limitations of this approach are
discussed. There is little analytical information in three and higher
dimensions,
but a variety of numerical results are available, mostly by taking advantage
of a mapping to growing surfaces.

The spin glass problem describes a mixture of ferromagnetic and
antiferromagnetic
bonds. The resulting sums in the high temperature expansion involve products
over a random mixture of positive and negative factors. The calculation of
moments is somewhat different from the case of purely positive random
bonds. However, we shall demonstrate that the scaling behavior of the
distribution is unchanged. A similar sum involving products of random
signs is encountered in calculating the probability of an electron tunneling
under a random potential. In the strongly localized limit, it is again
sufficient to
focus on the interference of the forward--scattering (directed) paths.
A magnetic field introduces {\it random phases} in the sum; while to describe
the
tunneling of an electron in the presence of spin--orbit scattering requires
examining the evolution of a two component spinor and keeping track of
products of {\it random matrices}. We shall argue that all these cases are in
fact
described by the {\it same universal probability distribution} which, however,
does retain some remnant of the underlying symmetries of the original
electronic
Hamiltonian.

Yet another class of directed paths has been introduced in the context of
light scattering in turbulent media. Assuming that  inelastic  scattering can
be neglected, the intensity of the beam is left unchanged, and the evolution
is unitary. Due to the constraint of unitarity the resulting directed paths
are described by a probability distribution belonging to a new universality
class.
We shall introduce a discrete matrix model that explicitly takes care of
the unitarity constraint. In this model, several properties
of the resulting sum over paths can be calculated exactly.

In the last sections of the course we shall go beyond the limitations
of directed paths. In a uniform system the sum over all paths is
calculated {\it approximately}, reproducing mean--field critical behavior.
In two dimensions, the sum can be performed {\it exactly}
for the Ising model. This leads to exact solutions for the pure
Ising model, or other uniformly frustrated two dimensional lattices. We
shall use this method of exact summation to develop an integer algorithm for
obtaining exact  partition functions for two dimensional random lattices
in polynomial time. Some results are described for the two dimensional
spin glass, and contrasted with those obtained from Monte Carlo
or transfer matrix methods.

The results described in these notes were the outcome of many collaborations.
In particular, I would like to express my thanks to E. Medina, L. Saul,
Y. Shapir, and Y.-C. Zhang. I am grateful to F. David, P. Ginsparg, and J.
Zinn--Justin
for organizing the Les Houches summer school, and to the many students who
helped
me with correcting the lecture notes. The work at MIT was supported by
the NSF through grants DMR-93-03667 and PYI/DMR-89-58061.

\section{High Temperature Expansions for the Ising Model}

Consider a $d$ dimensional hypercubic lattice of $N$ sites. On each site
there is an Ising variable $\sigma_i=\pm 1$, and the spins interact through
a Hamiltonian
\eqn\eHIsing{\CH=-\sum_{<ij>}J_{ij}\sigma_i\sigma_j\quad.}
The symbol $<ij>$ implies that the sum is restricted to the  $dN$
{\it nearest neighbor} bonds on the lattice. The bonds $\left\{ J_{ij} \right\}
$
are {\it quenched} random variables, independently chosen from
a probability distribution $p(J)$. For each realization of random
bonds, the partition function is computed as
\eqn\eZrIsing{Z[J_{ij}]=\sum_{\{\sigma_i\}}\exp\left( -\beta \CH \right)=
\sum_{\{\sigma_i\}}\prod_{<ij>}e^{K_{ij}\sigma_i\sigma_j}\quad,}
where the sums are over the $2^N$ possible configurations of spins,
$\beta=1/(k_BT)$ and $K_{ij}=\beta J_{ij}$.
To obtain a high temperature expansion,  it is more convenient to
organize the partition function in powers of $\tanh K_{ij}$.
Since $(\sigma_i \sigma_j)^2 = 1$,
the Boltzmann factor for each bond can be written as
\eqn\eIHTB{e^{K _{ij}{\sigma_i \sigma_j}} = {e^{K_{ij}} + e^{-K_{ij}} \over 2}
+
{e^{K_{ij}} - e^{-K_{ij}} \over 2} \sigma_i \sigma_j =
\cosh {K_{ij}}\, (1 + \tau_{ij} \sigma_i \sigma_j)\quad,}
where $\tau_{ij}\equiv\tanh K_{ij}$ is a good high temperature expansion
parameter. Applying this transformation to each
bond of the lattice results in
\eqn\eZIHTgen{Z[J_{ij}] = \sum_{\{\sigma_i\}}{e^{\sum_{\left\langle ij
\right\rangle} K_{ij}
{\sigma_i \sigma_j}}}= \overline{C}\,^{dN}
\sum_{\{\sigma_i\}}{\prod_{\left\langle ij \right\rangle}{(1 + \tau_{ij}
\sigma_i \sigma_j)}}\quad,}
where
$$\overline{C}\,^{dN}\equiv \left(\prod_{\left\langle ij
\right\rangle}\cosh{K_{ij}}\right)\quad. $$

The term $\overline{C}\,^{dN}$ is non-singular, and will be mostly ignored
henceforth.
The final product in eq.~\eZIHTgen\ generates $2^{dN}$
terms which can be represented diagrammatically by drawing a line
connecting sites $i$ and $j$ for each factor of $\tau_{ij} \sigma_i \sigma_j$.
Each site now obtains a factor of $\sigma_i^{p_i}$, where $0\leq p_i\leq 2d$ is
the number of bonds emanating from $i$. Summing over the two
possible values $\sigma_i=\pm 1$, gives a factor of 2 if $p_i$
is even and 0 is $p_i$ is odd. Thus the only graphs that survive
the sum have an even number of lines passing through each site.
The resulting graphs are collections of closed paths $\CG$ on the lattice.
The contribution of each graph is the product of $\tau_{ij}$ for the bonds
making up the graph, resulting in
\eqn\eHTIZ{Z[J_{ij}] = 2^N\times\overline{C}\,^{dN}
\sum_{\CG}\left(\prod_{\left\langle ij \right\rangle\in\CG}
\tau_{ij}\right)\quad.}

For a $d$-dimensional hypercubic lattice the smallest closed graph
is a square of 4 bonds and the next graph has 6 bonds. Thus,
\eqn\eHTZd{Z[J_{ij}] = 2^N\times\overline{C}\,^{dN}
 \left[1 + \sum_P\,\tau_{P1}\tau_{P2}\tau_{P3}\tau_{P4}
+\CO( \tau^6)+ \cdots \right]\quad,}
where the sum runs over the $Nd(d-1)/2$ plaquettes on the lattice and
$\tau_{P\alpha}$ indicate the four bonds along each plaquette.
A {\it quench averaged} free energy is now obtained as
\eqn\eHTlnZd{\overline{{\ln{Z}} \over N} = \ln{2} + d\, \overline{\ln{\cosh
{K}} }
+ {d (d-1) \over 2}\,\overline{ \tau}\,^4 + \cdots\quad ,}
where the over--lines indicate averages over the probability distribution
$p(J)$.

The same method can be used to obtain expansions for various spin operators.
For example the two spin correlation function is given by
\eqn\eCF{\left\langle \sigma_m \sigma_n \right\rangle =
 \sum_{ \{\sigma_i\}} {e^{\sum_{\left\langle
ij \right\rangle}K_{ij}{\sigma_i \sigma_j}}\over Z} \sigma_m \sigma_n =
{\overline{C}\,^{dN} \over Z}
\sum_{ \{\sigma_i\}}\sigma_m \sigma_n
\prod_{\left\langle ij \right\rangle}{(1 + \tau _{ij}\sigma_i \sigma_j)}\quad.}
The terms in the numerator involve an additional factor of
$\sigma_m \sigma_n$. To get a finite value after summing over
$\sigma_m=\pm1$ and $\sigma_n=\pm1$ we have to examine graphs with
an odd number of bonds emanating from these {\it external} sites.
After cancelling the common factors between the numerator and
denominator, we obtain
\eqn\eHTICF{\left\langle \sigma_m \sigma_n \right\rangle =
{\sum_{\CG_{mn}}\left(\prod_{\left\langle ij \right\rangle\in\CG_{mn}}
\tau_{ij}\right)\over \sum_{\CG}
\left(\prod_{\left\langle ij \right\rangle\in\CG}\tau_{ij}\right)}\quad.}
Whereas the graphs in $\CG$ have an even number of bonds going through
each site, those of  $\CG_{mn}$ have an odd number of bonds going through
the external points $m$ and $n$. This procedure can be generalized to multiple
spin correlation functions.

\section{Characteristic Functions and Cumulants}

As equations \eHTZd\ and \eCF\ indicate, the partition function and
correlation functions of the random system are themselves random
quantities, dependent on all the bonds $K_{ij}$. It may not be
sufficient to just characterize the mean value of $Z$ (or $\ln Z$),
since the full information about these fluctuating quantities is
only contained in their respective probability distributions $p(Z)$ and
$p\left(\langle\sigma_m\sigma_n\rangle\right)$.
It is thus important to learn to characterize and manipulate probability
distributions, necessitating the short digression taken in this section
to define various notations I shall use in describing such random
quantities.

Consider a continuous random variable $x$, whose outcome is a
real number $E$.

\bt {\it The cumulative probability function} (CPF) $P(x)$, is the
probability of an outcome with {\it any value} less than $x$, i.e.
$P(x)={\rm prob.}(E\leq x)$. $P(x)$ must be a monotonically
increasing function of $x$, with $P(-\infty)=0$ and $P(+\infty)=1$.

\bt {\it The probability density function} (PDF), is defined by
$p(x)\equiv dP(x)/dx$. Hence, $ p(x)dx={\rm prob.}(x<E<x+dx)$. As
a probability density, it is {\it positive}, and normalized such that
\eqn\epxnorm{P(\infty )=\int_{-\infty}^{\infty}dx\, p(x)=1\quad .}

\bt {\it The expectation value} of any function $F(x)$, of the random
variable is
\eqn\expectedF{\left\langle F(x)\right\rangle =\int_{-\infty}^{\infty}{dx\,
p(x)F(x)}\quad.}
The function $F(x)$ is itself a random variable, with an associated
PDF of $p_F(f)df={\rm prob.}(f<F(x)<f+df)$. There may be multiple
solutions $x_i$, to the equation $F(x)=f$, and
\eqn\echangextoF{p_F(f)df=\sum_{i}p(x_i)dx_i\quad\Rightarrow\quad
p_F(f)=\sum_{i}p(x_i)\left|{dx \over dF}\right|_{x=x_i}\quad.}
The factors of $\left|dx/dF\right|$ are the {\it Jacobians} associated with the
change of variables from $x$ to $F$.

\bt {\it Moments} of the PDF are expectation values for powers of
the random variable. The $n^{\rm th}$ moment is
\eqn\enthmoment{ \left\langle x^n\right\rangle =\int{dx p(x)\, x^n\quad}.}

\bt {\it The characteristic function}, is the generator of moments of
the distribution. It is simply the Fourier transform of the PDF, defined
by
\eqn\echaracteristic{{\tilde p}(k)=\left\langle e^{-ikx}\right\rangle =\int
dxp(x)\, e^{-ikx}\quad.}
Moments of the distribution can be obtained by expanding ${\tilde p}(k)$
in powers of $k$,
\eqn\egenxn{{\tilde p}(k)=\left\langle\sum_{n=0}^{\infty}{(-ik)^n \over
n!}x^n\right\rangle =
\sum_{n=0}^{\infty}{(-ik)^n \over n!}\left\langle x^n\right\rangle \quad .}

\bt {\it The cumulant generating function} is the logarithm of the
characteristic function. Its expansion generates the {\it cumulants} of
the distribution defined through
\eqn\ecumulants{\ln {\tilde p}(k)=
\sum_{n=1}^{\infty}{(-ik)^n \over n!}\left\langle x^n\right\rangle _c\quad .}
Relations between moments and cumulants can be obtained by expanding
the logarithm of ${\tilde p}(k)$ in eq.~\egenxn, and using
\eqn\elnexpansion{\ln(1+\epsilon)=\sum_{n=1}^\infty (-1)^{n+1}{\epsilon^n
\over n}\quad.}
The first four cumulants are called the {\it mean}, {\it variance},
{\it skewness}, and {\it curtosis} of the distribution respectively,
and are obtained from the moments as
\eqn\ecumulantsa{\eqalign{
\left\langle x\right\rangle _c=&\left\langle x\right\rangle \quad,\cr
\left\langle x^2\right\rangle _c=&\left\langle x^2\right\rangle -\left\langle
x\right\rangle ^2\quad,\cr
\left\langle x^3\right\rangle _c=&\left\langle x^3\right\rangle -3\left\langle
x^2\right\rangle \left\langle x\right\rangle +2\left\langle x\right\rangle
^3\quad,\cr
\left\langle x^4\right\rangle _c=&\left\langle x^4\right\rangle -4
\left\langle x^3\right\rangle \left\langle x\right\rangle -3\left\langle
x^2\right\rangle ^2+12\left\langle x^2\right\rangle \left\langle x\right\rangle
^2-6\left\langle x\right\rangle ^4\quad.
}}
The cumulants are usually the most compact way of describing a PDF.
An important theorem allows easy computation of moments in terms of
the cumulants: Represent the $n^{\rm th}$ cumulant graphically
as a {\it connected cluster} of $n$ points. The $m^{\rm th}$
moment is then
obtained by summing all clusters (connected or disconnected) of
$n$ points; the contribution of each cluster being the
product of the connected cumulants that it represents. Using this
result the first four moments are easily computed as
\eqn\emomentsa{\eqalign{
\left\langle x\right\rangle =&\left\langle x\right\rangle _c\quad,\cr
\left\langle x^2\right\rangle =&\left\langle x^2\right\rangle _c+\left\langle
x\right\rangle _c^2\quad,\cr
\left\langle x^3\right\rangle =&\left\langle x^3\right\rangle _c+3\left\langle
x^2\right\rangle _c\left\langle x\right\rangle _c+\left\langle x\right\rangle
_c^3\quad,\cr
\left\langle x^4\right\rangle =&\left\langle x^4\right\rangle _c+4\left\langle
x^3\right\rangle _c\left\langle x\right\rangle _c+3\left\langle
x^2\right\rangle _c^2+6\left\langle x^2\right\rangle _c\left\langle
x\right\rangle _c^2+\left\langle x\right\rangle _c^4\quad.
}}

\bt {\it The normal (Gaussian) distribution} describes
a continuous real random variable $x$, with
\eqn\eNormalPDF{p(x)={1 \over \sqrt{2\pi\sigma^2} }\exp
\left[ -\,{(x-\lambda)^2 \over 2\sigma^2} \right]\quad.}
The corresponding characteristic function also has a Gaussian form,
\eqn\eNormalCF{{\tilde p}(k)=\int_{-\infty}^\infty{dx~{1 \over
\sqrt{2\pi\sigma^2} }
\exp\left[-{(x-\lambda)^2 \over 2\sigma^2}-ikx\right]}=
\exp\left[ -ik\lambda-{k^2\sigma^2 \over 2} \right]\quad.}
Cumulants of the distribution can be identified from $\ln {\tilde p}(k)=
-ik\lambda-k^2\sigma^2 / 2$, using eq.~\ecumulants, as
\eqn\eNormalsCM{\left\langle x\right\rangle _c=\lambda\quad,\quad\left\langle
x^2\right\rangle _c=\sigma^2\quad,\quad
\left\langle x^3\right\rangle _c=\left\langle x^4\right\rangle
_c=\cdots=0\quad.}
The normal distribution is thus completely specified by its
two first cumulants. This makes the computation of moments using
the cluster expansion (eqs.~\emomentsa) particularly simple, and
\eqn\eNormalsMM{\eqalign{
\left\langle x\right\rangle =&\,\lambda\quad,\cr
\left\langle x^2\right\rangle =&\,\sigma^2+\lambda^2\quad,\cr
\left\langle x^3\right\rangle =&\,3\sigma^2\lambda+\lambda^3\quad,\cr
\left\langle x^4\right\rangle
=&\,3\sigma^4+6\sigma^2\lambda^2+\lambda^4\quad,\quad\cdots\quad.
}}

\bt {\it The central limit theorem} describes the probability distribution for
a
sum $ {\cal S} =\sum_{i=1}^{N}{x_i}$ over a large number of random variables.
Cumulants of the sum are given by,
\eqn\cCMsum{\left\langle {\cal S} \right\rangle _c=\sum_{i=1}^{N}\left\langle
x_{i}\right\rangle _c\quad,\quad \left\langle {\cal S} ^2\right\rangle _c=
\sum_{i,j}^{N}\left\langle x_ix_j\right\rangle _c\quad,\quad\cdots\quad.}
If the random variables are independent, $p({\bf x})=\prod{p_i(x_i)}$,
and ${\tilde p}_{\cal S} (k)=\prod {\tilde p}_i(k)$. The cross--cumulants in
eq.~\cCMsum\ vanish, and the $n^{\rm th}$ cumulant
of ${\cal S} $ is simply the sum of the individual cumulants,
$ \left\langle {\cal S} ^n\right\rangle _c=\sum_{i=1}^{N}\left\langle
x_i^n\right\rangle _c$. When all the $N$
random variables are independently taken from the same distribution
$p(x)$, this implies $ \left\langle {\cal S} ^n\right\rangle _c=N\left\langle
x^n\right\rangle _c$. For large values
of $N$, the average value of the sum is proportional to $N$ while
fluctuations around the mean, as measured by the standard deviation, grow
only as $\sqrt{N}$. The random variable $y=({\cal S} -N\left\langle
x\right\rangle _c)/\sqrt{N}$,
has zero mean, and cumulants that scale as $\left\langle y^m\right\rangle
_c\propto
N^{1-m/2}$. As $N\to\infty$, only the second cumulant survives and
the PDF for $y$ converges to the normal distribution,
\eqn\eCLT{\lim_{N\to\infty}p\left(y={\sum_{i=1}^{N}x_i-N\left\langle
x\right\rangle _c \over \sqrt{N}}\right)=
{1 \over \sqrt{2\pi\left\langle x^2\right\rangle _c}}\exp\left(- {y^2 \over
2\left\langle x^2\right\rangle _c} \right)\quad.}
The convergence of the PDF for the sum of many random variables to
a normal distribution is a most important result in the context of
statistical mechanics where such sums are frequently encountered.
The {\it central limit theorem} proves a more general form of this
result: It is not necessary for the random variables to be independent,
as the condition $\sum_{i_1,\cdots,i_m}^N\left\langle x_{i_1}\cdots
x_{i_m}\right\rangle _c\ll
\CO(N^{m/2})$, is sufficient for the validity of eq.~\eCLT.

\section{The One Dimensional Chain}

The graphical method provides a rapid way of solving the Ising
model at zero field in $d=1$. We can compare and contrast the
solutions on chains with open and closed (periodic) boundary
conditions.
\nextp
{\bf 1.} {\it An open chain} of $N$ sites has $N-1$ bonds. It is
impossible to draw any closed graphs on such a lattice, and hence
\eqn\eZopen{Z = 2^N\prod_{\alpha=1}^{N-1} \cosh{K_\alpha}
\times 1\quad\Longrightarrow\quad
{\overline{\ln Z}\over N}=\overline{\ln[2\cosh K]}-{\overline{\ln[\cosh
K]}\over N}\quad,}
where $K_\alpha\equiv K_{\alpha\,\alpha+1}$.
There is also only one graph that contributes to the two point correlation
function,
\eqn\eCFopen{\left\langle \sigma_m \sigma_n \right\rangle =
 \sum_{ \{\sigma_i\}} {e^{\sum_ i K_i{\sigma_i \sigma_{i+1}}}\over Z}
\sigma_m \sigma_n = \prod_{\alpha=m}^{n-1} \tau_\alpha
\quad.}

\nextp
{\bf 2.} {\it A closed chain} has the same number of sites and bonds, $N$.
It is now possible to draw a closed graph that circles the whole chain, and
\eqn\eZclosed{\eqalign{&Z=2^N\left(\prod_{\alpha=1}^N\cosh{K_\alpha} \right)
\left[1+\left(\prod_{\alpha=1}^N\tau_\alpha \right)\right]
\qquad\Longrightarrow\cr
&{\overline{\ln Z}\over N}\approx\,\overline{\ln[2\cosh K]}+{1\over N}
\overline{\tau}\,^N}\quad.}
The difference between the quenched free energies of closed and open chains
is a surface term of the order of $1/N$, and an exponential decay
reflecting the interaction between  edges, both vanishing
in the thermodynamic limit of $N\to \infty$.
The correlation function can again be calculated from eq.~\eCF.
There are two paths connecting the points $m$ and $n$, along the
two possible directions on the chain, giving
\eqn\eCFclosed{\left\langle \sigma_m \sigma_n \right\rangle =
{\left[\prod_{\alpha=m}^{n-1}\tau_\alpha+\prod_{\alpha=n}^{m-1}
\tau_\alpha\right]
\over \left[  1+\left(\prod_{\alpha=1}^N\tau_\alpha \right)\right]}\quad.}

Since the partition function of the open chain
is the sum of $N-1$ independent variables,
\eqn\eLnZopen{{\ln Z\over N} = \ln 2 + \sum\nolimits^{N-1}_{\alpha = 1}
{\ln \cosh K_\alpha\over N}\quad,}
we can use the central limit theorem to conclude that
as $N\to\infty$ the probability distribution
$p(\ln Z/ N)$, is a {\it gaussian} with mean
\eqn\eLnZmean{\overline {\ln Z} = N\left(\ln 2+ \overline { \ln \cosh
K}\right)\quad,}
and variance
\eqn\eLnZvar{\overline {(\ln Z)^2_c}  \equiv \overline {(\ln Z)^2} -
\overline {(\ln Z)}^2 = N\overline {(\ln \cosh K)^2_c}\quad.}
(Note that I have ignored the small difference between $N$ and $N-1$ in the
thermodynamic limit.) Similarly, for the correlation function of two points
separated
by a distance $t$, we have
\eqn\eLnCF{\ln\ \langle\sigma_0\sigma_t\rangle = \sum\nolimits^{t-1}_{\alpha =
0}
\ln \tau_\alpha\quad.}
As long as the random variables on the bonds are independently distributed,
the {\it cumulants} of $\ln\ \langle\sigma_0\sigma_t\rangle$ are given by,
\eqn\ecumCF{\left\{\eqalign {\overline {\ln\langle\sigma_0\sigma_t\rangle} & =
t\ \overline {\ln\tanh K}\cr
\overline {(\ln\langle\sigma_0\sigma_t\rangle)^2_c} & =
t\ \overline {(\ln \tanh K)^2_c}\cr
\vdots	&\qquad 	\vdots\cr
\overline {(\ln\langle\sigma_0\sigma_t\rangle)^p_c} & =
t\ \overline {(\ln\tanh K)^p_c}\cr}\right.\quad.}

In the following sections we shall try to obtain similar information
about probability distribution functions for the partition and correlation
functions in higher dimensions. To do this we shall employ the
{\it replica method} for calculating the moments of the distribution.
For example, the cumulants of the free energy are given by
\eqn\eZncum{\overline {Z^n} = \overline {e^{n\ln Z}} =
\exp\left[n \overline {\ln Z} +
{n^2\over 2}\ \overline {(\ln Z)^2_c} + \cdots +
{n^p\over p!} \overline {(\ln Z)^p_c} + \cdots\right]\quad,}
where we have taken advantage of  eq.~\ecumulants, replacing $(-ik)$
with $n$. Usually,  the moments on the left hand side of the above
equation are known only for integer $n$, while the evaluation of the
cumulants on the right hand side relies on an expansion around $n=0$.
This is one of the difficulties associated with the problem of deducing
a probability distribution $p(x)$, from the knowledge of its moments
$\overline {x^n}$. There is in fact a rigorous theorem that the
probability distribution cannot be uniquely inferred if
its $n^{th}$ moment increases faster than $n!$ \ref
\rNIK{N. I. Akhiezer, {\it The Classical Moment Problem} (Oliver and Boyd,
London, 1965).}.
Most of the distributions of interest to us (such as the above log-normal)
do not satisfy this condition! Similar problems are encountered in the
replica studies of spin glasses \ref
\rBeyond{M. M\'ezard, G. Parisi, and M. A. Virasoro, {\it Spin Glass
Theory and Beyond} (World Scientific, Singapore, 1987).}.
It turns out that many of the difficulties associated with a rigorous inversion
are related to the tail of the distribution. Most of the information of
interest to us is contained in the ``bulk" of the distribution which is easier
to investigate. Rather than taking a rigorous approach to the problem, we
shall illustrate the difficulties and their resolution by examining the one
dimensional case in detail since it actually presents the worst case
scenario for the inverting of moments.

We used the central limit theorem to deduce that the probability distribution
for $\langle\sigma_0\sigma_t\rangle$ is log--normal. Its moments are
computed from,
\eqn\eCFin{\overline {\langle\sigma_0\sigma_t\rangle^n} =
\prod\limits^{t-1}_{\alpha =0}
\overline {\tau^n_\alpha}
= \left( \overline {e^{n\ln \tau}}\right)^t
= \exp\left[{t\sum_p}{n^p\over p!} \overline {(\ln \tau)^p_c}\right]\quad.}
Let us consider a binary distribution in which $\tau$ takes two positive values
of
$\tau_1$ and $\tau_2>\tau_1$ with equal probability. Then
\eqn\ebintn{\overline {\tau^n} = {\tau^n_1 + \tau^n_2\over 2}\quad,}
and the generating function for the cumulants of the correlation function is
\eqn\ebincum{\eqalign {\ln \overline {\tau^n}  =
&n\ln \tau_1 + \ln \left({1 + \left({\tau_2/ \tau_1}\right)^n\over 2}\right)\cr
{=\atop n\to 0}\ &n\ln \tau_1 +
\ln \left({1+ 1 + n \ln \left({\tau_2/ \tau_1}\right) + {n^2/2}
\ln^2\left({\tau_2/ \tau_1}\right) + \cdots\over 2}\right)\cr
=&  	n \ln \tau_1 + \ln \left[ 1 + {n\over 2} \ln\left( {\tau_2\over
\tau_1}\right) +
{n^2\over 4}\ln^2 \left({\tau_2\over \tau_1}\right) + \cdots \right]\cr
= &n \ln \tau_1 + {n\over 2} \ln\left( {\tau_2\over \tau_1}\right) + {n^2\over
8}
\ln^2 \left({\tau_2\over \tau_1}\right) + \cdots\cr
= &n \ln (\sqrt{\tau_1\tau_2}) + {n^2\over 8} \ln^2 \left({\tau_2\over
\tau_1}\right)
 + \cdots\quad.\cr}}
Combining eqs.~\eCFin\ and \ebincum, the cumulants of the correlation function
are
given by
\eqn\eCFcums{\left\{\eqalign{
\overline {\ln \langle\sigma_0\sigma_t\rangle} =&t \ln (\sqrt{\tau_1\tau_2})\cr
\overline {\ln \langle\sigma_0\sigma_t\rangle^2_c} = &
{t\over 4} \ln^2\left({\tau_2/ \tau_1}\right)\cr
\vdots&}\right.\quad.}

While it is true that $\ln  \langle\sigma_0\sigma_t\rangle$ is normally
distributed
for large $t$, we should be careful about the order of limits in terminating
the power series in the exponent at the second order. If we do so, from
\eqn\eCFterm{\overline {\langle\sigma_0\sigma_t\rangle^n} \approx
\exp\left[t \left(na_1 + n^2a_2 \right)\right]\quad,}
we should not infer anything about the high moments ($n\to\infty$) and the
tail of the distribution. Otherwise (since $a_2>0$), we would conclude that
sufficiently large moments of $\langle\sigma_0\sigma_t\rangle$ diverge
with separation; a clearly false conclusion as $\langle\sigma_0\sigma_t\rangle$
is bounded by unity! The exact result is that
\eqn\eCFlargen{\lim_{n\to\infty}\langle\sigma_0\sigma_t\rangle^n ={
\tau_2^{nt}\over 2^t}\quad,}
i.e., the high moments are almost entirely dominated by the one exceptional
sample in which all bonds are equal to $\tau_2$.  We can summarize the
situation as
follows: The ``bulk" of the probability distribution for
$\ln  \langle\sigma_0\sigma_t\rangle$ is described by the small moments
($n\to 0$), while the tail of the distribution is governed by the large moments
($n\to\infty$). We should have a clear idea of the crossover point $n^*$ in
applying the replica method. For the above one dimensional example, an
estimate of $n^*$ is given by the ratio of the successive terms in the
expansion, i.e.
\eqn\elnstar{n^* = {1\over \ln(\tau_2/ \tau_1)}\quad.}
Note that as $\tau_2/\tau_1$ becomes large, $n^*$ decreases, possibly becoming
smaller than unity. This does not imply that we should conclude that
$\ln  \langle\sigma_0\sigma_t\rangle$ is not normally distributed, just that
the tail of the distribution is more prominent. Failure to appreciate this
point is
the source of some misunderstandings on the use of the replica method \ref
\rSZF{B. Z. Spivak, H. L. Zhao and S. Feng, preprint (1993).}.

Clearly, it is possible to come up with many different microscopic
distributions
$p(\tau)$, which result in the same first two cumulants in eqs.~\eCFcums, but
different higher cumulants. All these cases lead to the same universal
bulk probability distribution for $\ln\langle\sigma_0\sigma_t\rangle$ at
large $t$, but very different tails. Thus the non-uniqueness of the overall
probability in this example has to do with the rather uninteresting (and
nonuniversal) behavior of the tail of the distribution.
The correct interpretation of eqs.~\eCFcums\ is that the mean value for the
logarithm of the correlation function grows linearly with the separation $t$.
In analogy with pure systems, we can regard the coefficient of this decay
as the inverse correlation length, i.e. $\xi^{-1}=-\ln\sqrt{\tau_1\tau_2}$.
However,
due to randomness in the medium, correlations have different decays
between different realizations (and between different points in the same
realization).
The variations in this ``inverse correlation length" are scale dependent
and fall off as $1/\sqrt{t}$. In the next sections we shall attempt to
generalize these
results to higher dimensions.

\section{Directed Paths and the Transfer Matrix}

Calculation of the correlation function in higher dimensions is complicated by
the presence of an exponentially large number of paths connecting any pair
of points. On physical grounds we expect the high temperature phase to be
disordered, with correlations that decay exponentially as a function of the
separation $t$. The essence of this exponential decay is captured by the
lowest order terms in the high temperature expansion. The first term in the
series comes from the {\it shortest path} connecting the two points. Actually,
along a generic direction on a hypercubic lattice there are many paths
that have the same shortest length. (In two dimensions, the length of the
shortest
path connecting $(0,0)$ to $(t,x)$ is the ``Manhattan" distance $|t|+|x|$.)
The number of paths grows from a minimum of 1 along a lattice direction to
a maximum of $d$ per step along the diagonal. (The number of paths on the
square lattice is $(t+x)!/(t!x!)$.) Thus the
decay of correlations depends on orientation, a consequence of
the {\it anisotropy} of the hypercubic lattice. (Note that this anisotropy
is absent at distances {\it less than} the correlation length. We thus don't
have to worry about anisotropy effects in discretizing critical (massless)
theories on a lattice.)

In a uniform system these shortest paths are sufficient to capture the essence
of correlation functions of the high temperature phase: An exponential decay
with separation which is generic to all spin systems. As temperature is
reduced, more complicated paths (e.g. with loops and overhangs) start
contributing to the sum. Although the contribution of these paths decays
exponentially in their length, their number grows exponentially. Ultimately
at the critical point this ``entropic" increase in the number of paths
overcomes
the ``energetic" decrease due to the factors of $\tau<1$, and paths of all
length
become important below $T_c$. However, throughout the high
temperature phase, it is possible to examine the paths at the coarse grained
scale where no loops and overhangs are present. The scale of such structures
is roughly the correlation length $\xi$, and if we use $\xi$ as the unit of a
coarse--grained lattice, the paths contributing to the correlation function
are {\it directed}.

Let me define  ``directed paths'' more carefully: Between any
pair of points on the lattice we can draw an imaginary line which I shall
refer to as the ``time'' axis $t$. Transverse directions (perpendicular
to the $t$ axis) are indicated by $\vec{x}$. Directed paths are similar to
the worldlines of a particle $\vec{x}(t)$ in time; they exclude any path from
the initial to the final point that has steps opposite to the main time
direction.
The question of the validity of this approximation, and the importance of the
neglected loops continually comes up. This is possibly because it is more
common to think about the vicinity of the critical point where loops of all
sizes are present and equally important. Away from the critical point, we have
to distinguish between properties at scales smaller and larger than the
correlation length $\xi$; there is no other length scale (except for the
lattice spacing) in the problem. Limiting the sum for the correlation function
to directed paths is only useful for separations $t\gg \xi$. Loops, overhangs,
and additional structures occur up to size $\xi$ (the only length scale
present) and can be removed by coarse graining such that the lattice
spacing is larger than or equal to $\xi$. This is automatically the case in a
high temperature expansion since $\xi$ is initially less than a lattice
spacing.
By the same argument, we may also neglect the
closed loops (vacuum bubbles) generated by the denominator of eq.~\eHTICF.

In this section I shall demonstrate how sums over
directed paths {\it in the uniform system} can be calculated exactly by
transfer matrix methods. The method can also be generalized to random
systems, providing an algorithm for summing all paths in polynomial time.
For ease of visualization, I shall demonstrate this method in
two dimensions; the results are easily generalizable to higher dimensions.
Also to emphasize the general features of the transfer matrix method,
we shall compare and contrast the behavior of correlations along the axis
and the diagonal of the square lattice.

To calculate the correlation function $\langle\sigma_{0,0}\sigma_{0,t}\rangle$,
on
a non-random square lattice, we shall focus on directed paths oriented along
the main axis of the square. These paths are specified by a set of transverse
coordinates $(x_0, x_1, x_2, \cdots, x_t)$, with $x_0=x_t=0$. Of course, there
is only one shortest path with all $x_i$ equal to zero, but we would like to
explore the corrections due to longer directed paths.  Consider the set
of quantities
\eqn\edefWxt{\left\langle x,t |W| 0,0 \right\rangle=
{\rm sum~over~paths~from~}(0,0){\rm ~to~}(x,t)\equiv W(x,t)\quad.}
The calculation of $W(x,t)$ is easily accomplished by taking
advantage of its {\it Markovian} property: Each step of a path
proceeds from its last location and is  independent of the previous
steps. Hence $W$ can be calculated {\it recursively} from,
\eqn\eWaRR{\eqalign{W(x,t+1) & = \tau\left[W(x,t) + \tau\left(W(x-1,t) +
W(x+1,t)\right) +
\CO(\tau^2)\right]\cr
& \equiv \sum\nolimits_{x'}\ \langle x|T|x'\rangle\ W(x',t)\quad,}}
where we have introduced a {\it transfer matrix},
\eqn\eTMa{\langle x|T|x'\rangle = \tau\delta_{x,x'} +
\tau^2\left(\delta_{x,x'+1}
+ \delta_{x,x'-1}\right) +\CO(\tau^3)\quad.}

If we treat the values of $W$ at a particular $t$ as a vector, eq.~\eWaRR\ can
be iterated as,
\eqn\eWaTt{\underline W (t) = T\underline W(t-1) = \cdots = T^t\ \ \underline W
(0)\quad,}
starting from
\eqn\eWzero{\underline W(0) = \pmatrix {\vdots\cr \tau\cr 1\cr
\tau\cr\vdots\cr}\quad.}
The calculations are simplified by diagonalizing the matrix $T$, using the
Fourier basis
$ \left\langle x|q \right\rangle =  e^{iq \cdot x}/\sqrt{N}$, as
\eqn\eTMaq{ {T(q) = \tau  \left(1 + 2\tau\cos q  + \cdots\right)
= \tau \exp\left[{2\tau \left(1-{q^2\over 2} + \cdots\right)}\right]}\quad.}
In this basis, $W$ is calculated as
\eqn\eWta{\eqalign{
W(x,t) &= \langle x|T^t|0\rangle = \sum_q\langle x |q\rangle T(q)^t
\langle q|0\rangle\cr
& =  \tau^t e^{2\tau  t}\quad \int {dq\over 2\pi}
\exp\left[{iqx - q^2\tau  t+\cdots}\right]\cr
& =\exp\left[t\left(\ln\tau  + 2\tau  +\CO(\tau ^2)\right)\right]\times
{1\over\sqrt{4\pi\tau t}} \exp \left[ -{x^2\over 4\tau  t}\right]}\quad.}
The result is proportional to a gaussian form in $x$ of width $\sqrt{2\tau t}$.
The
exponential decay with $\xi^{-1}=\ln(1/\tau)-2\tau+\CO(\tau^2)$ at
$x=0$ is accompanied by a subleading $1/\sqrt{t}$.

The corresponding calculation of paths along the diagonal, contributing to
 $\langle\sigma_{0,0}\sigma_{0,t}\rangle$, is even simpler.
(Note that the $t$ and $x$ axes are rotated by 45$^\circ$ compared to the
previous example.) At each step the
path may proceed up or down, leading to the recursion relation
\eqn\eWdRR{W(x,t+1) = \tau  \left(W(x-1,t) + W(x+1,t)\right )\equiv
\sum_{x'}\langle x|T|x'\rangle W(x',t)\quad,}
with the transfer matrix
\eqn\eTMd{\langle x|T|x' \rangle=\tau \left(\delta_{x,x' +1} +
\delta_{x,x' -1}\right)\quad\Longrightarrow\quad T(q) = 2\tau \cos q\quad.}
The calculation of $W$ proceeds as before,
\eqn\eWtd{\eqalign{W(x,t) & = \langle x|T^t|0\rangle = \sum\nolimits_q\langle
x|q\rangle T
(q)^t\ \langle q|0\rangle\cr
& =	\int {dq\over 2\pi} (2\tau )^t (\cos q)^te^{iqx} \cr
&\approx  (2 \tau )^t \times{1\over\sqrt{2\pi t}} \exp\left[-{x^2\over
2t}\right]},}
where the final result is obtained by a saddle point evaluation of the
integral,
essentially replacing $\cos^t q$ with $\exp\left( -q^2t/2 \right)$.

The similarity between eqs.~\eWta\ and \eWtd\ is apparent. Note that in both
cases
the leading exponential decay is determined by $T(q=0)$, i.e.
\eqn\eGSD{W (0,t)\ \approx\  \lambda_{\rm max}^t \ =\ T(q = 0)^t\quad.}
This is an example of the dominance of the largest eigenvalue in the product of
a large number of matrices. There is a corresponding ground state dominance
in the evolution of quantum systems. The similarities become further apparent
by taking the {\it continuum limit} of the recursion relations, which are
obtained
by regarding $W(x,t)$ as a smooth function, and expanding in the derivatives.
{}From eq.~\eWaRR, we obtain
\eqn\eWaCL{W+{\partial W\over\partial t}+\cdots=\ \tau W + \tau^2
\left(2W + {\partial^2 W\over \partial x^2}  +\cdots\right)\quad,}
while eq.~\eWdRR\ leads to
\eqn\eWdCL{W+{\partial W\over\partial t}+\cdots= 2 \tau  W + \tau
{\partial^2W\over
\partial x^2} + \cdots\quad.}
For large $t$, the function $W$ decays slowly for adjacent points in the
$x$ direction, and it is justified to only consider the lowest order
derivatives with respect to $x$. The decay factor along the $t$ direction is,
however, quite big and we shall keep track of all derivatives in this
direction, leading to
\eqn\eWaCtx{e^{\partial_t}W=
\tau\exp\left[2\tau+\tau\partial_x^2+\cdots\right]W\quad,}
and
\eqn\eWdCtx{e^{\partial_t}W=2\tau\exp\left[{1\over2}\partial_x^2
+\cdots\right]W\quad,}
respectively. Both equations can be rearranged
(and generalized in higher dimensions) into the differential form,
\eqn\eWCL{{\partial W\over \partial t} = -{W\over \xi(\theta)} +
\nu (\theta) \nabla^2 W\quad,}
where $\xi(\theta)$ and $\nu(\theta)$ are the orientation dependent correlation
length and dispersion coefficient. Eq.~\eWCL\ can be regarded as a diffusion
equation in the presence of a sink, or an imaginary--time Schr\"odinger
equation.

It is of course quite easy to integrate this linear equation to reproduce the
results in eqs.~\eWta\ and \eWtd. However, it is also possible\rFeynman\
to express the solution in the form of a continuous {\it path integral}. The
solution is trivial in Fourier space,
\eqn\eWDtq{{\partial W(q)\over \partial t} = -(\xi^{-1} + \nu q^2) W\quad
\Longrightarrow\quad W(q,t+\Delta t) = e^{-(\xi^{-1} + \nu q^2)\Delta t}\
W(q,t)\quad,}
while in real space,
\eqn\eWDtx{\eqalign {W(x,t + \Delta t)	& =	\int {dq\over 2\pi} e^{iqx}
e^{-(\xi^{-1} + \nu q^2) \Delta t} W(q,t)\cr
& =	\int {dq\over 2\pi} e^{iqx}e^{(-\xi^{-1} - \nu q^2)\Delta t }
\int dx_t  e^{-iqx_t } W(x_t , t)\cr
& =	\int dx_t \exp\left[-{\Delta t\over \xi} -{\left(x-x_t \right)^2\over
4\nu\Delta t}\right]
\ W(x_t , t)\cr
& =	\int d x_t \exp\left[-{\Delta t\over \xi}+{\Delta t\over 4\nu}
{\left({ x_{t+\Delta t} - x_t\over\Delta t}\right)^2}\right]
W\left(x_t, t\right)}\quad,}
which is just a continuum version of eqs.~\eWaRR\ and \eWdRR. We
can subdivide the interval $(0,t)$ into $N$ subintervals of length $\Delta
t=t/N$.
In the limit of $N\to\infty$, recursion of eq.~\eWDtx\ gives
\eqn\eUPI{W(x,t) =\int_{(0.0)}^{(x,t)} \CD x(t')\exp\left[
\int^t_0 dt'\left(-{1\over \xi(\theta)} - {\dot x^2\over
4\nu(\theta)}\right)\right]\quad,}
where $\dot x= dx/dt'$, and the integration is over all functions $x(t')$.

It is instructive to compare the above path integral with the partition
function of
a string stretched between $(0,0)$ and $(x,t)$,
\eqn\eZstring{\eqalign{Z(x,t) & =	\int_{(0,0)}^{(x,t)} \CD x(t')
\exp\left[-\beta\sigma\int^t_0 dt'\sqrt{1 + \dot x^2} \right]\cr
& = \int_{(0,0)}^{(x,t)} \CD x(t') \exp\left[-\int^t_0 dt' \left(\beta\sigma +
{\beta\sigma\over 2} \dot x^2 + \cdots\right)\right]\cr}\quad,}
where $\sigma$ is the {\it line tension}. Whereas for the string
$\xi^{-1}=(2\nu)^{-1}=
\beta\sigma$, in general due to the anisotropy of the lattice these quantities
need not be equal. By matching solutions at nearby angles of $\theta$ and
$\theta+d\theta$, it is possible to obtain a relation between
$\xi^{-1}(\theta)$
and $\nu^{-1}(\theta)$. (For a similar relation in the context of interfaces of
Ising models, see ref.~\ref
\rMEF{M. E. Fisher, J. Stat. Phys. {\bf 34}, 667 (1984).}.)
However, $\xi(\theta)^{-1}$ calculated from the shortest paths only is
singular along the axis $\theta=0$. This is why to calculate the parameter
$\nu$ along this direction it is necessary to include longer directed paths.

\section{Moments of the Correlation Function}

We now return to the correlation functions in the presence of random bonds.
In the high temperature limit, we can still set
\eqn\eHTCF{W(x,t)\equiv \langle\sigma_{0,0}\sigma_{x,t}\rangle=
\sum_P \prod_{i=1}^t\tau_{Pi}\quad,}
where the sum is over all the diagonally oriented
directed paths $P$ from $(0,0)$ to $(x,t)$, and the
$\tau_{Pi}$ denote factors of $\tanh K$ encountered for the random bonds
along each path. The $\tau_i$ are random variables, independently chosen for
each bond. We shall assume that the probability distribution $p(\tau)$ is
narrowly distributed around a mean value $\overline{\tau}$ with width
$\sigma$. Clearly, $W(x,t)$ is itself a random variable and we would like
to find its probability distribution. Rather than directly calculating $p(W)$,
we shall first examine its moments $\overline{W^n}$.

Calculation of the first moment is trivial: Each factor of $\tau_i$ occurs at
most once in eq.~\eHTCF, and hence after averaging,
\eqn\eHTCFa{\overline{W(x,t)}\equiv \langle\sigma_{0,0}\sigma_{x,t}\rangle=
\sum_P\overline{\tau}\,^t\quad.}
This is precisely the sum encountered in a non-random system, with
$\overline{\tau}$ replacing $\tau$. For example, along the square diagonal,
\eqn\eaWl{\overline{W(x,t)} \approx  (2 \overline{\tau} )^t
\times{1\over\sqrt{2\pi t}} \exp\left[-{x^2\over 2t}\right]\quad,}
and in general, in the continuum limit,
\eqn\eaWlC{{\partial\overline W\over \partial t} = -{\overline W\over\xi} +
\nu\  {\partial^2\overline W\over \partial x^2}\quad.}

For the calculation of the second moment we need to evaluate
\eqn\eWWa{\overline {WW} =\sum\limits_{P,P'}
\prod\limits_{i=1}^t\overline{\tau_{Pi}\tau_{P'i}}\quad.}
For a particular $i$, there are two possible averages depending on whether or
not the two paths cross the same bond,
\eqn\etta{\overline{\tau_{Pi}\tau_{P'i}}=\cases{\overline {\tau}\,^2 &
if	$Pi\ne P'i$\cr \overline {\tau^2}	&	if	$Pi= P'i$\cr}\quad.}
Since $\overline{\tau^2} > \overline{\tau}\,^2$ there is an additional
weight for paths that intersect compared to those that don't. This can
be regarded as an attraction between the two paths, represented by a
Boltzmann weight,
\eqn\eAttract{U = {\overline {\tau^2}\over \overline{\tau}\,^2} =
{\overline{ \tau}\,^2 + \sigma^2\over \overline \tau\,^2} = 1 + {\sigma^2\over
\overline {\tau}\,^2}\approx e^{\sigma^2/ \overline {\tau}\,^2}\quad.}
Including the attraction, the recursion relation for $\overline {WW} $ is,
\eqn\eWWRR{W_2(x_1, x_2,t) \equiv\overline{W(x_1,t)W(x_2,t)}=
\sum\nolimits_{x'_1 x'_2}\langle x_1x_2|T_2|x_1'x'_2\rangle
\ W(x_1',x'_2, t-1)\quad,}
with the two body transfer matrix
\eqn\etwoTM{\langle x_1x_2|T_2|x_1'x'_2\rangle=\overline{\tau}^2
\left(\delta_{x_1,x_1' +1} +
\delta_{x_1,x_1' -1}\right)\left(\delta_{x_2,x_2' +1} +
\delta_{x_2,x_2' -1}\right)
\left( 1+(U-1) \delta_{x_1,x_2}\delta_{x_1',x_2'}\right).}

The significance of the attraction in eq.~\eAttract\ is as follows: In the
random
system the paths prefer to pass through regions with particularly favorable
values of $\tau$. After performing the quench averaging the paths go
through a uniform medium. The tendency for the original paths to bunch up
through
favorable spots is instead mimicked by the uniform attraction which tends
to bundle together the paths representing the higher moments.

In the continuum limit, eq.~\eWWRR\ goes over to a differential equation of the
form,
\eqn\eWWC{{\partial W_2(x_1,x_2,t)\over \partial t}= -{2W_2\over\xi}
+ \nu{\partial^2W_2\over \partial x_1^2}+\nu{\partial^2W_2\over\partial x^2_2}
+
u\ \delta(x_1-x_2)W_2\equiv -H_2W_2\quad,}
with $u\approx \sigma^2/\overline\tau\,^2$.
Alternatively, we could have obtained eq.~\eWWC\ from the continuum version
of the path integral,
\eqn\eWWPI{\eqalign{W_2(x_1,x_2,t)=\int_{(0.0)}^{(x_1,t)} \CD x_1(t')
\int_{(0.0)}^{(x_2,t)} \CD x_2(t')
\exp\left[\int^t_0 dt'u\delta\left(x_1(t')-x_2(t')\right)\right]&\cr
\exp\left[\int^t_0 dt'\left(-{1\over \xi} - {\dot x_1^2\over
4\nu}\right)\right]
\exp\left[\int^t_0 dt'\left(-{1\over \xi} - {\dot x_2^2\over
4\nu}\right)\right]&
}\quad.}
Formally integrating eq.~\eWWC\ yields $W_2\propto\exp(-tH_2)$, which can be
evaluated in the basis of eigenvalues of $H_2$ as
\eqn\eWWe{W_2(x_1,x_2,t)= \langle x_1 x_2|T_2^t|00\rangle
=\sum_m\langle x_1 x_2|m\rangle e^{-\epsilon_m t}\langle m|00\rangle
\,{\approx\atop t\to\infty}\,e^{-\varepsilon_0t}\quad,}
where $ \left\{ \varepsilon_m \right\}$ are the eigenenergies of $H_2$,
regarded
as a quantum Hamiltonian. The exponential growth of $W_2$ for $t\to\infty$
is dominated by the ground state $\varepsilon_0$.

The two body Hamiltonian depends only on the relative separation of the
two particles. After transforming to the center of mass coordinates,
\eqn\eCOMc{\left\{\eqalign{r=&x_1-x_2\cr R=&(x_1+x_2)/2}\right.
\qquad\Longrightarrow\qquad
\partial_1^2 + \partial^2_2 = {1\over 2}\partial_R^2 + 2 \partial^2_r\quad,}
the Hamiltonian reads,
\eqn\eHCOM{H_2 = {2\over\xi} -{\nu\over 2}\partial_R^2 - 2\nu\partial_r^2 -
u\delta(r)\quad.}
The relative coordinate describes a particle in a delta--function potential,
which
has a ground state wavefunction
\eqn\ePsit{\psi_0(r,R) \propto e^{-\kappa |r|}\quad.}
The value of $\kappa$ is obtained by integrating $H_2\psi_0$ from $-\epsilon$
to $\epsilon$, and requiring the discontinuity in the logarithmic derivative of
$\psi_0$ to match the strength of the potential; hence
\eqn\eKappa{-2\nu(-\kappa -\kappa) = u\qquad	\Longrightarrow\qquad
\kappa = {u\over 4\nu} \approx {\sigma^2\over 2 \overline \tau\,^2}\quad.}
The ground state energy of this two particle system is
\eqn\eGStwo{\varepsilon_0 = + {2\over \xi} - 2\nu \kappa^2
\approx {2\over \xi} - {u^2\over 8\nu}\quad.}
The inequality,
\eqn\eWWv{\overline {W^2(t)} =\exp\left[ -{2t\over \xi} + {u^2t\over 8
\nu}\right] =
\overline {W(t)}\,^2\exp\left({u^2t\over 8\nu}\right) \gg\ \overline
{W(t)}\,^2\quad,}
implies that the probability distribution for $W(t)$ is quite broad, and
becomes
progressively wider distributed as $t\to\infty$.

Higher moments of the sum are obtained from
\eqn\eWWn{\overline {W^n} =\sum\limits_{P_1,\cdots,P_n}
\prod\limits_{i=1}^t\overline{\tau_{P_1i}\cdots\tau_{P_ni}}\quad.}
At a particular ``time" slice there may or may not be intersections amongst
the paths. Let us assume that $\tau$ is Gaussian distributed with a mean
$\overline{\tau}$, and a {\it narrow width} $\sigma$; then,
\eqn\etaum{\eqalign {\overline {\tau^m} \approx &\int {dx\ x^m\over
\sqrt{2\pi\sigma^2}}
\exp\left[-{(x-\overline \tau)^2\over 2\sigma^2}\right]\qquad\qquad
({\rm set}\,\,x = \overline \tau + \epsilon\,\,{\rm and~expand~in~}\epsilon)\cr
\hfill \approx &\int {d\epsilon\over \sqrt{2\pi\sigma^2}}
\left(\overline \tau\,^m + m \,\overline \tau\,^{m-1}\epsilon +
{m(m-1)\over 2}\, \overline \tau\,^{m-2}\epsilon^2 + \cdots\right)
\exp\left[-{\epsilon^2\over 2\sigma^2}\right]\cr
\hfill \approx &\, \overline \tau\,^m + {m(m-1)\over 2} \, \overline
\tau\,^{m-2}\sigma^2
+ \cdots\approx \, \overline \tau\,^m \left(1+ {m(m-1)\over 2}
{\sigma^2\over \, \overline \tau\,^2}+\cdots\right)\cr\approx&
\, \overline \tau\,^m \exp\left[{m(m-1)\over 2} u\right]\cr}\quad.}
Since there are $m(m-1)/2$ possibilities for pairing $m$ particles,
the above result represents the Boltzmann factor for a pairwise attraction of
$u$ for particles in contact. Since $\tau$ is bounded by unity, the
approximations
leading to eq.~\etaum\ must break down for sufficiently large $m$.  This
implies the
presence of three and higher body interactions. Such interactions are usually
of higher order and can be safely ignored. For a discussion of these higher
order interactions in a similar context see ref.~\ref
\rLassig{M. L\"{a}ssig, Phys. Rev. Lett. {\bf 73}, 561 (1994).}.

The continuum version of the resulting path integral is
\eqn\ePIn{\eqalign{W_n(x_1,\cdots,x_n,t)\equiv&\overline {W(x_1,t)\cdots
W(x_n,t)} =
\int^{(x_1,\cdots,x_n,t)}_{(0,0,\cdots,0)} \CD x_1(t')\cdots \CD x_n(t')\cr
&\exp\left[-{nt\over\xi} - \int^t_0 dt' \left( \sum_\alpha {\dot
x^2_\alpha\over 4\nu}
- {u\over 2}\sum_{\alpha \ne \beta}\delta\left(x_\alpha (t') - x_\beta
(t')\right)
\right)\right]}\quad,}
and evolves according to
\eqn\eWndt{{\partial W_n\over \partial t} = -{nW_n\over\xi} +
\nu\sum_{\alpha = 1}^n\ {\partial^2W_n\over\partial x_\alpha^2} +
{u\over2}\sum_{\alpha\ne\beta}
\delta\left(x_\alpha(\tau)-x_\beta(\tau)\right)W_n\equiv -H_nW_n\quad.}
The asymptotic behavior of $W_n$ at large $t$ is controlled by the ground state
of
$H_n$. The corresponding wavefunction is obtained by a simple
{\it Bethe Ansatz} \ref
\rBethe{H. B. Thacker, Rev. Mod. Phys. {\bf 53}, 253 (1981).},
which generalizes eq.~\ePsit\ to
\eqn\eBAn{\psi_0(x_1,\cdots,x_n) \propto\exp\left[-{\kappa\over
2}\sum_{\alpha\ne\beta}
\left|x_\alpha - x_\beta \right|\right]\qquad{\rm with}\quad
\kappa = {u\over 4\nu}\quad.}
For each ordering of particles on the line the wave function can be expanded as
\hbox{$\psi_0 \propto \exp\left[{\kappa_\alpha x_\alpha}\right]$}, with the
``momenta"
$\kappa_\alpha$ getting permuted for different orderings. For example, if
$x_1< x_2 < \cdots < x_n$, the momenta are
\eqn\enstring{\kappa_\alpha=\kappa\left[ 2\alpha-(n+1) \right]\quad,}
forming a so called $n$--{\it string}. The kinetic energy is proportional to
\eqn\esumns{\eqalign{S = &\sum\nolimits_{\alpha = 1}^n \big[ 2 \alpha - (n+1)
\big]^2
 = \sum\nolimits_{\alpha = 1}^n \left[(n+1)^2 -
4\alpha(n+1) + 4\alpha^2\right ]\cr
= &	n(n+1)^2 - 4(n+1) \cdot {n(n+1)\over 2} + {4\ n(n+1)(2n+1)\over 6}\cr
=	&	n(n+1)\left[-(n+1) + {2(2n+1)\over 3}\right] = {n(n+1)(n-1)\over 3}},}
leading to the ground state energy
\eqn\eGSn{\varepsilon_0 = {n\over\xi} -\nu\sum_{\alpha = 1}^n\kappa_\alpha^2
= {n\over\xi} - {\nu \kappa^2\over 3} n(n^2-1)\quad.}
Thus the asymptotic behavior of moments of the sum has the form
\eqn\eWnasy{\lim_{t\to\infty}\overline {W^n(t)} =\exp\left[- {nt\over \xi} +
{n(n^2-1)\nu\kappa^2t\over 3 }\right] =
\overline {W(t)}\,^n\exp\left( {n(n^2-1)u^2t\over 48 \nu}\right) \quad.}

\section{The Probability Distribution in Two Dimensions}

It is tempting to use eq.~\eWnasy\ in conjunction with
\eqn\elnWcum{\lim_{n\to0}\ln\left(\overline {W^n(t)} \right)=
n\, \overline {\ln W} +
{n^2\over 2}\, \overline {(\ln W)^2_c} + \cdots +
{n^p\over p!}\, \overline {(\ln W)^p_c} + \cdots\quad,}
to read off the
cumulants for the probability distribution for $\ln W$. The key point is the
absence of the $n^2$ term and the presence of the $n^3t$ factor in the
exponent of eq.~\eWnasy, suggesting a third cumulant, and hence
fluctuations in $\ln W$ that grow as $t^{1/3}$ \ref
\rMKnpb{M. Kardar, Nucl. Phys. B {\bf 290}, 582 (1987).}.
However, as discussed
before, there are subtleties in trying to deduce a probability distribution
from
the knowledge of its moments which we need to consider first. Since $W(t)$
is bounded by unity, eq.~\eWnasy\ cannot be valid for arbitrarily large $n$.
Our first task is to identify the crossover point $n^*$ beyond
which this result is no longer correct.

Eq.~\eGSn\ is obtained for the ground state of $n$ particles subject
to a two body interaction in the continuum limit.
A simple argument can be used to understand the origin of the $n^3$
term in the energy, as well as the limitations of the continuum approach.
Let us assume that the $n$ particles form a bound state of size $R$. For
large $n$, the energy of such a state can be estimated as
\eqn\eGSest{\varepsilon\approx{ n\over\xi}+{\nu n \over R^2}-{u n^2\over
R}\quad.}
A variational estimate is obtained by minimizing the above expression
with respect to $R$, resulting in $R\propto \nu/(u n)$ and $\varepsilon
\propto u^2 n^3/\nu$. The size of the bound state decreases with
increasing $n$, and the continuum approximation breaks down when
it becomes of the order of the lattice spacing for $n^*\propto \nu/u\approx
\overline{\tau}\,^2/\sigma^2$. For $n\gg n^*$ all the paths collapse
together and
\eqn\eWlargen{\lim_{n\to\infty}\overline{W^n(t)}\simeq
\left( 2\overline{\tau^n}\, \right)^t\quad.}
This asymptotic behavior is {\it non-universal} and depends on the extreme
values of the local probability distribution for $\tau$.
Depending on the choice of parameters, $n^*$ can be large or small.
However, as discussed in the context of the one dimensional problem,
its value controls only the relative importance of the tail and the bulk
of the probability distribution for $\ln W$. The behavior of the bulk of
the distribution is expected to be {\it universal}. The crossover at $n^*$
is explicitly demonstrated in a simpler model in ref.~\ref
\rMKMoments{E. Medina and M. Kardar, J. Stat. Phys. {\bf 71}, 967 (1992).}.

Another important consideration is the order of limits. Eq.~\eWnasy\ is
obtained
by taking the $t\to\infty$ limit at fixed $n$, while the cumulant series in
eq.~\elnWcum\ relies on an expansion around $n\to0$ for fixed $t$. The two
limits do not commute. In fact, we would naively deduce from eq.~\eWnasy\
that only the third cumulant of $\ln W$ is non-zero. This is incorrect as
it is impossible to have a probability distribution with only a third cumulant
\ref\rFell{W. Feller, {\it An Introduction to Probability Theory and its
Applications} (Wiley, 1971).}.
The correct procedure \rMKMoments\
is to {\it assume} that the {\it singular} behavior associated
with $n\to0$ and $t\to\infty$ is described by a scaling function of the
form $g_s(nt^\omega)$. (This is similar to the singularity of the free
energy at a critical point with $n^{-1/\omega}$ playing the role of
a correlation length.)
If $t\to\infty$ at fixed $n$, extensivity of the free energy of
the $n$ particle system forces $\ln\overline{W^n(t)}$ to be proportional to
$t$.  At the other limit of $n\to0$ at fixed $t$, the result is a power series
in $n$, i.e.
\eqn\ecumlim{\ln\overline{W^n(t)}=ant+g_s(nt^\omega)=\cases{
ant+\rho n^{1/\omega}t& for $t\to\infty$ at fixed $n$\cr
ant+g_1nt^\omega+g_2\left( nt^\omega \right)^2+\cdots&
for fixed $t$ as $n\to0$}.}
Note that I have included a {\it non-singular} term, $ant$.
Similar considerations have been put forward in ref.~\ref
\rFY{R. Friedberg and Y.-K. Yu, Phys. Rev. E {\bf 49}, 4157 (1994).}.
Comparison  with eq.~\eWnasy\ gives $\omega=1/3$, and we can read off the
cumulants of $\ln W$ as
\eqn\elncumt{\left\{\eqalign{
\overline{\ln W(t)}=&at+g_1\,t^{1/3}\cr
\overline{\ln W^2(t)_c}=&2g_2\,t^{2/3}\cr
\vdots&\cr
\overline{\ln W^p(t)_c}=&p!g_p\,t^{p/3}\cr
}\right.\quad.}

The existence of $t^{1/3}$ corrections to the quench averaged value of
$\ln W(t)$ was first suggested by Bouchaud and  Orland \ref
\rBO{J.-P. Bouchaud and H. Orland, J. Stat. Phys. {\bf 61}, 877 (1990).}\
and has been numerically verified \ref
\rMKqie{E. Medina and M. Kardar, Phys. Rev. B {\bf 46}, 9984 (1992).}.
The $t^{2/3}$ growth of the variance of the probability distribution was
obtained by Huse and Henley \ref
\rHH{D. A. Huse and C. L. Henley, Phys. Rev. Lett. {\bf 54}, 2708 (1985).}\
in the context of interfaces of Ising models at zero temperature where an
optimal path dominates the sum. The results remain valid at finite temperatures
\ref\rKcomment{M. Kardar, Phys. Rev. Lett. {\bf 55}, 2923 (1985).}.
Simulations are performed by implementing the transfer matrix
method numerically. For example, along the diagonal of the square
lattice, the recursion relation
\eqn\eTMran{W(x,t+1)=\tau_{x,t,-}W(x-1,t)+\tau_{x,t,+}W(x+1,t)\quad,}
is iterated starting from $W(x,0)=\delta_{x,0}$.
The random numbers $\tau_{x,t,\sigma}$
are generated as the iteration proceeds.  The memory requirement (the arrays
$W(x)$) depend on the final length $t$; each update requires $t$ operations,
and the total execution time grows as $t^2$. Thus for a given realization
of randomness, {\it exact} results are obtained in {\it polynomial} time.
Of course the results have to be averaged over many realizations of
randomness. The typical values of $t$ used in the transfer matrix simulations
range from $10^3$ to $10^4$, with $10^2$ to $10^3$ realizations.
Calculating higher cumulants becomes progressively more difficult. The
existence
of the third cumulant was verified by Halpin--Healy \ref
\tTHHcum{T. Halpin--Healy, Phys. Rev. A {\bf 45}, 638 (1992);
J.  Krug, P. Meakin and T. Halpin--Healy, Phys. Rev. A, {\bf 45} 638 (1992).}.
A fourth cumulant, growing as $t^{4/3}$ was observed by Kim et al.\ \ref
\rKMB{J. M. Kim, M. A. Moore, and A. J. Bray, Phys. Rev. A {\bf 44}, 2345
(1991).}.
Starting from the replica result,
Zhang \ref\rZ{Y.-C. Zhang, Phys. Rev. Lett. {\bf 62}, 979 (1989);
Europhys. Lett. {\bf 9}, 113 (1989); J. Stat. Phys.  {\bf 57}, 1123 (1989).}\
proposed an analytical form, $p(\ln W,t)\sim
\exp(-a|\ln W-\overline{\ln W} |^{3/2}/t^{1/2})$.
While this form captures the correct scaling of free energy
fluctuations, it is symmetric about the average value precluding
the observed finite third cumulant. This
deficiency was remedied by Crisanti et al.\ \ref
\rCPSV{A. Crisanti, G. Paladin, H.-J. Sommers, and A. Vulpiani,
J. Phys. I (France) {\bf 2}, 1325 (1992).}\
who generalized the above probability to one with different coefficients
$a_\pm$ on the two sides of the mean value.

So far, we focused on the asymptotic behavior of $W(x,t)$ at large $t$,
ignoring the dependence on the transverse coordinate. For the pure
problem, the dependence of $W$ on the transverse coordinate is a
Gaussian, centered at the origin, with a width that grows as $t^{1/2}$.
The full dependence is obtained in the pure problem by including
the band of eigenvalues with energies close to the ground state.
Unfortunately, determining the appropriate eigenvalues for the
interacting problem is rather difficult. In addition to the eigenvalues
obtained by simply multiplying eq.~\eBAn\ by $\exp\left[ iq\left(
x_1+\cdots+x_n\right) \right]$, there are other states with broken
replica symmetry \ref
\rParisi{G. Parisi, J. Physique (France) {\bf 51}, 1595 (1990).}.
A treatment by Bouchaud and Orland \rBO\ includes some of the
effects of such excitations but is not fully rigorous. It does predict
that the extent of transverse fluctuations grows as $t^\zeta$ with
$\zeta=2/3$ as observed numerically\refs{\rHH,\rKcomment}. There is
in fact a relation between the exponents $\zeta$ and $\omega$
which follows from simple physical considerations \rHH: By analogy with
a string, the energy to stretch a path by a distance $x$ grows as
$x^2/t$. The path wanders away from the origin, only if the cost of
this stretching can be made up by favorable configurations of
bonds. Since the typical fluctuations in (free) energy at scale $t$
grow as $t^\omega$, we have
\eqn\eHHid{{x^2\over t}\propto t^\omega\qquad\Longrightarrow\qquad
\omega=2\zeta-1\quad.}
This relation remains valid in higher dimensions and has been verified
in many numerical simulations. The first (indirect) proof of $\omega=1/3$
was based on a replica analysis of a problem with many interacting paths \ref
\rKNelson{M. Kardar and D. R. Nelson, Phys. Rev. Lett. {\bf 55}, 1157 (1985).}.
It was soon followed by a more direct proof \ref
\rFHH{D. A. Huse, C. L. Henley, and D. S. Fisher, Phys. Rev. Lett. {\bf 55},
2924 (1985).}\
based on a completely different
approach: the Cole--Hopf transformation described in next section.

\section{Higher Dimensions}

The approach described in the previous sections is easily generalized to
higher dimensions. The directed path in $d=D+1$ is described by $\vec{x}(t)$,
where $\vec{x}$ is a $D$ dimensional vector. The recursion relation of
eq.~\eTMran\ is generalized to
\eqn\eTMrand{W(\vec{x},t+1)=\sum_{i=1}^d
\tau_{\vec{x }-\vec{e}_i,t}W\left(\vec{x }-\vec{e}_i,t\right)\quad,}
where $\vec{e}_i$ are unit vectors. The recursion relation is easily iterated
on a computer, but the memory requirement and execution time now grow
as $t^D$ and $t^{D+1}$ respectively. The continuum limit of this recursion
relation is
\eqn\edWdt{{\partial W(\vec{x}, t)\over \partial t} =
-{W\over \xi} + \nu \nabla^2 W+\mu(\vec{x},t)W\quad,}
where $\mu(\vec{x},t)$ represents the fluctuations of $\tau(\vec{x},t)$ around
its average. Thus it has zero mean, and a variance
\eqn\evarmu{\overline{\mu(\vec{x},t)\mu(\vec{x}\,',t')}=\sigma^2\delta^D\left(
 \vec{x}-\vec{x}\,'\right)\delta(t-t')\quad.}
(In a more general anisotropic situation, eq.~\edWdt\ has to be generalized to
include different diffusivities $\nu_\alpha$ along different directions. Such
anisotropy is easily removed by rescaling the coordinates $x_\alpha$.)

Eq.~\edWdt\ can be regarded as the imaginary time Schr\"odinger equation
for a particle in a random {\it time dependent } potential. It can be
integrated
to yield the continuous path integral
\eqn\ePId{W(\vec{x}, t)=\int^{(\vec x,t)}_{(0,0)} \CD\vec x(t')
\exp\left[ -\int^t_0dt'\left({1\over\xi}+{ \dot {\vec x}\,^2\over 4\nu}
-\mu\left(\vec{x}(t'),t'\right)\right) \right]\quad,}
describing the fluctuations of a directed polymer in a random medium (DPRM)
\ref\rKZhang{M. Kardar and Y.-C. Zhang, Phys., Rev. Lett. {\bf 58},
2087 (1987).}.
The $n^{\rm th}$ moment of $W$ is computed by replicating the above path
integral and averaging over $\mu(\vec{x},t)$. It generalizes eq.~\eWWPI\ to
\eqn\eWnPI{\eqalign{W_n(\left\{  \vec{x}_\alpha \right\},t)=&
\int_{(\left\{  \vec{0} \right\},0)}^{(\left\{  \vec{x}_\alpha \right\},t)} \CD
\vec x_1(t')
\cdots \CD \vec x_n(t')\cr
&\exp\left[-\int^t_0dt'\left(\sum_{\alpha} {1\over\xi}+{\dot {\vec
x}_\alpha^2\over 4\nu}
- {u\over 2}\sum_{\alpha\ne\beta}\delta^D\left(\vec x_\alpha(t') -
\vec x_\beta(t')\right)\right)\right]}\quad,}
with $u\propto \sigma^2$. The differential equation governing the evolution of
$W_n(t)$ is,
\eqn\edWndt{{\partial W_n\over \partial t} =
-{n\over\xi}W_n + \nu\sum_\alpha  \nabla^2_\alpha W_n
+ {u\over 2}\sum_{\alpha\ne\beta}
\delta^D(\vec x_\alpha - \vec x_\beta) W_n \equiv - H_n W_n\quad.}

Evaluating the asymptotic behavior of $W_n(t)$ requires knowledge of the ground
state energy of the Hamiltonian $H_n$. Unfortunately, the exact dependence of
the
bound state energy on $n$ is known only for $D=0$ ($\varepsilon\propto n(n-1)$)
and $D=1$ ($\varepsilon\propto n(n^2-1)$).  As discussed earlier, these two
results can then be used to deduce the behavior of the bulk of the probability
distribution for $\ln W(t)$. Elementary results from quantum mechanics tell us
that
an arbitrarily small attraction leads to the formation of a bound state in
$D\leq 2$,
but that a finite strength of the potential is needed to form a bound state in
$D>2$. Thus, in the most interesting case of $2+1$ dimensions we expect a
non-trivial probability distribution, while the replica analysis provides no
information
on its behavior. In higher dimensions, there is a phase transition between weak
and
strong randomness regimes. For weak randomness there is no bound state and
asymptotically $\overline{W^n(t)}= \overline{W(t)}\,^n$, indicating a sharp
probability distribution. This statement has also been established by more
rigorous methods \ref
\rImbrie{J. Imbrie and T. Spencer, J. Stat. Phys. {\bf 52}, 609 (1988).}.
There is another phase for strong randomness where the probability distribution
for $W(t)$ becomes broad. The resulting bound state has been analytically
studied in a $1/D$ expansion valid for large $D$ \ref
\rGold{Y. Y. Goldschmidt, Nucl. Phys. B {\bf 393}, 507 (1993).}.
The ground state wavefunction is rather complex, involving replica symmetry
breaking. Note that the phase transition in the probability
distribution of the correlation function occurs in the high temperature phase
of the
random Ising model. The implications of this phase transition for bulk
properties are not known. As the stiffness associated with line tension
decreases
on approaching the order/disorder phase transition of the Ising model, close
to this transition the probability distribution for $W(t)$ is likely to be
broad.

As one of the simplest models of statistical mechanics in random systems
(a ``toy" spin glass), the problem of DPRM
has generated considerable interest \ref
\rFisherH{D. S. Fisher and D. A. Huse, Phys. Rev. B {\bf 43}, 10728 (1991).}.
The model has been generalized to manifolds of arbitrary internal dimensions
in random media \ref
\THHrbf{T. Halpin-Healy, Phys. Rev. Lett. {\bf 62}, 442 (1989).},
and treated by functional RG methods \ref
\rBalentsF{L. Balents and D. S. Fisher, Phys. Rev. B {\bf 48}, 5949 (1993).}.
The same model has also been studied by a variational approach that
involves replica symmetry breaking \ref
\rMezardP{M. M\`ezard and G. Parisi, J. Physique I {\bf 1}, 809 (1991);
J. Phys. A {\bf 23}, L1229 (1990); J. Physique I {\bf 2}, 2231 (1992).}.
The latter is also exact in the $D\to \infty$ limit. Directed paths have
been examined on non-Euclidean lattices: In particular, the
problem can be solved exactly on the Cayley tree \ref
\rDerridaS{B. Derrida and H. Spohn, J. Stat. Phys. {\bf 51}, 817 (1988).},
where it has a transition between a ``free" and a glassy state.
There are also quite a few treatments based on a position space renormalization
group scheme \ref
\rDerridaPSRG{B. Derrida and R. B. Griffiths, Europhys. Lett. {\bf 8}, 111
(1989); J. Cook and B. Derrida, J. Stat. Phys. {\bf 57}, 89 (1989).}\
which becomes exact on a {\it hierarchical}  lattice \ref
\rHierarchical{See A. N. Berker S. Ostlund, J. Phys. C {\bf 12}, 4961 (1979);
and references therein.}. This lattice has no loops, and at the $m+1^{\rm th}$
level is constructed by putting together $2^{D}$ branches, each containing
two lattices of the $m^{\rm th}$ level. Starting from a set of random bonds at
the first level, the values of the sum $W(m=\log_2 t)$ are constructed
recursively
from
\eqn\eHier{W(m+1,\beta)=\sum_{\alpha=1}^{2^D}W(m,\alpha_1)W(m,\alpha_2)\quad,}
where the greek indices are used to indicate specific bonds for a particular
realization. Alternatively, these recursion relations can be used to
study the evolution of the probability distribution for $W$ \ref
\rRHSLP{S. Roux, A. Hansen, L. R. di Silva, L. S. Lucena, and R. B. Pandey,
J. Stat. Phys. {\bf 65}, 183 (1991).}.
The exponent $\omega\approx0.30$ for $D=1$ is not too far off from the
exact value of 1/3.

Additional information about the higher dimensional DPRM is obtained by
taking advantage of a mapping to the nonequilibrium problem of kinetic
roughening of growing interfaces. Using the Cole--Hopf transformation \ref
\rCole{E. Hopf, Comm. Pure Appl. Math. {\bf 3}, 201 (1950);
J. D. Cole, Quart. Appl. Math. {\bf 9}, 225 (1951).},
\eqn\eCole{W(\vec{x},t)=\exp\left[ -{\lambda h(\vec{x},t)\over 2\nu}
\right]\quad,}
eq.~\edWdt\ is transformed to the Kardar, Parisi, Zhang (KPZ) \ref
\rKPZ{M. Kardar, G. Parisi, and Y.-C. Zhang, Phys. Rev. Lett. {\bf 56}, 889
(1986).}\
equation,
\eqn\eKPZ{ {\partial {h}\over \partial t}={2\nu\over\lambda\xi}+\nu
\nabla^2 h-{\lambda\over 2}\left( \nabla h
\right)^2-{2\nu\over\lambda}\mu(\vec{x},t)\quad,}
describing the fluctuations in height $h(\vec{x},t)$ of a growing interface.
A dynamical renormalization group (RG) analysis at the one--loop level
\nref\rFNS{D. Forster, D. R. Nelson, and
M. J. Stephen, Phys. Rev. A {\bf 16}, 732 (1977).}\nref
\rMHKZ{E. Medina, T. Hwa, M. Kardar, and Y.-C. Zhang, Phys. Rev. A
{\bf 39}, 3053 (1989).}\refs{\rFNS,\rMHKZ}\
of this equation indicates that the effective coupling constant
$g=4\sigma^2/\nu$,
satisfies the rescaling relation
\eqn\egRG{{d{g} \over d\ell}=(2-D)g+{K_D(2D-3)\over D}g^2\quad,}
where $K_D$ is the $D$ dimensional solid angle divided by $(2\pi)^D$. The RG
equation merely confirms the expectations based on the replica analysis: there
is flow to strong coupling for $D\leq 2$, while there is a transition between
weak
and strong coupling behavior in higher dimensions.
The RG equation has been recently extended to two loops \nref
\rSunP{T. Sun and M. Plischke, Phys. Rev. E {\bf 49}, 5046 (1994).}\nref
\rFrey{E. Frey and U. C. T\"auber, Phys. Rev. E {\bf 50}, 1024 (1994).}
\refs{\rSunP,\rFrey}.
According to one calculation \rFrey, there is no stable strong coupling fixed
point for
$D\geq 2$, and the nature of this phase remains a mystery.

Since there are several comprehensive reviews of the KPZ equation \ref
\rKPZrevs{See, e.g., {\it Dynamics of Fractal Surfaces},
edited by F. Family and T. Vicsek (World Scientific,
Singapore, 1991); J. Krug and H. Spohn, in {\it Solids
Far From Equilibrium: Growth, Morphology and Defects}, edited by
 C. Godreche (Cambridge University Press, Cambridge, 1991).
A very enjoyable review article by T. Halpin--Healy and Y.-C. Zhang
is currently under press; and a book by A. L. Barabasi and H. E. Stanley
is also in preparation.},
I will not discuss its properties in any detail here. It suffices to say
that there are many numerical models of growth that fall in the
universality class of this equation. They are in complete agreement
with the exactly known results for $D=1$.
The estimates for the exponent $\zeta$ in higher dimensions are
$\zeta=0.624\pm 0.001$ for $D=2$ \ref
\rForrestT{B. M. Forrest and L.-H. Tang, Phys. Rev. Lett. {\bf 64}, 1405
(1990).}\
and $\zeta\approx 0.59$ for $D=3$ \ref
\rKBMa{J. M. Kim, A. J. Bray, and M. A. Moore, Phys. Rev. A {\bf 44}, 2345
(1991).}.
The numerical results in higher dimensions are consistent with an exponent
$\zeta$ that gets closer to 1/2 as $D\to\infty$. It is not presently known
whether
there is a finite upper critical dimension
\nref\r{M. A. Moore, T. Blum, J. P. Doherty, J.-P. Bouchaud, and P. Claudin,
preprint (1994).}\refs{\THHrbf,\r}\
beyond which $\zeta=1/2$ exactly.

\section{Random Signs}

So far we focused on nearest neighbor bonds $ \left\{ K_{ij} \right\}$, which
though
random, were all positive. For such couplings the ground state is uniform and
ferromagnetic. The study of low temperature states is considerably more
complicated
for the random {\it spin glass} which describes a mixture of ferromagnetic and
antiferromagnetic bonds. The competition between the bonds leads to
{\it frustration} effects, resulting in quite complicated landscapes for the
low
energy states\rBeyond. Here we shall explore the high temperature properties
of spin glass models. To focus on the effects of the randomness in sign, we
study
a simple binary probability distribution in which negative and positive bonds
{\it of equal} magnitude occur with  probabilities $p$ and $1-p$ respectively.

The computation of the high temperature series for the correlation function
(along the diagonal) proceeds as before, and
\eqn\eCFpm{W(\vec  x,t)\equiv \langle\sigma_{0,0}\sigma_{\vec  x,t}\rangle=
\tau^t\sum_P \prod_{i=1}^t\eta_{Pi}\quad,}
where $\tau$ indicates the fixed magnitude of $\tanh K$, while $\eta_{Pi}=\pm1$
are random signs. Since the elements of the sum can be both positive and
negative, the first question is whether the system maintains a coherence in
sign (at least for small $p$), i.e. what is the likelihood that the two spins
separated
by a distance $t$ have a preference to have the same sign. This question
can be answered only in one and high dimensions.

For the one dimensional chain the moments of $W(t)$ are easily calculated as
\eqn\eWnlpm{\overline{W^n(t)}=\tau^{nt}\times
\cases{(1-2p)^t& for $n$ odd,\cr 1& for $n$ even.}}
As all odd moments asymptotically decay to zero, at large distances
$W(t)$ is equally likely to be positive or negative. This is because the sign
of the
effective bond depends only on the product of the intermediate bonds and
the possibility of a few negative bonds is sufficient to remove any information
about the overall sign. From eq.~\eWnlpm, we can define a characteristic sign
correlation length $\xi_s =- {1/\ln(1-2p)}$.

There is also a ``mean-field" type of approach to the sign coherence problem
\ref\rSOb{See the contribution by S. Obukhov in {\it Hopping Conduction in
Semiconductors}, by M. Pollak and B.I. Shklovskii (North Holland, 1990).}\
which is likely to be exact in high dimensions. For paths along the diagonal
of the hypercubic lattice, the mean value of $W(t)$ is
\eqn\emWpm{\overline{W(t)}\approx \left[ d\tau(1-2p) \right]^t\quad.}
Calculating the variance of $W$ is complicated due to the previously
encountered
problem of intersecting paths. We can {\it approximately} evaluate it by
considering
a subset of paths contributing to the second moment as,
 \eqn\evarWpm{\eqalign{\overline{W^2}\approx &\left[ d\tau(1-2p) \right]^{2t}+
(d\tau^2)\left[ d\tau(1-2p) \right]^{2(t-1)}+(d\tau^2)^2\left[ d\tau(1-2p)
\right]^{2(t-2)}
+\cdots+\left( d\tau^2 \right)^t \cr
=&\tau^{2t}\,{\left[ d(1-2p) \right]^{2(t+1)}-d^{t+1}\over\left[ d(1-2p)
\right]^{2}-d
}\quad.}}
The first term in the above sum comes from two distinct paths between the end
points;
the second term from two paths that have their first step in common and then
proceed
independently. The $m^{\rm th}$ term in the series describes two paths that
take
$m$ steps together before becoming separated. The underlying assumption is that
once the two paths have separated they will not come back together again. This
Independent Path Approximation (IPA) is better justified in higher dimensions
and
leads to
\eqn\eWWratio{{ \overline{W^2(t)}\over \overline{W(t)}\,^2}=
{d(1-2p)^{2}-\left[ d (1-2p)^2 \right]^{-t}
\over d(1-2p)^{2}-1}\quad.}

For small $p$ such that $d(1-2p)^2>1$, the above ratio converges to a constant
as $t\to\infty$;
the distribution is asymptotically sharp and the correlations preserve sign
information.
However, if the concentration of negative bonds exceeds $p_c=\left(
1-1/\sqrt{d} \right)/2$,
the ratio diverges exponentially in $t$, indicating a broad distribution.
This has been interpreted \ref
\rNSS{V. L. Nguyen, B. Z. Spivak, and B. I. Shklovskii, Pis'ma Zh. Eksp. Teor.
Fiz. {\bf 41}, 35 (1985) [JETP Lett. {\bf 41}, 42 (1985)]; Zh. Eksp. Teor.
Fiz. {\bf 89}, 11 (1985) [JETP Sov. Phys. {\bf 62}, 1021 (1985)].}\
as signalling a ``sign transition". This argument suggests that there is a
finite $p_c$ for all $d>1$. However, it is important to note that the IPA
ignores important correlations between the paths. Shapir and Wang \ref
\rSW{Y. Shapir and X.-R. Wang, Europhys. Lett. {\bf 4}, 1165 (1987).}\
criticize the assumption of independent paths and suggest that
as intersections are important for $d\leq 3$, there should be no
phase transition in these dimensions. However, the identification of
the {\it lower critical dimension} for the sign transition is not completely
settled.
Numerical simulations based on the transfer matrix method for $t$ of
up to 600\rMKqie, as well as exact enumeration studies \ref
\rWSMK{X.-R. Wang, Y. Shapir, E. Medina, and M. Kardar, Phys. Rev. B
{\bf 42}, 4559 (1990).}\
for $t\leq 10$, fail to find a phase transition in $d=2$.  The results suggest
that if there is a phase transition in $d=2$ it occurs for $p_c<0.05$.
The phase diagram of a generalized model with complex phases has
also been studied in higher dimensions \nref
\rCookDb{J. Cook and B. Derrida, J. Stat. Phys. {\bf 61}, 961 (1990).}\nref
\rGoldBlum{Y. Y. Goldschmidt and T. Blum, J. Physique I {\bf 2}, 1607 (1992);
T. Blum and Y. Y. Goldschmidt, J. Phys. A: Math. Gen. {\bf 25}, 6517 (1992).}%
\refs{\rCookDb,\rGoldBlum}.

For $p>p_c$ the information on sign is lost beyond a coherence length
$\xi_s$.  If the system is coarse grained beyond this scale, the
effective bonds are equally likely to be positive or negative. Thus we
shall concentrate on the symmetric case of $p=1/2$ in the rest of
this section. This corresponds to the much studied $\pm J$ Ising spin
glass \ref
\toul{G. Toulouse, Commun. Phys. {\bf 2}, 115 (1977).}\
which will be discussed in more detail later on.
We performed \rMKqie\ transfer matrix computations on systems of up to size
$t=2000$, and averaged over $2000$ realizations of randomness, on a
VaxStationII. The random numbers ($+1$ or $-1$) were generated by a well
tested random number generator \ref
\rNrec{W. H. Press, B.P. Flannery, S. A. Teukolsky, and W. T. Vetterling,
{\it Numerical Recipes}, (Cambridge University Press 1986).}.
Since $W$ grows exponentially in $t$,
$\ln |W|$ has a well defined  probability distribution; we examined its
mean $\overline{\ln |W(t)|}$, and variance $\overline{\ln |W(t)|^2}
{}~-~\overline{\ln |W(t)|}\, ^2$, for $p=1/2$ (both signs equally
probable). We also computed the typical excursions of the paths in the
lateral direction as defined by
\eqn\eBxa{\overline{ \left[x(t)^2\right]_{av}} \equiv\overline{ {\sum_{x} x^2
|W(x,t)|^2\over  \sum_{x} |W(x,t)|^2}}\quad ,}
and
\eqn\eBxb{\overline{ \left[x(t)\right]_{av}^2} \equiv\overline{
\left({\sum_{x} x |W(x,t)|^2\over \sum_{x} |W(x,t)|^2}\right)^2}\quad, }
where  $[\cdot]_{av}$ denotes an average over the lateral coordinate
at a fixed $t$, using a weight $|W(x,t)|^2$.

We confirmed that the average of $\ln |W(t)|$ is extensive
($\overline{\ln |W(t)|} =(0.322\pm 0.001)t$), while its fluctuations
satisfy a power law growth $t^\omega$, with $\omega = 0.33\pm0.05$.  For
several choices of $t$ we also checked in detail that $W(t)$ is positive
or negative with equal probability.
For lateral excursions,  we examined simulations with
$t=4000$, and with $200$ realizations of randomness (reasonable data for
fluctuations of $\ln |W(t)|$ are only obtained from higher averaging).
The results for $\overline{ [x^2]_{av}}$ and $\overline{ [x]_{av}^2}$
appear to converge to a common asymptotic limit; fitted
to a power law $t^{2\zeta}$ with $\zeta=0.68\pm0.05$.
The scaling properties of $|W(x,t)|$ thus appear identical to those
of directed polymers with positive random weights!
It should be noted, however, that using a similar procedure, Zhang \ref
\rZ{Y.-C. Zhang, Phys. Rev. Lett. {\bf 62}, 979 (1989); Europhys. Lett.
{\bf 9}, 113 (1989); J. Stat. Phys. {\bf 57}, 1123 (1989).}\
concluded from fits to his numerical results a value of $\zeta=0.74\pm 0.01$.
Using a variety of theoretical arguments\rZ, he suggests $\omega=1/2$ and
$\zeta=3/4$. The exponent $\omega=1/2$ is clearly inconsistent with our data,
while $\zeta=3/4$ can be obtained if one fits only to $\overline{
[x]^2_{av}}$. Two subsequent, rather extensive, numerical studies
\nref\rMPG{M. P. Gelfand, {\it Physica A\/} {\bf 177}, 67 (1991).}%
\nref\rGB{Y. Y. Goldschmidt and T. Blum, Nucl. Phys. B {\bf 380}, 588 (1992).}%
\refs{\rMPG,\rGB}\
shed more light on this problem. Both simulations seem to equivocally point
to the importance of including corrections to scaling in the fits. In
1+1 dimensions they indeed find $\omega=1/3$ for the variance, and
$\zeta=2/3$ (with a large correction to scaling term) for transverse
fluctuations.

The similarity in the probability distributions of random weight and random
sign problems can be understood by examination of the moments.
The terms in $W^n$ correspond
to the product of contributions from $n$ independent paths. Upon
averaging, if $m$ paths cross a particular bond ($0\leq m\leq n$), we
obtain a factor of $[1+(-1)^m]/2$, which is $0$ or $1$ depending on the
parity of $m$. For odd $n$ there must be bonds with $m$ odd, and hence
$\overline {W^{2n+1}}=0$; which of course implies and follows from
the symmetry $p(W)=p(-W)$. For even moments $\overline {W^{2n}}$,
the only configurations that survive averaging are those in which the $2n$
replicated paths are arranged such that each bond is crossed an even number
of times. The simplest configurations satisfying this constraint correspond
to drawing $n$ independent paths between the end points and assigning two
replica indices to each. The above constraint is also satisfied by forming
groups of four or higher even numbers, but such configurations are
entropically unlikely and we shall henceforth consider only paired
paths. There is an important subtlety in calculating $\overline{W^{2n}}$
from the $n$ paired--paths: After two such paths cross, the outgoing pairs
can either carry the same replica labels as the ingoing ones, or they
can exchange one label (e.g. $(12)(34) \rightarrow (12)(34),~
(13)(24),~~{\rm or} ~(14)(23)$). Therefore, after summing over all possible
ways of labelling the paired paths, there is a multiplicity of three
for each intersection.  The $n$ paired paths attract each other
through the exchange of replica partners!

Although the origin of the attraction between paths is very different from
the case of random weights, the final outcome is the same. The even
moments in $1+1$ dimension are related by an expression similar to
eq.~\eWnasy,
\eqn\eWnasypm{\lim_{t\to\infty}\overline {W^{2n}(t)} =
\overline {W(t)^2}\,^{n}\exp\left[ \rho n(n^2-1)t\right] \quad,}
and the conclusions regarding $\ln W(t)$ are the same as before.
If, rather than having only one possible value for the magnitude of
the random bond, we start with a symmetric distribution $p(\tau)$, there will
be an additional attraction between the paired paths coming
from the variance of $\tau^2$. This increases the bound state
energy (and the factor $\rho$) in eq.~\eWnasypm\ but does not
affect the universal properties.

\section{Other Realizations of DPRM}

So far we focused on sums over DPRM as encountered in high
temperature series of Ising models. In fact several other realizations of
such paths have been discussed in the literature, and many more are
likely to emerge in the future. One of the original motivations was to
understand the domain wall of an Ising model in the presence of
random bond impurities \rHH. As mentioned in the previous section,
if all the random bonds are ferromagnetic, in the ground state
all spins are up or down. Now consider a mixed state in which a
domain wall is forced into the system by fixing spins at opposite
edges of the lattice to $+$ and $-$.  Bonds are broken at the
interface of the two domains, and the total energy of the defect is
twice the sum of all the $K_{ij}$ crossed by the interface. In the
Solid--On--Solid
approximation, configurations of the domain wall are restricted to directed
paths.
The resulting partition function $Z(t)$, can be computed by exactly
the same transfer matrix method used to calculate $W(t)$. Rather
than looking at the finite temperature partition function, Huse
and Henley \rHH\ worked directly with the zero temperature
configuration of the interface.

Denoting by $E(x,t)$ the minimum in the energy  of all paths connecting $(0,0)$
to
$(x,t)$, oriented along the diagonal of the square lattice,
it is possible to construct the recursion relation,
\eqn\eERR{E(x,t+1)=\min\left\{ E(x-1,t)-2J_{x-1,t}\,,\,
E(x+1,t)-2J_{x+1,t}\right\}\quad,}
closely related to eq.~\eTMran. To find the actual configuration of
the path, it is also necessary to store in memory one bit of information at
each point $(x,t)$, indicating whether the minimum in eq.~\eERR\ comes
from the first or second term. This bit of information indicates the
direction of arrival for the optimal path at $(x,t)$. After the recursion
relations
have been iterated forward to ``time" step $t$, the optimal location is
obtained
from the minimum of the array $\left\{ E(x,t) \right\}$. From this location
the optimal path is reconstructed by stepping backward along the
stored directions. This is how the pictures of optimal paths in
refs.~\refs{\rKZhang,\rMHKZ}\ were constructed.  These optimal paths have
a beautiful ultrametric structure that resembles the deltas of river basins,
and many other natural branching patterns. Finding the optimal interface
is reminiscent of the travelling salesman problem. However, in this case,
although the number of possible paths grow as $2^t$, their directed
nature allows us to find the best solution in polynomial time.

The statistics of the $E(x,t)$ at $T=0$ are identical to those of $\ln W(x,t)$:
the optimal path wanders as $t^{2/3}$, while the fluctuations in $E(t)$ scale
as $t^{1/3}$ \rHH. It is frequently assumed that these fluctuations also
set the scale of energy barriers that the interface must cross from one
optimal state to another. Since such barriers grow with $t$, any activated
process is slowed down to a logarithmic crawl \rHH.

It has been suggested that optimal paths are also relevant to fracture and
failure phenomena \ref
\rDisFrac{{\it Disorder and Fracture}, edited by J. C. Charmet, S. Roux, and
E. Guyon, (Plenum Press, New York, 1990).}.
Imagine a two dimensional elastic medium with impurities, e.g. a network of
springs of different strengths and extensions \ref
\Herrmann{See, e.g., the article by H. J. Herrmann and L. de Arcangelis, in
ref.~\rDisFrac; and references therein.}.
If the network is subjected to external shear, a complicated stress field
is set up in the material. It is possible that non-linear effects in
combination
with randomness enhance the stress field along particular paths in the
medium. Such bands of enhanced stress are visible by polarized light in
a sheet of plexiglass.
The localization of deformation is nicely demonstrated in a two dimensional
packing of straws \ref
\rPABT{C. Poirier, M. Ammi, D. Bideau, and J.-P. Troadec, Phys. Rev. Lett.
{\bf 68}, 216 (1992).}.
The roughness of the localization band is characterized by the exponent
$\zeta=0.73\pm 0.07$, not inconsistent with the value of 2/3 for DPRM.
The experiment was inspired by random fuse models \ref
\rRandFuse{B. Kahng, G. G. Batrouni, S. Redner, L. de Arcangelis, and
H. J. Herrmann, Phys. Rev. B {\bf 37}, 7625 (1988).}\
which apply a similar procedure to describe the failure of an electrical
network.
Hansen et al.\ \ref
\rHHRoux{A. Hansen, E. L. Hinrichsen, and S. Roux, Phys. Rev. Lett. {\bf 66},
2476 (1991).}\
suggest that at the threshold in all such models, failure occurs along an
optimal path with statistics similar to a DPRM. Their numerical results
obtain a roughness exponent of $\zeta=0.7$ for the crack interface
with a precision of about $10\% $.

In fact, the minimal directed path was proposed in 1964 \ref
\rTyde{P. A. Tydeman and A. M. Hiron, {\it B. P. \& B.I.R.A. Bulletin} {\bf
35}, 9 (1964).}\
as a model for tensile rupture of paper. The variations in brightness of a
piece of
paper held in front of a light source are indicative of nonuniformities in
local thickness and
density $\rho({\bf x})$. Tydeman and Hiron suggested that rupture occurs along
the weakest line for which the sum of $\rho({\bf x})$ is minimum. This is
clearly just
a continuum version of the optimal energy path in a random medium. (Since the
average of $\rho({\bf x})$ is positive, the optimal path will be directed.)
This model
was tested by Kert\'esz et al.\ \ref
\rPaper{J. Kert\'esz, V. K. Horv\'ath, and F. Weber, Fractals {\bf 1}, 67
(1993).}\
who used a tensile testing machine to gradually tear apart many sheets of
paper.
They found that the resulting rupture lines are self--affine, characterized by
$0.63<\zeta<0.72$.

The three dimensional DPRM was introduced \rKZhang\ as a model for a
polyelectrolyte in a gel matrix. Probably a better realization is
provided by defect lines, such as dislocations or vortices, in a medium
with impurities. There has been a resurgence of interest in this problem
since it was realized that flux lines in high temperature ceramic
superconductors are highly flexible, and easily pinned by the oxygen
impurities that are usually present in these materials \ref
\rDRN{D. R. Nelson, Phys. Rev. Lett. {\bf 60}, 1973 (1988);
D. R. Nelson and H. S. Seung, Phys. Rev. B {\bf 39}, 9153 (1989).}.
Pinning by impurities is in fact crucial for any application, as otherwise
the flux lines drift due to the Lorentz force giving rise to flux flow
resistivity \ref\rBlatter{See, for example,
G. Blatter et. al., ETH preprint, and references therein.}.

\section{Quantum Interference of Strongly Localized Electrons}

The wavefunctions for non-interacting electrons in a regular solid are
{\it extended} Bloch states. In the presence of disorder and impurities,
gradually more and more of these states become {\it localized}.
This was first pointed out by Anderson \ref
\rPWAa{P. W. Anderson, Phys. Rev. {\bf 109}, 1492 (1958).}\
who studied a random tight--binding Hamiltonian
\eqn\eATH{{\cal H} = \sum _{i} \varepsilon _{i} a^{+} _{i} a _{i} +
\sum _{<ij>} V_{ij} a^{+} _{i} a _{j}\quad.}
Here $\varepsilon_{i}$ are the site energies and $V_{ij}$ represent the nearest
neighbor couplings or transfer terms. For simplicity we shall focus on
$$V_{ij}= \cases{V &if $i,j$ are nearest neighbors\cr
0  &otherwise\cr}\quad,$$
so that all the randomness is in the site energies. This is just a discretized
version of the continuum Hamiltonian $H = \nu\nabla^2+\varepsilon(\vec x)$, for
a quantum particle in a random potential $\varepsilon(\vec x)$.
For a uniform $\varepsilon$,
the Hamiltonian is diagonalized by extended Fourier modes
$a_{\vec q}^\dagger =\sum_{\vec x} \exp\left({i\vec q\cdot \vec x}\right)
a_{\vec x}^\dagger /\sqrt N$, resulting in a band of energies
$\varepsilon(\vec q) = \varepsilon+2V(\cos q_1 + \cos q_2 +\cdots + \cos q_d)$.
(The lattice spacing has been set to unity.) As long as the fermi energy falls
within
this band of excited states the system is metallic.

In the random system the wave functions become distorted, and localized to
the vicinity of low energy impurities \rPWAa. This localization starts with the
states
at the edge of the band and proceeds to include all states as randomness is
increased. In fact in $d \le 2$, as the diffusing path of a non--localized
electron will
always encounter an impurity, {\it all} states are localized by even weak
randomness.
The original ideas of Anderson localization\rPWAa, and a heuristic
scaling approach by Thouless \ref
\rDJTa{D. J. Thouless, Phys. Rep. {\bf 13}, 93 (1974); Phys. Rev. Lett.
{\bf 39}, 1167 (1977).},
have been placed on more rigorous footing by perturbative RG studies \nref
\rAALR{E. Abrahams, P. W. Anderson, D. C. Licciardello, and T. V.
Ramakrishnan, Phys. Rev. Lett. {\bf 42}, 673 (1979).}\nref
\rGLK{L. P. Gor'kov, A. I. Larkin, and D. E. Khmel'nitskii, Zh. Eksp. Teor.
Fiz. Pis'ma Red. {\bf 30}, 248 (1979) [JETP Lett. {\bf 30}, 248 (1979).].}\nref
\rFZW{F. J. Wegner, Z. Phys. B {\bf 35}, 207 (1979).}\refs{\rAALR{--}\rFZW}.
The perturbative approach emphasizes the importance
of quantum interference effects in the weakly disordered metal.
{\it Weak localization} phenomena include the effects of magnetic fields,
spin--orbit (SO)
scattering  (corresponding respectively to interactions breaking time
reversal and spin space symmetries) on the conductivity \ref
\rLR{P. A. Lee and T. V. Ramakrishnan, Rev. Mod. Phys. {\bf 57}, 287 (1985).},
as well as predicting a universal value of the order of
$e^2/\hbar$ for conductance fluctuations \nref
\rALTb{B. L. Altshuler, JEPT Lett. {\bf 41}, 648 (1985).}\nref
\rLS{P. A. Lee and A. D. Stone Phys. Rev. Lett. {\bf 55}, 1622 (1985).}%
\refs{\rALTb,\rLS}.
These phenomena can be traced to the quantum interference of
time reversed paths in {\it backscattering\/} loops and their suppression
by magnetic fields and SO \ref
\rGB{G. Bergmann, Phys. Rep. {\bf 107}, 1 (1984).}:
In the of absence SO, a magnetic field causes an increase in the
localization length, and a factor of 2 decrease in the conductance
fluctuations; with SO, it has the opposite effect of decreasing the
localization length, while still reducing the conductance fluctuations \nref
\rAS{B. L. Altshuler and B. I. Shklovskii, Zh. Eksp. Teor. Fiz. {\bf 91}, 220
(1986) [Sov. Phys. JETP {\bf 64}, 127 (1986)].}\nref
\rLSF{P. A. Lee, A. D. Stone, and H. Fukuyama, Phys. Rev. B {\bf 35}, 1039
(1987).}\refs{\rAS,\rLSF}.
An alternative description
of these phenomena is based on the theory of random matrices \ref
\rZP{N. Zannon and J.-L. Pichard, J. Phys. (Paris) {\bf 49}, 907 (1988).},
where the only input is the symmetries of the underlying Hamiltonian and
their modification by a magnetic field. Mesoscopic devices at low temperature
have provided many experimental verifications of {\it weak localization}
theory \nref\rWW{R. A. Webb and S. Washburn, Physics Today {\bf 41}, No. 12, 46
(1988).}\refs{\rGB,\rWW}\
and there are many excellent reviews on the subject \nref
\rLA{P. A. Lee and B. L. Altshuler, Physics Today {\bf 41}, No. 12, 36
(1988).}\refs{\rLR\rGB,\rLA}.

When the electronic states at the fermi energy are localized, the material
is an insulator and there is no conductivity at zero temperature.
However, at finite temperatures there is a small conductivity that
originates from the quantum tunneling of electrons between localized
states, described by Mott's  variable range hopping (VRH) process \ref
\rNMa{N. F. Mott, J. Non-Cryst. Solids {\bf 1}, 1 (1968).}:
The probability for tunneling a distance $t$ is the product of two factors
\eqn\eMott{p(t)\propto \exp\left(-{2t\over \xi}\right)\times
\exp\left(-{\delta \varepsilon\over k_B T}\right)\quad.}
The first factor is the quantum tunneling probability and assumes that the
overlap
of the two localized states decays with a characteristic {\it localization
length} $\xi$. The second factor recognizes that the different localized
states must have different energies $\delta \varepsilon$ (otherwise
a new state is obtained by their mixture using degenerate perturbation
theory). The difference in energy must be provided by inelastic
processes such as phonon scattering, and is governed by the
Boltzmann weight at temperature $T$. The most likely tunneling sites
must be close in energy. If there is a uniform density of states
$N(\varepsilon_f)$ in the vicinity of the fermi energy, there are
roughly $N(\varepsilon_f)t^d$ candidate states in a volume of linear
size $t$ in $d$ dimensions, with the smallest energy difference of the
order of $ \delta \varepsilon\propto \left(N(\varepsilon_f)t^d \right)^{-1}$.
Thus the two exponential factors in eq.~\eMott\ oppose each other,
encouraging the electron to travel shorter and longer distances
respectively. The optimal distance scales as
\eqn\eTVRH{t\approx \xi(T_0/T)^{1 \over d+1}\quad,}
with $T_0\propto  \left(k_B N(\varepsilon_f) \xi^d\right)^{-1}$, diverging at
zero temperature.

In the strongly localized regime, the optimal hopping length is  many times
greater
than the localization length $\xi$. The localized sites are then
assumed to be connected by a classical random resistor network \ref
\rMA{A. Miller and E. Abrahams,  Phys. Rev. {\bf 120}, 745 (1960).}.
Since the individual resistors are taken from a very wide distribution,
it is then argued \ref
\rAHL{V. Ambegaokar, B. I. Halperin, and J. S. Langer, Phys. Rev. B {\bf 4},
2612 (1971).}\
that the resistance of the whole sample is governed by the critical
resistor that makes the network percolate. This leads to a dependence
\eqn\eCVRH{\sigma(T)=\sigma_0\exp[-(T_0/T)^{{1\over d+1}}]\quad,}
for the conductivity. This behavior has been verified experimentally both
in two and three dimensions \ref
\rPS{M. Pollak and B. I. Shklovskii, {\it Hopping Conduction in
Semiconductors} (North Holland, 1990).}.
Due to the difficulty of measuring variations in the much smaller
conductivities
of insulators, there have been relatively few studies of the conductivity
and its fluctuations for {\it strongly localized\/} electrons.
Nonetheless, recent experiments \ref
\rFO{O. Faran and Z. Ovadyahu, Phys. Rev. {\bf B38}, 5457 (1988).}\
find a {\it positive MC} in $Si$-inversion layers, $GaAs$ and $In_2O_{3-x}$
films. Furthermore, the observed reproducible conductance fluctuations are
quite suggestive of quantum interference (QI) effects. However, the
magnitudes of these fluctuations grow with lowering temperature, and
are about 100 times larger than $e^2/\hbar$ at the lowest temperature.

Clearly a different theory is needed to account for QI effects in the
{\it strong localization} regime. The most natural candidate is the quantum
overlap factor in eq.~\eMott.  Nguyen, Spivak, and Shklovskii (NSS)\
have proposed a model that accounts for QI of multiply scattered tunneling
paths in the hopping probability: In between the phonon assisted tunneling
events the electron preserves its phase memory. However, at
low temperatures it tunnels over very large distances according to eq.~\eTVRH,
and encounters many impurities. The overall tunneling amplitude is then
obtained from the sum over all trajectories between the initial and
final sites. NSS emphasized that since the contribution of each trajectory
is exponentially small in its length, the dominant contributions to the
sum come from the shortest or {\it forward scattering} paths. The
traditional explanations of weak localization phenomena which rely on the
QI of {\it back scattering} paths are therefore inappropriate to this
regime. This picture is clearly reminiscent of the directed paths and
will be developed more formally in the next section.

\section{The Locator Expansion and Forward Scattering Paths}

The overlaps in the insulating regime can be studied by
performing a ``locator'' expansion\rPWAa; valid in the limit
$|V _{ij}|=V\ll(E - \varepsilon _{i}),$  where $E$ is the electron energy.
Indeed, for $V=0$, the eigenfunctions are just the single site states,
and the localization length is zero (no transfer term).
For $V /(E - \varepsilon _{i}) \ll 1$, various  quantities can be obtained
perturbatively around this solution, as expressed by the Lippman--Schwinger
equation \ref
\rLipp{K. Gottfried, {\it Quantum Mechanics} (Addison-Wesley, 1990).}\
\eqn\eALSa{|\Psi^+\rangle=|\Phi\rangle+
{1\over E-H_0+i\delta}{\cal V}|\Psi^+\rangle\quad.}
The bare Hamiltonian
$$H_0 = \sum _i\varepsilon_{i}a^{+}_{i}a_{i}\quad,$$
has no nearest-neighbor coupling, while the perturbation
$${\cal V}=\sum _{<ij>}V_{ij}a^{+}_{i}a_{j}\quad,$$
describes the small transfer terms.  $|\Phi\rangle$ represents the state with
a localized electron at the initial site (or incident wave), $|\Psi^+\rangle$
the state where a localized electron is at the final site. In the coordinate
representation, the wavefunctions are exponentially localized around the
impurity sites and there are no propagating waves since electrons can only
tunnel under a potential barrier. (This situation was first addressed in detail
by Lifshits and Kirpichenko \ref
\rLK{I. M. Lifshits and V. Ya. Kirpichenko, Sov. Phys. JETP {\bf 50}, 499
(1979).}.)
We can now iterate this implicit equation to obtain an expansion in powers
of the ratio ${\cal V}/(E-\varepsilon_i)$ as
\eqn\eALSb{|\Psi^+\rangle=|\Phi\rangle+{1\over E - H_0
 +i\delta}{\cal V}|\Phi\rangle+{1\over
E - H_0 +i\delta} {\cal V}{1\over E - H_0 +i\delta}{\cal V}|\Phi\rangle
+\cdots\quad.}
Acting with $\langle\Psi^+|$ on the left and taking $\delta$ to zero, we obtain
the overlap between the two states
\eqn\eALSc{\langle\Psi^+|\Psi^+\rangle=\langle\Psi^+|\Phi\rangle+
\langle\Psi^+|{1\over E - H_0}{\cal
 V}|\Phi\rangle+\langle\Psi^+|{1\over E - H_0} {\cal V}{1\over E - H_0}{\cal
V}|\Phi\rangle +\cdots\quad.}
For a more general transfer term ${\cal V}$ connecting all sites, the
first term represents an electron starting from the initial site and
ending at the final site without scattering
(the overlap $\langle\Psi^+|\Phi\rangle$);
the second term represents electrons scattering {\it once} off
intermediate sites, the third, scattering {\it twice}, {\it etc.}.  The
operator ${\cal V}$ acting on $|\Phi\rangle$ produces a factor $V$ for
each segment crossed, and $H_0$ acting on a particular site $i$ results in
$\varepsilon_i$, the bare site energy. Thus we finally arrive at
a simple expression for the amplitude or the Green's function between the
initial and final states as
\eqn\eAGFa{\langle\Psi^+|\Psi^+\rangle=\langle\Phi|G(E)|\Psi^+\rangle=
V\sum_{\Gamma}\prod_{i_{\Gamma}}
{V\over E-\varepsilon_{i_{\Gamma}}}\quad.}

The terms in the above perturbation series correspond to all paths
$\Gamma$ connecting the end points; $i _{\Gamma}$ label the sites along each
path. Except that the random variables appear on the sites rather than
the bonds of the lattice, this sum over paths is quite reminiscent of the
corresponding one for the correlation functions of the random bond
Ising model. There is, however, one complication that distinguishes
the localization problem: The energy denominators in eq.~\eAGFa\
may accidentally be zero, invalidating the perturbation series.
Physically, this corresponds to intermediate sites that are at the
same energy as the external points. Presumably in this
case a degenerate perturbation theory has to be used to construct the
wavefunction. NSS \rNSS\ circumvent this issue by considering
initial and final sites of approximately the same energy
$\varepsilon_F=E=0$, while the intermediate sites have
energies $\varepsilon _{i}=\pm U$ with equal probability.
All the energy denominators in eq.~\eAGFa\  now
contribute the same finite magnitude $U$, but random signs
$\eta_{i_{\Gamma}}=\varepsilon_{i_{\Gamma}}/U$.
The justification is that the Mott argument implicitly assumes
that the lowest energy $\delta\varepsilon$ occurs at a
distance $t$, and that there are no intermediate sites that are
more favorable. However, it is not clear that due to the very
same considerations, we should not include some dependence of
the effective energy gap $U$ on $t$. We shall set aside
such considerations and focus on the properties of the NSS
model in  the remainder.

A path of length $\ell$ now contributes an
amplitude $U(V/U)^\ell$ to the sum, as well as an overall sign. In the
localized regime the sum is rapidly  convergent, dominated  by its lowest
order terms \rPWAa. In general, the sum is bounded by one in which all terms
make a {\it positive} contribution, i.e. by a lattice random walk which is
convergent for $z(V/U)<1$, where $z$ is the lattice coordination number.
This provides a lower bound for the delocalization transition, and the
series is certainly convergent for smaller values of $V/U$. As in the Ising
model we expect loops to become important only after the transition, while
in the localized phase typical paths are directed beyond the localization
length $\xi$. For $(V/U)\ll 1$,
the localization length is less than a single lattice spacing, and only
directed ({\it forward scattering}) paths need to be considered. Loops
({\it back scattering} paths) are irrelevant in the renormalization
group sense.  For sites separated by a distance $t+1$ along a
diagonal of the square lattice, eq.~\eAGFa\  is now simplified to
\eqn\eAGFb{\langle i|G(E)|f\rangle =V\left({V \over U}\right)^{t} \,
\sum_P \prod_{i=1}^t\eta_{Pi}\quad,}
which is identical to eq.~\eCFpm\ with $(V/U)$ replacing $\tau$.
The diagonal geometry  maximizes possible
interference by having a large number of shortest paths. For tunneling
along the axes rather than the diagonal of a square lattice there is
only one shortest path. Then, including longer paths with kinks is
essential to the interference phenomena. However, the analogy to previous
results suggests that the universal behavior is the same in the two cases
while the approach to asymptotic behavior is much slower in the latter.

Using the equivalence to eq.~\eWnasypm, in conjunction with eq.~\ecumlim,
results in
\eqn\eBLnG{\lim_{t\to\infty}\overline{ \ln |\langle i|G|f\rangle |^2}=
\ln\left[ 2\left( {V \over U} \right)^2 \right]t-\rho t
\equiv -2t\left( \xi^{-1}_0 + \xi^{-1}_g \right)\quad, }
where we have defined {\it local} and {\it global} contributions to
the effective localization length, respectively given by
\eqn\eBLL{\xi_0=\left[\ln \left({U\over \sqrt{2}V}\right)\right]^{-1}\quad,
\quad{\rm and}\quad\xi^{-1}_g={\rho\over 2}\quad.}
The QI information is encoded in
$2\xi_g^{-1}=\rho$. Numerical
estimates indicate that for the NSS model $\xi_g\approx 40$, and
confirm that the width of the distribution scales as
\eqn\eBdG{\delta\ln |\langle i|G|f\rangle |\sim \left|{t\over \xi_g}
\right|^{1/3}\quad.}
Since $t\propto T^{-1/3}$ in Mott VRH, we expect fluctuations in
log--conductivity to grow as $T^{-1/9}$ for $T\to0$, in qualitative
agreement with the experimental results of ref.~\rFO.
(A quantitative test of this dependence has not yet been performed.)

\section{Magnetic Field Response}

All that is needed to include a magnetic field $B$ in the tight binding
Hamiltonian of eq.~\eATH\ is to multiply the transfer elements $V_{ij}$
by $\exp\left( A_{ij} \right)$, where $A_{ij}$ is the line integral of
the gauge field along the bond from $i$ to $j$. Due to these factors,
the Hamiltonian becomes complex and is no longer time reversal
symmetric ($H^*\ne H$).  In the parlance of random matrix theory \rZP,
the Hamiltonian with $B=0$ belongs to the {\it orthogonal} ensemble,
while a finite field places it in the {\it unitary} ensemble. Actually,
random matrix theory recognizes a third ({\it symplectic}) ensemble
of Hamiltonians which are time reversal symmetric, but not invariant
under rotations in spin space. Up to this point we had not mentioned the
spin of the electron: The states of eq.~\eATH\ are thus doubly degenerate
and can be occupied by (non-interacting) up or down spin states.
We can remove this degeneracy by including {\it spin-orbit} (SO) scattering,
which rotates the spin of the electron as it moves through the lattice.

The generalized tight binding Hamiltonian that includes both the effects of
SO scattering and magnetic field is
\eqn\eHSO{ H=\sum_{i,\sigma}\varepsilon_i
a^\dagger_{i,\sigma}a_{i,\sigma} +\sum_{<ij>,\sigma \sigma'}
V_{ij,\sigma \sigma'} e^{iA_{ij}}a^\dagger_{i,\sigma}a_{j,\sigma'}\quad.}
The constant, nearest-neighbor only hopping, elements $V$ in eq.~\eATH\
are no longer diagonal in spin space. Instead, each is multiplied by
${\cal U}_{ij}$, a randomly chosen $SU(2)$ matrix which describes the spin
rotation due to strong SO scatterers on each bond \rZP.
Eq.~\eAGFa\ for the overlap of wavefunctions at the two end-points
must now include the initial and final spins, and
eq.~\eAGFb\ for the sum of directed paths generalizes to
\eqn\eAGso{{\cal A}=\langle i\sigma|G(0)|f\sigma'\rangle =V(V/U)^tJ(t)
\quad;\quad J(t)=
\sum_{P} \prod_{j=1}^t \eta_{Pj} e^{iA_{Pj,P(j+1)}}{\cal U}_{Pj,P(j+1)}\quad.}
After averaging over the initial spin, and summing over the final spin, the
tunneling probability is
\eqn\eAT{T={1\over2}\tr({\cal A}^\dagger {\cal A})=V^2
(V/U)^{2t} I(t) \quad;\quad I(t)={1\over2}\tr(J^\dagger J)\quad.}

We numerically studied the statistical properties of $I(t)$, using a transfer
matrix method to exactly calculate $I$ up to $t=1000$, for over 2000
realizations of the random Hamiltonian. We again found that the
distribution is broad (almost log--normal), and that the appropriate variable
to consider is $\ln I(t)$. In all cases the mean of ${\ln I(t)}$
scaled linearly with $t$, while its fluctuations scaled as $t^\omega$
with $\omega\approx 1/3\,$ \nref
\rMKSWa{E. Medina, M. Kardar, Y. Shapir, and X.-R. Wang,
Phys. Rev. Lett. {\bf 62}, 941 (1989).}\nref
\rMKSWb{E. Medina, M. Kardar, Y. Shapir, and X.-R. Wang,
Phys. Rev. Lett. {\bf 64}, 1816 (1990).}%
\nref\rMKa{E. Medina, and M. Kardar, Phys. Rev. Lett. {\bf 66}, 3187 (1991).}%
\refs{\rMKSWa{--}\rMKa}.
For the sake of comparison with experiments we define a
log--magnetoconductance (MC) by
\eqn\edefMC{MC(t,B)\equiv\overline{\ln I(t,B)}-\overline{\ln I(t,0)}\quad.}
We find numerically that the magnetic field always causes an enhancement
in tunneling (a positive MC), but that the asymptotic behavior is quite
distinct in the presence or absence of SO scattering.
\item {\it (1)} In the absence of SO, the MC is unbounded and grows linearly
with $t$. This can be interpreted as an increase in the global contribution to
the localization length. The numerical results indicate that for small $B$, the
change in slope is proportional to $B^{1/2}$. Indeed the data for
different $t$ and $B$ can be collapsed together, using the fit
\eqn\eMCnSO{MC(t,B)=(0.15\pm0.03)\left( {\phi \over \phi_0} \right)^{1/2}
t\quad,}
where $\phi=Ba^2$ is the flux per plaquette, and $\phi_0$ is the elementary
flux quantum.

\item {\it (2)} In the presence of SO, the MC quickly saturates with $t$
and there is no change in the localization length. The data can still be
collapsed, but by using $Bt^{3/2}$ as the scaling argument, and we find
\eqn\eMCSO{MC_{SO}(t,B) =
\cases{cB^2t^3 &if $B^2t^3<1$\cr C\approx 0.25 &if $B^2t^3>1$\cr}\quad.}

We can gain some analytic understanding of the distribution function for
$I(t,B)$ by examining the moments $\overline{ I(t)^n}$. From eqs.~\eAGso\
and \eAT\ we see that each $I(t)$ represents a forward path from $i$ to $f$,
and a time reversed path from $f$ to $i$. For $\overline{ I(t)^n}$,
we have to average over the contributions of $n$ such pairs of paths.
Averaging over the random signs of the site energies forces a {\it pairing}
of the $2n$ paths (since any site crossed by an odd number of paths leads
to a zero contribution) \rMKSWa. To understand the MC, it is useful to
distinguish two classes of pairings: {\bf (1)} {\it Neutral paths} in which
one member is selected from $J$ and the other from $J^\dagger$. Such pairs
do not feel the field since the phase factors of $e^{i A}$ picked
up by one member on each bond are canceled by the conjugate factors
$e^{-iA}$ collected by its partner. {\bf (2)} {\it Charged paths} in which both
elements are taken from $J$ or from $J^\dagger$. Such pairs couple to the
magnetic field like particles of charge $\pm 2e$. In the presence of SO,
we must also average over the random $SU(2)$ matrices.
{}From the orthogonality relation for group representations \ref
\rGOT{H. F. Jones, {\it{Groups, Representations and Physics}}
(Adam Hilger, Bristol and New York, 1990).},
we have
\eqn\eBOR{\int \Gamma^k(g)_{ij}^*\Gamma^{k'}
(g)_{i'j'}W(\alpha_1,\cdots,\alpha_n) d\alpha_1\cdots d\alpha_n
={\delta_{ii'}\delta_{jj'} \delta_{kk'}\over \lambda_k}\int
W(\alpha_1,\cdots,\alpha_n)d\alpha_1\cdots d\alpha_n,}
where $\Gamma^k(g)_{ij}$ is the $ij$ matrix element of a representation
of the group element $g$, $W(\alpha_1,\cdots,\alpha_n)$ is an appropriate
weight function so that the matrix space is sampled uniformly as the
continuous parameters $\alpha_1,\cdots,\alpha_n$ vary (e.g. Euler angles
for a representation of $SU(2)$). Finally $\lambda_k$ is the order of
the representation $k$. Choosing the Euler angle parametrization of
$SU(2)$ it can be shown that the only nonzero paired averages are
\eqn\eBC{\overline{ \CU _{\alpha\beta}\,\CU _{\alpha\beta}^*}={1\over 2}\quad,
\qquad\overline{ \CU _{\uparrow \uparrow}\, \CU _{\downarrow \downarrow}}=
{1\over 2}\quad,\qquad \overline{ \CU _{\uparrow \downarrow} \,
\CU _{\downarrow \uparrow}} =-{1\over 2}\quad,}
and their complex conjugates. Thus SO averaging forces neutral paths to
carry parallel spins, while the spins on the two partners of charged paths
must be antiparallel.

We next consider the statistical weights associated with the intersections
of paths. These weights depend crucially on the symmetries of the Hamiltonian
in eq.~\eHSO:
For $B=0$ and without SO, the Hamiltonian has {\it orthogonal\/} symmetry. All
pairings are allowed and the attraction factor is 3, since an incoming
(12)(34) can go to (12)(34), (13)(24), or (14)(23). Note that even
if both incoming paths are neutral, one of the exchanged configurations
is charged. A magnetic field breaks time reversal
symmetry, discourages charged configurations, and reduces the exchange
attraction. The limiting case of a `large' magnetic field is mimicked
by replacing the gauge factors with random phases. In this extreme, the
Hamiltonian has {\it unitary\/} symmetry and only neutral paths are
allowed. The exchange factor is now reduced to 2; from $(11^*)(22^*)
\rightarrow (11^*)(22^*),\quad{\rm or}\quad(12^*)(21^*)$. With SO averaging, we
must also take into account the allowed spin exchanges: Two neutral
paths entering the intersection can have indices $(\alpha\alpha),
(\alpha\alpha)$ or
$(\alpha\alpha),(\overline{\alpha}\overline{\alpha})$; there are 2
possibilities for the first ($\alpha
= \uparrow~{\rm or}~\downarrow$) and two for the second ($\overline{\alpha})$
is antiparallel to $\alpha$).
In the former case, however, there are
two exchanges preserving neutrality, while in the latter only one exchange
is possible satisfying this constraint.
Hence an overall multiplicity of
$[2\times 2+ 2\times 1]\times (1/2)^2=3/2$ is obtained,
where the $(1/2)^2$ comes from the
averages in eq.~\eBC. Thus the intersection of two paired paths results in an
exchange attraction of 3/2; a signature of the {\it symplectic symmetry}.

Based on the above symmetry dependent statistical attraction factors, we can
provide an understanding of the numerical results for MC. The sum over
$n$ attracting paths again leads to
\eqn\eBIn{\langle I(t)^n\rangle =A(n) 2^{nt} \exp [\rho n(n^2-1)t]\quad,}
where we have also included an overall $n$--dependent amplitude.
Without SO, the magnetic field {\it gradually}
reduces the attraction factor from 3 to 2 leading to the increase in slope.
Addition of SO to the Hamiltonian has the effect of {\it suddenly}
decreasing the attraction factor to 3/2. Why does the addition of the  magnetic
field
lead to no further change in $\rho$ in the presence of SO? Without SO, the
origin of the continuous change in the attraction factor is a charged bubble
that may appear in between successive intersections of two neutral paths.
In the presence of SO, from the averages in eq.~\eBC\ we find the contribution
of such configurations to be zero. To produce
intermediate charged paths (with their antiparallel spins), the entering
pair must have indices of the type $(i\,i), (\bar i\,\bar i)$ (where
$\bar{\downarrow}=\uparrow,\quad{\rm and}\quad\bar{\uparrow}=\downarrow$).
Within the bubble we can have intermediate sites labeled $(j\bar j)$
and $(k\bar k)$ which must be summed over due to matrix contractions.
It is easy to check that, independent of the choice of $j$, if the
incoming and outgoing spins ($i$ and $m$) are the same on a branch
it contributes a positive sign, while if they are opposite the overall
sign is negative.  However, for any choice of $i$ and
$m$, one may choose similar (e.g. $i\rightarrow m$ on both branches),
or opposite (e.g. $i\rightarrow m$ on top and $i\rightarrow {\bar m}$
on lower branch) connections. The difference in sign between the two
choices thus cancels their overall contributions. Hence the neutral paths
traverse the system without being affected by charged segments.
In a magnetic field, their attraction factor stays at 3/2 and
$\rho=\xi_g^{-1}$ is unchanged.

The smaller positive MC observed in the simulations is due to changes in
the amplitude $A(n)$ in eq.~\eBIn. This originates from the charged paths
that contribute to tunneling at small $B$ but are quenched at higher $B$.
However, due to their lack of
interactions, we may treat the charged and neutral paths as independent.
At zero field any of the pairings into charged and neutral paths
is acceptable, while at finite fields only neutral pairs survive.
This leads to a reduction in the amplitude
$A(n)$ for $n\ge 2$, but an increase in $\ln I$ (a positive MC). The
typical value of $\ln I$ thus increases by a $t$ independent amount.
This behavior is similar to the predictions of IPA, and is indeed
due to the independence of charged and neutral paths. Since
the typical scale of decay for charged paths depends on
the combination $Bt^{3/2}$ (typical flux through a random walk of length $t$),
we can explain the scaling obtained numerically in eq.~\eMCSO.

The exchange attraction between neutral paths can also be computed for
(unphysical) $SU(n)$ impurities and equals $1+1/n$, which reproduces $2$ for
$U(1)$
or random phases, and $3/2$ for $SU(2)$ or SO scattering. The attraction
vanishes in the $n\to \infty$ limit, where the paths become independent.
The statistical exchange factors are thus universal numbers, simply related
to the symmetries of the underlying Hamiltonian. The attractions in turn
are responsible for the formation of bound states in replica space, and
the universal scaling of the moments in eq.~\eBIn. In fact, since the single
parameter $\rho$ completely characterizes the distribution, the variations
in the mean and variance of $\ln I(t)$ should be perfectly correlated.
This can be tested numerically by examining respectively coefficients
of the mean and the variance  for different cases. All results
do indeed fall on a single line, parametrized by $\rho$. The largest
value corresponds to the NSS model for $B=0$ and no SO (orthogonal symmetry,
exchange attraction 3). Introduction of a field gradually reduces $\rho$
until saturated at the limit of random phases (unitary symmetry, exchange
attraction 2). SO scattering reduces $\rho$ further (symplectic symmetry,
exchange attraction 3/2).

\section{Unitary Propagation}

We can put together the results discussed so far by generalizing
eq.~\edWdt\ to allow for complex (and matrix valued) parameters.
In the originally encountered directed polymer (DP),
the parameters $\nu>0$ and $\mu$ appearing in this equation
were both real. To discuss the wavefunction in a magnetic
field, we have to allow $\mu$ to take
complex values. Finally, SO scattering is included by generalizing
$W$ to a two component spinor, and using matrix valued $\mu$.
We found that in all these cases the statistical behavior of $\ln W({\bf x},t)$
is the same.  Is this a general property of eq.~\edWdt, independent of the
choice of parameters? A special limit of this equation is when both
$\mu\to-i\mu$ and $\nu\to-i\nu$ are purely imaginary. Then eq.~\edWdt\
reduces to the Schr\"odinger equation
\eqn\eschro{i \frac{\partial W}{\partial t} = \left[\nu \nabla^{2}+\mu(\xvec,t)
\right]W\quad,}
for a particle in a random {\it time dependent} potential.
This equation has been considered in the context of particle
diffusion in crystals at finite temperature \nref
\rJPB{J.-P. Bouchaud, Europhys. Lett. {\bf 11}, 505 (1990).}\nref
\rOE{A. A. Ovchinnikov and N. S. Erikhman, Sov. Phys. JETP {\bf 40}, 733
(1975).}\nref
\rJK{A. M. Jayannavar and N. Kumar, Phys. Rev. Lett. {\bf 48}, 553 (1982).}%
\refs{\rJPB{--}\rJK},
and to model the environment of a light test particle in a gas of
much heavier particles \ref
\rGFZ{L. Golubovic, S. Feng, and F. Zeng, Phys. Rev.
Lett. {\bf 67}, 2115 (1991).}.
Several authors \nref
\rRD{ R. Dashen, J. Math Phys. {\bf 20}, 894 (1979).}\nref
\rRLV{ H. de Raedt, A. Lagendijk, and P. de Vries, Phys. Rev. Lett.
{\bf 62}, 47 (1988).}\nref
\rFGZ{S. Feng, L. Golubovic, and Y.-C. Zhang, Phys. Rev.
Lett. {\bf 65}, 1028 (1990).}\refs{\rRD{--}\rFGZ}\
have also suggested that the diffusion of {\it directed} wave
fronts in disordered media are described by eq.~\eschro.

The path-integral solution to eq.~\eschro\ is \nref
\rRT{R. Tao, Phys. Rev. A {\bf 43}, 5284 (1991).}\refs{\rFeynman,\rRT}
\eqn\ePathInt{W(\xvec, t) = \int_{({\bf 0},0)}^{(\xvec,t)}
        {\cal D} \xvec(t') \exp
        \left\{-i \int_{0}^{t} dt'
            \left[ \frac{1}{4\nu}
            \left( \frac{d \xvec}{d t'}  \right)^{2}
            +\mu(\xvec(t', t')
            \right]
        \right\}\quad,}
where $\xvec(t')$ now describes a path in $d-1$ dimensions.
In writing eq.~\ePathInt, we have chosen the standard initial condition
that at time $t=0$, the ``wave function" is localized at the origin.  The
beam positions \xSq\ and \xx\ characterize the transverse fluctuations
of a directed {\it beam} $W$ about the forward path of least scattering.
Here we use $<\!\!\cdots\!\!>$ to indicate an average with the weight
$\mid\!\! W(x,t)\!\! \mid^{2}$ for a given realization, and $\overline{\cdots}$
to indicate quenched averaging over all realizations of randomness.
Roughly speaking, \xx\ describes the wandering of the beam center,
while \xSqxx\ provides a measure of the beam width.

A special property of eq.~\eschro\ which is valid only for real $\nu$ and
$\mu$ is {\it unitarity}, i.e. the norm $\int d\xvec |W(\xvec,t)|^2$ is
preserved at all times. (In the DP and tunnelling problems, the norm clearly
decays with the length $t$.) This additional conservation law sets
apart the random directed wave problem from DP, and in a sense makes its
solution
more tractable. This unitarity is of course a natural consequence of
particle conservation for the Schr\"odinger equation, but has no
counterpart for directed wave propagation. It is likely that a beam of
light propagating in a random medium will suffer a loss of intensity, due
to either back--reflection, inelastic scattering, or localization phenomena
\ref\rSJ{For a review of localization of light, see S. John, Physics Today, 32
(May 1991).}.

A number of efforts at understanding unitary propagation
in random media have focused on the scaling
behavior of the beam positions \xSq\ and \xx\ at large
$t$.   {\it Lattice} models have been used here with some
success. It has been shown using density-matrix
techniques, for instance, that \xSq\ scales linearly in
time as a consequence of unitarity\rOE; recent
numerical simulations \nref
\rMKS{E. Medina, M. Kardar, and H. Spohn, Phys.
Rev. Lett. {\bf 66}, 2176 (1991).}\nref
\rBTS{J.-P. Bouchaud, D. Touati, and D. Sornette,
Phys. Rev. Lett. {\bf 68}, 1787 (1992).}\refs{\rMKS,\rBTS}\
also support this view. The scaling behavior of \xx\ at large
$t$, however, is somewhat more complicated.  An early
numerical simulation in this area in ref.~\rFGZ, employed
a discretization procedure in
which the norm of the wave function was {\it not}
strictly preserved.  In $2d$, they found that \xxSqrt\
grew superdiffusively as $t^{\zeta}$ with $\zeta \approx
3/4$, while in $3d$, they found a phase
transition separating regimes of weak and strong
disorder. Subsequent numerical studies \rMKS\ on directed waves in
$2d$ cast doubt on the validity of these results when the
time evolution is strictly unitary, indicating that \xx\ scales subdiffusively
in
$2d$  with $\zeta  \approx 0.3$.

Somewhat surprising is the fact that a {\it continuum}
formulation of the wave problem leads to different
results.  An exact treatment of the continuum
Schr\"odinger equation\eschro\ has been given
by Jayannavar and Kumar \rJK. They show that
for a random potential $\delta$--correlated in time,
\xSq\ $\sim t^{3}$ as $t \rightarrow \infty$. This
behavior is modified when there are short-range
correlations in time\rGFZ, but the motion
remains non-diffusive in that the particle is
accelerated indefinitely as $t \rightarrow \infty$.
Lattice models introduce a momentum cutoff $p_{max}
\sim a^{-1}$, where $a$ is the lattice spacing, and
therefore do not exhibit this effect.
The momentum cutoff generated by the lattice discretization
is in some sense artificial.  Nevertheless, in a real fluctuating
medium, we do expect on large time scales to recover the lattice
result, i.e. normal diffusion.  The reason is that
dissipative effects do generate an effective momentum
cutoff in most physical systems.  (Strictly speaking,
even in the absence of dissipation, relativistic
constraints lead to a velocity cutoff $v=c$.)
The presence of such a cutoff
for the wave propagation problem, and hence the physical
relevance of lattice versus continuum models, is still a
matter of debate. While there is no underlying lattice, one
suspects on physical grounds that there does exist an
effective momentum cutoff for propagating waves, related
to the speed of light in the background medium.

Previous numerical investigations of this problem start with a discretization
of the parabolic wave equation in eq.~\eschro. By contrast, we assume that the
path integral representation  is more fundamental and provide a
direct discretization of eq.~\ePathInt\ that preserves unitarity \ref
\rSKR{L. Saul, M. Kardar, and N. Read, Phys. Rev. A {\bf 45}, 8859 (1992).}.
For concreteness, we introduce the model in $2d$.  A discussion of its
generalization to higher dimensions is taken up later.  As usual, we identify
the time axis with the primary direction of
propagation and orient it along the diagonal of the square lattice.
The wave function is defined on the {\it bonds }of this lattice.
We use $W_{\pm}(x,t)$ to refer to the amplitude for arriving at the site
$(x,t)$ from the $\pm x$ direction.  At $t=0$, the wave function is
localized at the origin, with $W_{\pm}(0,0) =1/{\sqrt{2}}$.  Transfer matrix
techniques are now used to simulate diffusion in the presence of disorder.
At time $t$, we imagine that a random scattering event occurs at each site
on the lattice at which either $\PsiPlus(x,t)$ or  $\PsiMinus(x,t)$ is
non-zero.
We implement these events by assigning to each scattering site a $2 \times 2$
unitary matrix $S(x,t)$.  The values of the wave function at time
$t+1$ are then computed from the  recursion relation:
\eqn\eU{
\left( {\PsiMinus(x+1,t+1) \atop \PsiPlus(x-1,t+1)} \right)=
\left(\matrix{S_{11}(x,t)&S_{12}(x,t)\cr S_{21}(x,t)&S_{22}(x,t)}\right)
\left( {\PsiMinus(x,t) \atop \PsiPlus(x,t)} \right) \quad.}
The $S$-matrices are required to be unitary in order to
locally preserve the norm of the wave function. As a particular
realization, we may consider the rotation matrix
\eqn\eSrotmat{S(\theta,\phi)=
\left(\matrix{\cos\left({\theta/ 2}\right)e^{i\phi}&
\sin\left({\theta/2}\right)e^{-i\phi}\cr
-\sin\left({\theta/ 2}\right)e^{i\phi}&
\cos\left({\theta/ 2}\right)e^{-i\phi}}\right)\quad.}
A physical realization of this model is obtained by placing semi polished
mirrors of variable thickness, parallel to the $t$ axis, on the sites of
a square lattice. Within this framework,
it should be clear that the value of $W_{\pm}(x,t)$ is obtained by
summing the individual amplitudes of all directed paths which start
at the origin and arrive at the point $(x,t)$ from the $\pm x$ direction.
We thus have a unitary discretization of the path integral in eq.~\ePathInt\
in which the phase change from the potential $\mu(x,t)$ is replaced by an
element of the matrix $S(x,t)$. A lattice $S$-matrix approach for the study
of electron localization and the quantum Hall effect has been used by
Chalker and Coddington \ref
\rCK{J. T. Chalker and P. D. Coddington, J. Phys. {\bf C21}, 2665 (1988).}.
A related model has also been recently proposed \ref
\rVSS{C. Vanneste, P. Sebbah, and D. Sornette, Europhys. Lett. {\bf 17}, 715
(1992).}\
to investigate the localization of wave packets in random media. These
models also include back scattering and hence involve a larger matrix
at each site.

\section{Unitary Averages}

A particularly nice feature of unitary propagation is that the weights $W(x,t)$
are
automatically normalized. In particular, we are interested in the beam
positions
\eqn\exSq{\overline{<\!\!x^{2}(t)\!\!>} = \sum_{x}\overline{P(x,t)}\
x^{2}\quad,}
and
\eqn\exx{\overline{<\!\!x(t)\!\!>^{2}}= \sum_{x_{1},x_{2}}\overline{
P(x_{1},t)\
P(x_{2},t)}\ x_{1} x_{2} \quad,}
where $P(x,t)$ is the probability distribution function on the lattice at time
$t$,
defined by
\eqn\ePDF{P(x,t) = \mid \PsiPlus(x,t) \mid^{2} + \mid \PsiMinus(x,t)
\mid^{2}\quad.}
(Defining the weights directly on the bonds does not substantially
change the results.)
Note that unlike the DP problem, $P(x,t)$ is properly
normalized, i.e.
$$\sum_{x} P(x,t) = 1\quad,$$
and eqs.~\exSq\ and \exx\ are not divided by normalizations such
as $\sum_{x} P(x,t)$.  This simplification is a consequence of
unitarity and makes the directed wave problem tractable.

The average $\overline{\cdots}$, in eqs.~\exSq\ and \exx\ is to be performed
over
a distribution of $S$-matrices that closely resembles the corresponding
distribution for $\mu$ in the continuum problem. However, by analogy to
the DP problem we expect any disorder to be relevant. Hence, to
obtain the asymptotic scaling behavior, we consider the extreme limit of
strong scattering in which each matrix $S(x,t)$ is an {\it independently}
chosen random element of the group $U(2)$. With such a distribution
we lose any pre--asymptotic behavior associated with weak scattering\rGFZ.
The results are expected to be valid over a range of length scales
$a\ll x \ll \xi$, where $a$ is a distance over which the change of phase
due to randomness is around $2\pi$, and $\xi$ is the characteristic length for
the decay of intensity and breakdown of unitarity.
In the language of path integrals, the quantity $\overline{P(x,t)}$ represents
the average over a conjugate pair of paths (from $W_\pm$ and $W^*_\pm$
respectively.) As in the random sign problem, the paths must be exactly
paired to make a non-zero contribution (since $\overline{S_{\alpha\beta}}=0$.
In the strong disorder limit, each step along the paired paths contributes
a factor of 1/2. (It can be easily checked from eq.~\eSrotmat\ that
$\overline{\left|S_{\alpha\beta}\right|^2}=\overline{\cos^2(\theta/2)}=
\overline{\sin^2(\theta/2)}=1/2$.)
Thus, in this limit, the effect of an impurity at $(x,t)$
is to redistribute the  incident probability
flux $P(x,t)$ at random in the $+x$  and $-x$ directions. On average,
the flux is scattered symmetrically so that the disorder-averaged probability
describes the event space of a classical random walk, i.e.
\eqn\eBinomial{\overline{P(x,t)}= \frac{t!}{(\frac{t-x}{2})!(\frac{t+x}{2})!}
\quad.}
Substituting this into eq.~\exSq, we find $\overline{<\!\!x^{2}(t)\!\!>}= t$,
in agreement with previous studies\rOE.

Consider now the position of the beam center $\overline{<\!\!x(t)\!\!>^{2}}$,
given by eq.~\exx. Unlike $\overline{P(x,t)}$, the correlation function
$\overline{P(x_{1},t)P(x_{2},t)}$ does not have a simple form. It involves a
sum over
four paths, collapsed into two pairs by randomness averaging. The center of
mass
coordinate $R=(x_1+x_2)/2$, performs a random walk with $\overline{R^2}=t/2$.
Let us define a new correlation function for the relative coordinate
$r=x_2-x_1$, as
\eqn\eW{W_2(r,t) = \sum_{R} \overline{P(R-r/2,t)P(R+r/2,t)}\quad, }
with the initial condition
\eqn\eICW{W_2(r,t=0) = \delta_{r,0}\quad. }
The value of $W_2(r,t)$ is the disorder-averaged probability that two paired
paths,
evolved {\it in the same realization of randomness}, are separated by a
distance
$r$ at time $t$, and can be computed as a sum over all configurations
that meet this criteria. Consider now the evolution of two such pairs
from time $t$ to $t+1$.  Clearly, at times when $r \neq 0$, the two pairs
behave as independent random walks.   On the other hand, when $r=0$,
there is an increased probability that the paths move together as a result of
participating in the same scattering event.
An event in which the pairs stay together is enhanced (since
$\overline{\left|S_{\alpha\beta}\right|^4}=\overline{\cos^4(\theta/2)}=
\overline{\sin^4(\theta/2)}=3/8$), while one in which the pairs separate is
diminished (since $\overline{\sin^2(\theta/2)\cos^2(\theta/2)}=1/8$).
These observations lead to the following recursion relation for the evolution
of $W_2(r,t)$,
\eqn\eWevolve{W_2(r,t+1) =
 \left( \frac{1 + \epsilon \delta_{r,0}}{2} \right)  W_2(r,t) +
 \left( \frac{1 - \epsilon \delta_{r,2}}{4} \right)  W_2(r-2,t) +
 \left( \frac{1 - \epsilon \delta_{r,-2}}{4} \right) W_2(r+2,t).}
The parameter $\epsilon\geq0$ measures  the tendency of the paths to stick
together on contact.  (If the $S$--matrix is uniformly distributed over the
group $U(2)$, then $\epsilon =1/4$.) Note that $\sum_r W_2(r)$ is preserved,
as required by unitarity.

Using eq.~\eWevolve, we evolved $W_2(r,t)$ numerically for various values of
$0\,<\,\epsilon\,<\,1$ up to $t\leq15000$.  The position of the beam center was
then calculated from
\eqn\eWxx{\overline{<\!\!x(t)\!\!>^{2}}=
\sum_{R,r} \left(R^2-{r^2\over 4}\right)\overline{P(R-r/2,t)P(R+r/2,t)}=
{t\over 2} - {1\over 4} \sum_{r} W_2(r,t)\ r^{2} \quad.}
The results suggest quite unambiguously that $\overline{<\!\!x(t)\!\!>^{2}}$
scales  as $t^{2\zeta}$, with $\zeta =1/4$.  We emphasize here the utility of
the $S$-matrix model for directed waves in random media.   Not only does
our final algorithm compute averages over disorder in an exact way, but it
requires substantially less time to do so than simulations which perform
averages
by statistical sampling as in DPRM.  We have in fact confirmed our $2d$ results
with these slower methods on smaller lattices ($t<2000$).

The model is easily extended to higher dimensions.  The wave function takes its
values on the bonds of a lattice in $d$ dimensions.  Random $d \times d$
dimensional
$S$-matrices are then used to simulate scattering events at the sites of the
lattice. When the matrices $S(\xvec,t)$ are distributed uniformly over the
group $U(d)$, the same considerations as before permit one to perform
averages over disorder in an exact way.  In addition, one obtains the general
result for $d \geq 2$ that \xSq\ scales linearly in time.
The computation of \xx\ in $d>2$, of course, requires significantly more
computer
resources.  We have computed \xx\ on a $d=3$ body-centered cubic lattice,
starting from the appropriate generalization of eq.~\eWevolve.  The results
for $t<3000$,  indicate that \xx\ scales logarithmically in time.

The above numerical results can be understood by appealing to some
well-known properties of random walks.  Consider a random walker on a $D=d-1$
dimensional hypercubic lattice.  We suppose, as usual, that the walker starts
out at the origin, and that at times $t=0$, 1, 2, $\cdots$, the walker has
probability
$0 < p\leq 1/2D$ to move one step in any lattice direction and probability $1 -
2Dp$
to pause for a rest.  The mean time $t_{0}$ spent by the walker at the origin
grows
as \ref
\rBG{J.-P. Bouchaud and A. Georges, Phys. Rep. {\bf 195}, 128 (1990).}
\eqn\eAtorigin{t_0 \sim \left\{\eqalign{ t^{1\over2} & \qquad (D=1)
\cr \ln t & \qquad(D=2) \cr {\rm constant} & \qquad (D=3) }\right.\quad .}
The numerical results indicate a similar scaling for the wandering of the beam
center \xx\ in $d=D+1$ dimensions, for $d=2$ and $d=3$.  We now show
that this equivalence is not coincidental; moreover, it strongly suggests that
$d_{u} = 3$ is a critical dimension for directed waves in random media.
To this end, let us consider a continuum version of eq.~\eWevolve, which
in general dimensions takes the form
\eqn\ecWevolve{W_2({\bf r},t+1)=W_2({\bf r},t)+\nabla^2
\left[W_2\left(1-\epsilon\delta^D({\bf r})+\cdots\right)\right]\quad.}
The asymptotic solution for $\epsilon=0$ is just a gaussian packet of width
$\overline{r^2}=2t$. We can next perform a perturbative calculation in
$\epsilon$.
However, simple dimensional analysis shows the corrections scale as powers of
$\epsilon/r^D\sim\epsilon t^{-D/2}$, and thus
\eqn\eWtwolim{\lim_{t\to\infty}W_2({\bf r},t)={1\over (4\pi t)^{D/2}}
\exp\left(-{r^2\over 4t}\right)\left[1+\CO\left(\epsilon
t^{-(d-1)/2}\right)\right]\quad.}

Applying the above results to the continuum version of eq.~\eWevolve, gives
\eqn\eNewt{\eqalign{\overline{\langle x\rangle_{t+1}^2} - \overline{\langle
x\rangle_{t}^2}
=& {1\over2}- {1\over 4}\sum_r [W_2(r,t+1) - W_2(r,t)]\ r^2 \cr
\simeq& {1\over2}- {1\over 4}\int d^D{\bf r}\nabla^2
\left[W_2\left(1-\epsilon\delta^D({\bf r})+\cdots\right)\right]\cr
\simeq& {1\over2}- {1\over 2}\int d^D{\bf r}W_2\left(1-\epsilon\delta^D({\bf
r})\right)
= \epsilon W_2(0,t)\quad.}}
Summing both sides of this equation over $t$, one finds
\eqn\eContacts{\overline{<\!\!x(t+1)\!\!>^{2}} =
\epsilon \sum_{t'=0}^{t} W_2(0,t')\approx \int_0^t dt'(4\pi t')^{-D/2} \quad.}
The final integral is proportional to the time a random walker spends at the
origin, and reproduces the results in eq.~\eAtorigin.

We can also regard $W_2(r,t)$ as a probability distribution function for
the relative coordinate between two interacting random walkers. In this
interpretation, the value of $\epsilon$ in eq.~\eWevolve\ parametrizes the
strength of a contact interaction between the walkers.  If $\epsilon=0$, the
walkers do not interact at all; if $\epsilon=1$, the walkers bind on contact.
According to eq.~\eContacts, the wandering of the beam center
$\overline{<\!\!x(t)\!\!>^{2}}$ is proportional to the mean number of times
that
the paths of these walkers intersect during time $t$. If $\epsilon=0$, the
number of intersections during  time $t$ obeys the scaling law in
eq.~\eAtorigin,
since in this case, the relative coordinate between the walkers performs a
simple
random walk.  Numerical results indicate that the same scaling law applies
when $0 <\epsilon < 1$: the contact attraction does {\it not} affect the
asymptotic
properties of the random walk.   In summary, three classes
of behavior are possible in this model. For $\epsilon=0$, i.e.
no randomness, the incoming beam stays centered at the origin, while
its width grows diffusively. For $0<\epsilon<1$, the beam center, \xx,
also fluctuates, but with a dimension dependent behavior as in eq.~\eAtorigin.
In the limit of $\epsilon=1$, interference phenomena disappear completely.
In this case, the beam width is zero and the beam center performs a
simple random walk.

We conclude by comparing the situation here to that of DPRM. In the replica
approach to DPRM, the $n$-th moment of the weight $W(x,t)$ is obtained from
the statistics of $n$ directed paths. Disorder--averaging produces an
attractive
interaction between these paths with the result that they may form a bound
state.
In $d\leq2$, any amount of randomness (and hence attraction) leads to the
formation of a bound state. The behavior of the bound state energy can then
be used to extract an exponent of $\zeta=2/3$ for superdiffusive wandering.
By contrast, the replicated paths encountered in the directed wave problem
(such as the two paths considered for eq.~\eW), although interacting, cannot
form a bound state \rMKS, as it is inconsistent with  unitarity.  This result
also
emerges in a natural way from our model of directed waves.  In $d=2$, for
instance,
it is easy to check that $W_2({\bf r}) \sim (1 - \epsilon \delta_{r,0})^{-1}$
is
the eigenstate of largest eigenvalue for the evolution of the relative
coordinate.
Hence, as $t \rightarrow \infty$, for randomness $\delta$-correlated in space
and time, there is no bound state. This result holds in $d \geq 2$ and is
not modified by short-range correlations in the randomness. The
probability-conserving nature of eq.~\eWevolve\ is crucial
in this regard \ref
\rFI{F. Igloi, Europhys. Lett. {\bf 16}, 171 (1991).}\
as it precludes a $u\delta^D({\bf r})$ attraction in eq.~\ecWevolve.
Small perturbations that violate the conservation of probability lead to the
formation of a bound state.  In the language of the renormalization group,
the scaling of directed waves in random media
is governed by a fixed point that is unstable with respect to changes that do
not preserve a strictly unitary evolution.

Subsequently, a number of authors have obtained additional results
from the random $S$-matrix model.
Following our work, Friedberg and Yu \ref
\rFriedberg{R. Friedberg and Y.-K. Yu, Phys. Rev. E {\bf 49}, 5755 (1994).}\
calculated the leading terms in the scaling laws for the beam center
in $d\geq 2$, and also the next-order corrections.
The analytical results are in agreement with those presented
above. Cule and Shapir \ref
\rCule{D. Cule and Y. Shapir, Phys. Rev. B {\bf 50}, 5119 (1994).}\
extended the
methods of this section to compute the higher moments of
the probability distribution for directed waves in
random media.  If this probability distribution
is multifractal, as claimed in ref.~\rBTS,
the higher moments should obey new scaling laws whose exponents are not
simply related to those of the lower moments.
Within the framework of the $S$-matrix model,
Cule and Shapir did not find evidence for multifractal scaling,
while suggesting that certain aspects of the scaling behavior
may be sensitive to details of the unitary time evolution.

\section{Summing all Paths in High Dimensions}
In the next few sections we shall return to the {\it non-random} Ising model.
The high temperature series can be {\it approximately}
summed so as to reproduce mean--field (gaussian) behavior. This  correspondence
provides a better understanding of why such behavior
is applicable in high dimensions, and also prepares the way for the
exact summation of the series in two dimensions (next section).
We shall then use these methods to look at two dimensional
random Ising models. The high temperature series for the partition
function of the {\it non-random} Ising
model on a $d$--dimensional hypercubic lattice is
\eqn\eHTZd{Z=\sum_{\{ \sigma_i \}}{e^{K \sum_{\left\langle ij \right\rangle}
{\sigma_i \sigma_j}}}=2^N\cosh^{dN}K\times S\quad,}
where $S$ is the sum over all allowed graphs on the lattice, each
weighted by $\tau$ raised to the number of   bonds in the graph, with
$\tau\equiv\tanh K$.
The allowed graphs have an even number of bonds per site. The simplest
graphs have the topology of a single closed loop. There are also
graphs composed of {\it disconnected} closed loops. Keeping in mind
cumulant expansions, we set
\eqn\eonel{\Xi={\rm sum~over~contribution~of~all~graphs~with~one~loop}\quad,}
and introduce another sum,
\eqn\eSce{\eqalign{
S'= \exp \left( \Xi \right) =& 1 + \Xi + {1 \over 2}\left( \Xi \right)^2 + {1
\over 6}\left( \Xi \right)^3 + \cdots \cr
=& 1 +\left( {\rm 1~loop~graphs} \right)+\left( {\rm 2~loop~graphs}
\right)+\left( {\rm 3~loop~graphs} \right)+\cdots\quad.}}

Despite their similarities, the sums $S$ and $S'$ are not identical in
that $S'$ includes additional graphs where a particular bond
contributes more than once. (In the original sum $S$, each lattice
bond contributes a factor of 1 or $\tau$. After raising $\Xi$ to a power
$n$, a particular bond may contribute up to $\tau^n$.) In a similar
approximation, we shall allow additional closed paths in $\Xi$
in which a particular bond is traversed more than once in completing
the loop. Qualitatively, $S$ is the partition function of a gas of
{\it self--avoiding} polymer loops with a monomer fugacity of $\tau$.
The self--avoiding constraint is left out in the partition function
$S'$, which thus corresponds to a gas of {\it phantom} polymer loops.
The corresponding free energy is
\eqn\efpp{\eqalign{
\ln S'= & { {\rm sum~ over~ all~ closed ~ random ~ walks ~ on ~
the ~ lattice} \times\tau^{{\rm length ~ of ~ walk}}} \cr
              =& N \sum_\ell {{\tau^\ell \over \ell}\,({\rm number~ of ~
closed ~ walks~of~}\ell{\rm ~steps~ starting ~ and ~ ending ~ at ~ }
{\bf 0})},}}
where extensivity is guaranteed since (up to boundary effects) the
same loop can be started from any point on the lattice. The additional
factor of $1/\ell$ accounts for the $\ell$ possible starting
points for a loop of length $\ell$.  To count
the number of paths we introduce a set of $N\times N$ matrices,
\eqn\eWlij{\left\langle {\bf i} |W(\ell)| {\bf j} \right\rangle\equiv
{\rm number~of~walks~from~}{\bf j}{\rm ~to~}{\bf i}{\rm ~in~}
\ell{\rm ~steps}\quad.}
Then
\eqn\eflinW{{\ln S'\over N} = \sum_\ell {{\tau^\ell \over \ell} \left\langle
{\bf 0} |W(\ell)| {\bf 0} \right\rangle}\quad.}

Similarly, the spin--spin correlation function
\eqn\esscorr{\left\langle \sigma({\bf 0}) \sigma({\bf r}) \right\rangle =
{1 \over Z} \sum_{\left\{ \sigma_i \right\}}{\sigma({\bf 0}) \sigma({\bf r})}
\prod_{\left\langle ij \right\rangle} {(1 + \tau\sigma_i \sigma_j)}\quad,}
is related to the sum over all paths connecting the points {\bf 0}
and {\bf r} on the lattice. In addition to the simple paths that
directly connect the two points, there are {\it disconnected} graphs
that contain additional closed loops.
In the same approximation of allowing
all intersections between paths, the partition function $S'$ can be
factored out of the numerator and denominator of eq.~\esscorr, and
\eqn\esscora{\left\langle \sigma({\bf 0}) \sigma({\bf r} \right\rangle
\approx \sum_\ell \tau^\ell \left\langle {\bf r} |W(\ell)| {\bf 0}
\right\rangle\quad.}

As the walks are {\it Markovian}, their number can
be calculated {\it recursively}. First note that any walk from
{\bf 0} to {\bf r} in $\ell$ steps can be regarded as a walk from
{\bf 0} to some other point ${\bf r'}$ in $\ell-1$ steps, followed
by a single step from ${\bf r'}$ to {\bf r}. Summing over all possible
locations of the intermediate point leads to
\eqn\erecW{\eqalign{
\left\langle {\bf r} |W(\ell)| {\bf 0} \right\rangle =& \sum_{\bf r'}
{\left\langle {\bf r} |W(1)| {\bf r'} \right\rangle \times \left\langle {\bf
r'} |W(\ell - 1)| {\bf 0} \right\rangle} \cr
=& \left\langle {\bf r} |T W(\ell - 1)| {\bf 0} \right\rangle\quad,}}
where the sum corresponds to the product of two matrices and we
have defined $T\equiv W(1)$. The recursion process can be continued and
\eqn\eWlTl{W(\ell) = T W(\ell - 1) = T^2 W(\ell - 2)^2 = \cdots = T^\ell\quad.}
Thus all lattice random walks are generated by the {\it transfer matrix\/}
$T$, whose elements are
\eqn\eRWTM{\left\langle {\bf r} |T| {\bf r'} \right\rangle =
\cases{
1 & if ${\bf r}$ and ${\bf r'}$ are nearest neighbors \cr
0 & otherwise
}\quad.}
For example in $d=2$,
\eqn\etwodTM{\left\langle x,y |T| x',y' \right\rangle = \delta_{y,y'}
(\delta_{x,x'+1} + \delta_{x,x'-1}) +  \delta_{x,x'}(\delta_{y,y'+1}
+\delta_{y,y'-1})\quad,}
and successive actions of $T$ on a walker starting at the origin,
$|x,y>=\delta_{x,0}\delta_{y,0}$, generate the patterns
$$\matrix{
0 & 0 & 0 \cr
0 & 1 & 0 \cr
0 & 0 & 0 \cr
}
\longrightarrow
\matrix{
0 & 1 & 0 \cr
1 & 0 & 1 \cr
0 & 1 & 0 \cr
}
\longrightarrow
\matrix{
0 & 0 & 1 & 0 & 0 \cr
0 & 2 & 0 & 2 & 0 \cr
1 & 0 & 4 & 0 & 1 \cr
0 & 2 & 0 & 2 & 0 \cr
0 & 0 & 1 & 0 & 0 \cr
}
\longrightarrow
\cdots\quad.
$$
The value at each site is the number of walks ending at that point
after $\ell$ steps.

Various properties of random walks can be deduced from diagonalizing
the matrix $T$. Due to the translational symmetry of the lattice, this
is achieved in the  Fourier basis
$ \left\langle {\bf r}|{\bf q} \right\rangle =
e^{i{\bf q} \cdot {\bf r}}/\sqrt{N}$. For example, in $d=2$ starting
from eq\etwodTM, it can be checked that
\eqn\eTMeq{\eqalign{
\left\langle x,y |T|q_x, q_y \right\rangle=&\sum_{x',y'}
\left\langle x,y |T| x',y' \right\rangle \left\langle x',y' | q_x, q_y
\right\rangle\cr
=& {1 \over \sqrt{N}} \left[ e^{i q_y y} \left( e^{i q_x (x+1)} +e^{i q_x
(x-1)} \right) + e^{i q_x x} \left( e^{i q_y (y+1)} + e^{i q_y (y-1)} \right)
\right] \cr
=& {1 \over \sqrt{N}} e^{i(q_x x + q_y y)} \left[ 2 \cos q_x + 2 \cos q_y
\right] = T(q_x,q_y)  \left\langle x,y | q_x, q_y \right\rangle\quad.}}
The generalized eigenvalue for a $d$--dimensional hypercubic lattice is
\eqn\eTqd{T(\bq) = 2\sum_{\alpha=1}^{d}{\cos{q_\alpha}}\quad.}

The correlation function in eq.~\esscora\ is now evaluated as
\eqn\essph{\eqalign{
\left\langle \sigma({\bf r}) \sigma({\bf 0}) \right\rangle &\approx
\sum_{\ell}^{\infty} {\tau^\ell \left\langle {\bf r} |W(\ell)| {\bf 0}
\right\rangle} = \sum_{\ell}^{\infty}{\left\langle {\bf r} |(\tau T)^\ell| {\bf
0} \right\rangle}\cr
=& \left\langle {\bf r} \left| {1 \over 1-\tau T} \right| {\bf 0} \right\rangle
= \sum_\bq \left\langle {\bf r}|{\bf q} \right\rangle {1 \over 1 - \tau T(\bq)}
\left\langle {\bf q}|{\bf 0} \right\rangle \cr
=& N \Intq\,{e^{i{\bf q} \cdot {\bf r}} \over N}\, {1 \over 1 - 2\tau
\sum_{\alpha = 1}^{d} {\cos {q_\alpha}}} = \Intq \,{e^{i{\bf q} \cdot
{\bf r}} \over 1 - 2\tau\sum_{\alpha} {\cos {q_\alpha}}}\quad.  }}
For $\tau\rightarrow 0$, the shortest path costs least energy and $\left\langle
\sigma({\bf 0}) \sigma({\bf r}) \right\rangle \sim \tau^{|{\bf r}|}$.  As
$\tau$ increases, larger paths dominate the sum because they are more numerous
(i.e. entropically favored).  Eventually there is a singularity for $1 - \tau
T({\bf 0}) = 0$, i.e. at $2d\times \tau_c =1$,
when arbitrarily long paths become important. For $\tau<\tau_c$, the
partition function is dominated by small loops, and a polymer
connecting two far away points is stretched by its line tension.
When the fugacity exceeds $\tau_c$, the line tension vanishes and
loops of arbitrary size are generated. Clearly the neglect of
intersections (which leads to a finite density) is no longer
justified in this limit. This transition is the manifestation
of Ising ordering in the language of paths
representing the high temperature series. On approaching the
transition from the high temperature side, the sums are dominated
by very long paths. Accordingly, the denominator of eq.~\essph\
can be expanded for small $\bq$ as
\eqn\eprop{1 - \tau T({\bq}) = 1 - 2\tau  \sum_{\alpha = 1}^{d}
{{\cos {q_\alpha}}} \approx (1 - 2d\tau ) + \tau q^2 + \CO (q^4) \approx
\tau_c(\xi^{-2}+q^2 +\CO (q^4)),}
where
\eqn\eHTxi{\xi\equiv\left( {1-2d\tau  \over \tau_c} \right)^{-1/2}\quad.}
The resulting correlation functions, $\left\langle \sigma({\bf 0}) \sigma({\bf
r}) \right\rangle\propto  \Intq e^{i{\bf q} \cdot {\bf r}} /
( q^2 + \xi^{-2})$, are identical to those obtained from a
free (gaussian) field theory, and
\eqn\esscorc{\left\langle \sigma({\bf 0}) \sigma({\bf r})
\right\rangle\propto\left\{\eqalign{
{1 \over r^{d-2}} & {\rm \qquad for~} r \ll \xi \cr
{e^{-r/\xi} \over r^{(d-1)/2}} & {\rm \qquad for~} r\gg\xi
}\right. \quad.}
The correlation length in eq.~\eHTxi\ diverges as $\xi \sim (\tau_c -
t)^{-\nu}$, with the
exponent $\nu = 1/2$.

We can also calculate the free energy in eq.~\eflinW\ as
\eqn\ephfe{\eqalign{
{\ln S' \over N} =&   \sum_\ell^{\infty} {{\tau^\ell \over \ell}} {\left\langle
{\bf 0} |W(\ell)| {\bf 0} \right\rangle} = \left\langle {\bf 0} \left|
\sum_\ell^{\infty}
{\tau^\ell T^\ell \over \ell} \right| {\bf 0} \right\rangle \cr
                              =& -\left\langle {\bf 0} |\ln (1-\tau T)| {\bf 0}
\right\rangle = -N \Intq  \left\langle {\bf 0}|{\bf q} \right\rangle
\ln\left(1-\tau T(\bq)\right)\left\langle {\bf q}|{\bf 0} \right\rangle \cr
=& - \Intq \ln\left(1-2\tau  \sum_{\alpha = 1}^{d} \cos q_\alpha\right)\quad.
}}
In the vicinity of the critical point at $\tau_c=1/(2d)$ the argument of
the logarithm is proportional to $(q^2+\xi^{-2})$ from eq.~\eprop.
This is precisely the free energy of a free field theory, and scales as
\eqn\efsing{f_{\rm sing}\propto \xi^{-d}\propto (\tau_c-\tau )^{d/2}.}
The singular part of the heat capacity, obtained after taking two
derivatives,  is governed by the exponent
$\alpha=2-d/2$. Note that in evaluating the sums appearing in
eqs.~\essph\ and \ephfe\ the lower limit for $\ell$ is not treated
carefully. The series in eq.~\essph\ is assumed to start from $\ell=0$,
and that of eq.~\ephfe\ from $\ell=1$. In fact the first few terms
of both series may be zero because the number of steps is not sufficient
to reach {\bf 0} from {\bf r}, or to from a closed loop. This is not
a serious omission in that the {\it singular} behavior of a series is
not effected by its first few terms. Treating the first few terms
properly can only add analytic corrections to the singular forms
calculated in eqs.~\essph\ and \ephfe.

The equivalence of these results to a free field theory is a
manifestation of field--particle duality. In a field theoretical
description, (imaginary) time appears as an additional dimension,
and the two point correlations describe the probability of
propagating a particle from one point in space--time to another.
In a wave description, this probability is calculated by evolving the
wave function using the Schr\"odinger equation. Alternatively,
the probability can be calculated as the sum over all (Feynman)
paths connecting the two points, each path weighted with the
correct action. The second sum is similar to the above calculation of
$\left\langle \sigma({\bf r}) \sigma({\bf 0}) \right\rangle$.

This approach provides an interesting geometrical interpretation
of the phase transition. The establishment of long range order
implies that all parts of the system have selected the same
state. This information is carried by the bonds connecting nearest
neighbors, and can be passed from the origin to a point ${\bf r}$
through all paths connecting these two points. The fugacity $\tau$
is a measure of the reliability of information transfer between
neighboring sites. Along a one dimensional chain, unless $\tau=1$,
the transferred information decays at large
distances and it is impossible to establish long range order.
In higher dimensions there are many more paths, and by accumulating
the information from all paths, it is possible to establish order
at $\tau_c<1$. Since the number of paths of length $\ell$ grows as
$(2d)^\ell$ while their information content decays as $\tau^\ell$,
the transition occurs at $\tau_c=1/(2d)$. (A better approximation is
obtained by including some of the constraints by noting that the
random walk cannot back track. In this case the number of walks
grows as $(2d-1)^\ell$.) The total information from
paths of length $\ell$ is weighted by $(2d\tau )^\ell$, and decays
exponentially for $\tau<\tau_c$. The characteristic path length,
$\overline{\ell}=-1/\ln(2d\tau )$, diverges as $(\tau_c-\tau )^{-1}$ on
approaching the transition. For paths of size $\ell\ll\overline{\ell}$
there is very good information transfer. Such paths execute random
walks on the lattice and cover a distance $\xi\approx
\overline{\ell}^{~1/2}$. The divergence of $\nu$ with an exponent of
1/2 is thus a consequence of the random walk nature of the paths.

Why does the classical picture fail for $d\leq4$? Let us focus on the
dominant paths close to the phase transition. Is it justified to
ignore the intersections of such paths? Random walks can be regarded
as geometrical entities of fractal (Hausdorf) dimension $d_f=2$.
This follows from the general definition of dimension relating
the mass and extent of an object by $M\propto R^{d_f}$. The size
of a random walk ($R\propto\xi$) is the square root of its
length ($M\propto \ell$). Two geometrical entities of dimensions
$d_1$ and $d_2$ will generally intersect in $d$--dimensional space
if $d_1 + d_2 \geq d$. Thus our random walkers are unlikely to
intersect in $d\geq d_u=2+2=4$, and the  results
obtained by neglecting intersections are asymptotically valid.
Below the upper critical dimension of 4, random walks have frequent
encounters and their intersections must be treated correctly. The
diagrams obtained in the perturbative calculation of the propagator
in a $\phi^4$ theory correspond precisely to taking into account the
intersections of paths. (Each vertex corresponds to one
intersection.) It is now clear that the constraint of self-avoidance
will swell the paths beyond their random walk size leading to an
increase in the exponent $\nu$. Below the transition, the length of dominant
paths grows without bound and the self--avoiding constraint is always
necessary.

\section{The Ising Model on a Square Lattice}

As indicated in eq.~\eHTZd, the Ising partition function is related
to a sum $S$, over collections of paths on the lattice. The allowed
graphs for a square lattice have 2 or 4 bonds per site. Each bond
can appear only once in each graph, contributing a factor of
$\tau\equiv\tanh K$. While it is tempting to replace $S$ with the
exactly calculable sum $S'$, of all loops of
random walks on the lattice, this leads to an
overestimation of $S$. The differences between the two sums arise
from intersections of random walks and can be divided into
two categories:
\item{{\bf (a)}} There is an over-counting of graphs which intersect
at a {\it site}, i.e. with 4 bonds through a point. Consider
a graph composed of two loops meeting at a site. Since a walker
entering the intersection has three choices, this graph can be
represented by {\it three distinct random walks}. One choice leads
to two disconnected loops; the other two are single loops with or
without a self--crossing in the walker's path.

\item{{\bf (b)}} The independent random walkers in $S'$ may go
through a particular lattice {\it bond\/} more than once.

Including these constraints amounts to introducing interactions
between paths. The resulting interacting random walkers are
non--Markovian, as each step is no longer independent of previous
ones and of other walkers.
While such interacting walks are not in general amenable to exact
treatment, in two dimensions an interesting topological property
allows us to make the following assertion:
\eqn\eSassert{\eqalign{S =
\sum & \,  {\rm collections ~  of ~ loops ~of~random~walks~}
with~no~{\rm U}~turns\cr
&\times \tau^{\rm number~of~bonds}
\times (-1)^{\rm number~of ~ crossings}\quad.}}
The negative signs for some terms reduce the overestimate and render
the exact sum.

\noindent{\bf Proof:} We shall deal in turn with the two problems
mentioned above
\nextp
{\bf (a)} Consider a graph with many intersections and focus on a
particular one. A walker must enter and leave such an intersection
twice. This can be done in three ways, only one of which
involves the path of the walker crossing itself (when the walker
proceeds straight through the intersection). This configuration
carries an additional factor of (-1) according to eq.~\eSassert.
Thus, independent of other crossings, these three configurations sum up
to contribute a factor of 1. By repeating this reasoning at each
intersection we see that the over-counting problem is removed, and
the sum over all possible ways of tracing the graph leads to the
correct factor of one.
\nextp
{\bf (b)} Consider a bond that is crossed by two walkers (or twice
by the same walker). We can imagine the bond as an avenue with
two sides. For each configuration in which the two paths enter and
leave on the same side of the avenue, there is another one in which
the paths go to the opposite side. The latter involves a crossing
of paths and hence carries a minus sign with respect to the former.
The two possibilities
thus cancel out! The reasoning can be generalized
to multiple passes through any bond. The only exception is when the
doubled bond is created as a result of a U--turn. This is why
such backward steps are explicitly excluded from eq.~\eSassert.

Let us label random walkers with no U--turns, and weighted by
$(-1)^{\rm number~of ~ crossings}$, as RW$^*$s.  Then as in eq.~\eSce\
the terms in $S$ can be organized as
\eqn\eSsqgc{\eqalign{
S =& \sum({\rm RW^*s~with~1~loop})+\sum({\rm RW^*s~with~2~loops})+
\sum({\rm RW^*s~with~3~loops})+ \cdots \cr
   =& \exp\left[ \sum({\rm RW^*s~with~1~loop})\right]\quad.}}
The exponentiation of the sum is justified since the only interaction
between RW$^*$s is the sign related to their crossings. As two
RW$^*$ loops always cross an even number of times, this is equivalent
to no interaction at all. Using eq.~\eHTZd, the full Ising free energy
is calculated as
\eqn\esqlnZ{\ln Z=N\ln2+2N\ln \cosh K+\sum\left({\rm RW^*s~with~1~loop}
\times \tau^{\rm \# ~ of ~ bonds}\right)\quad.}
Organizing the sum in terms of the number of bonds, and taking
advantage of the translational symmetry of the lattice (up to
corrections due to boundaries),
\eqn\esqlnZN{{\ln Z \over N} = \ln (2 \cosh^2 K) +
\sum_\ell^{\infty}{{\tau^\ell \over \ell}}
\left\langle {\bf 0}|W^* (\ell)|{\bf 0} \right\rangle\quad,}
where
\eqn\eWzero{\eqalign{\left\langle {\bf 0}|W^*(\ell)|{\bf 0}\right\rangle=
&{\rm number~of~closed~loops~of~\ell~steps,~with~no~U~turns,~from~
{\bf 0}~to~{\bf 0}}\cr &\times(-1)^{\#~{\rm of~crossings}}\quad.}}

The absence of U--turns, a local constraint, does not complicate the
counting of walks. On the other hand, the number of crossings depends
on all sites previously crossed by the walker and is
a non--Markovian property. Fortunately, in two dimensions it is
possible to obtain the {\it parity\/} of the number of crossings
from local considerations. The first step is to make the walks
{\it oriented\/} by placing an arrow along the direction
that the path is traversed. Since any path can be traversed in two
directions,
\eqn\eWzero{\left\langle {\bf 0}|W^*(\ell)|{\bf 0}\right\rangle=
{1\over2}\sum oriented~{\rm RW^*~loops~of~\ell~steps,~no~U~turns,~from~
{\bf 0}~to~{\bf 0}}\times(-1)^{n_c}\,,}
where $n_c$ is the number of self--crossings of the loop. We can
now take advantage of the following topological result \ref
\whit{H. Whitney, Comp. Math. {\bf 4}, 276 (1937).}:
\nextp
{\it Whitney's Theorem:} The number of self--crossings of a planar
loop is related to the total angle $\Theta$, through which the
tangent vector turns in going around the loop by
\eqn\eWT{(n_c)_{{\rm mod~2}} = \left( 1 + {\Theta \over 2\pi}
\right)_{{\rm mod~2}}\quad.}
This theorem can be checked by a few examples. A single loop
corresponds to $\Theta=\pm2\pi$, while a single intersection
results in $\Theta=0$.

Since the total angle $\Theta$ is the sum of the angles through
which the walker turns at each step, the parity of crossings can
be obtained using {\it local\/} information alone, as
\eqn\eparity{(-1)^{n_c} = e^{i \pi n_c} = \exp\left[{i\pi} \left(1 +
{\Theta \over 2\pi} \right)\right]= -e^{{i \over 2}
\sum_{j=1}^\ell {\theta_j}}\quad,}
where $\theta_j$ is the angle through which the walker turns on the
$j^{\rm th}$ step. Hence
\eqn\eWlocal{\eqalign{\left\langle {\bf 0}|W^*(\ell)|{\bf 0}\right\rangle=
-&{1\over2}\sum oriented~{\rm RW^*~loops~of~\ell~steps,~with~no~U~turns,
{}~from~ {\bf 0}~to~{\bf 0}}\cr
\times&\exp\left({1\over2}\sum{{\rm~local~change~of~angle~by~the
{}~tangent~vector}} \right)\quad.}}
The angle turned can be calculated at each site if we keep track of
the directions of arrival and departure of the path. To this end,
we introduce a label $\mu$ for the 4 directions {\it going out} of
each site, e.g. $\mu=1$ for right, $\mu=2$ for up, $\mu=3$ for left,
and $\mu=4$ for down. We next introduce a set of $4N \times 4N$ matrices
generalizing eq.~\eWlij\ to
\eqn\eWslij{\eqalign{
\left\langle x_2 y_2,\mu_2|W^*(\ell)| x_1 y_1, \mu_1 \right\rangle=
\sum {\rm oriented~random~walks~of~\ell~steps,~with~no~U~turns,}\cr
{\rm departing}~(x_1,y_1)~{\rm along~\mu_1,~}
{\rm proceeding~along~\mu_2~after~reaching}~(x_2,y_2)\times
e^{{i \over 2} \sum_{j=1}^\ell {\theta_j}} }\quad.}
Thus $\mu_2$ specifies a direction taken {\it after} the walker reaches
its destination. It serves to exclude some paths (e.g., arriving
along -$\mu_2$), and leads to an additional phase.
As in eq.~\erecW, due to their Markovian property, these matrices can
be calculated recursively as
\eqn\erecWs{\eqalign{
&\left\langle x_2 y_2,\mu_2|W^*(\ell)| x_1 y_1, \mu_1 \right\rangle=\cr
& \sum_{x' y', \mu'} \left\langle x_2 y_2, \mu_2 |T^*| x' y', \mu'
\right\rangle \left\langle x' y', \mu'|W^* (\ell - 1)| x_1 y_1 , \mu_1
\right\rangle= \cr
& \left\langle x_2 y_2, \mu_2 |T^* W^* (\ell - 1)| x_1 y_1, \mu_1 \right\rangle
= \left\langle x_2 y_2, \mu_2 |T^{* \ell}| x_1 y_1, \mu_1 \right\rangle\quad,}}
where $T^*\equiv W^*(1)$ describes one step of the walk. The direction
of arrival uniquely determines the nearest neighbor from which the
walker departed, and the angle between the two directions fixes
the phase of the matrix element. We thus generalize eq.~\etwodTM\
to a $4\times4$ matrix that keeps track of both connectivity and
phase between pairs of sites, i.e.
\eqn\eTMs{\eqalign{&\left\langle x' y'|T^*| x y\right\rangle =\cr
&\left [\matrix{
\left\langle x' y' | x+1 y \right\rangle & \left\langle x' y' | x+1 y
\right\rangle e^{{i\pi \over 4}}& 0 & \left\langle  x' y' | x+1 y \right\rangle
e^{-{i\pi \over 4}}\cr
\left\langle x' y' | x y+1 \right\rangle e^{-{i\pi \over 4}}& \left\langle x'
y' | x y+1 \right\rangle & \left\langle  x' y' | x y+1 \right\rangle e^{{i\pi
\over 4}}& 0 \cr
0 & \left\langle x' y' | x-1 y \right\rangle e^{-{i\pi \over 4}}& \left\langle
x' y' | x-1 y \right\rangle & \left\langle  x' y' | x-1 y
\right\rangle e^{{i\pi \over 4}}\cr
\left\langle x' y' | x y-1 \right\rangle e^{{i\pi \over 4}}& 0 & \left\langle
x' y' | x y-1 \right\rangle e^{-{i\pi \over 4}}& \left\langle x' y' | x y-1
\right\rangle
}\right ]
}\quad,}
where $<x y|x' y'>\equiv\delta_{x,x'}\delta_{y,y'}$.

Because of its translational symmetry, the $4N \times 4N$ matrix
takes a {\it block diagonal} form in the Fourier basis
$\left\langle xy | q_x q_y\right\rangle=e^{i(q_x x+q_y y)}/\sqrt{N}$, i.e.
\eqn\eFTbd{\sum_{xy} \left\langle x' y', \mu' |T^*| x y, \mu \right\rangle
\left\langle xy | q_x q_y \right\rangle = \left\langle {\bf \mu'} |T^*(\bq)|
{\bf \mu} \right\rangle \left\langle x'y' | q_x q_y \right\rangle .}
Each $4\times4$ block is labelled by a wavevector $\bq=(q_x,q_y)$,
and takes the form
\eqn\eTMq{T^*(\bq) =
\left [\matrix{
e^{-iq_x} & e^{-i(q_x-{\pi \over 4})} & 0 & e^{-i(q_x+{\pi \over 4})} \cr
e^{-i(q_y+{\pi \over 4})} & e^{-iq_y} & e^{-i(q_y-{\pi \over 4})} & 0 \cr
0 & e^{i(q_x-{\pi \over 4})} & e^{iq_x} & e^{i(q_x+{\pi \over 4})} \cr
e^{i(q_y+{\pi \over 4})} & 0 & e^{i(q_y-{\pi \over 4})} & e^{iq_y}
}\right ]\quad.}

To ensure that a path that starts at the origin completes a loop
properly, the final arrival direction at the origin must coincide
with the original one. Summing over all 4 such directions, the
total number of such loops is obtained from
\eqn\esloops{\left\langle {\bf 0} |W^*(\ell)| {\bf 0} \right\rangle = \sum_{\mu
= 1}^4 \left\langle 00, \mu |T^{*\ell}| 00, \mu \right\rangle = {1 \over N}
\sum_{xy, \mu} \left\langle xy, \mu |T^{*\ell}| xy, \mu \right\rangle={1 \over
N} \tr (T^{*\ell})\quad.}
Using eq.~\esqlnZN, the free energy is calculated as
\eqn\elnZsum{\eqalign{
{\ln Z\over N} =& \ln (2\cosh^2 K) - {1 \over 2} \sum_\ell {\tau^\ell \over
\ell} \left\langle {\bf 0} |W^*(\ell)| {\bf 0} \right\rangle = \ln (2\cosh^2 K)
- {1 \over 2N}\tr \left[ \sum_\ell {T^{*\ell} \tau^\ell \over \ell}\right] \cr
=& \ln (2\cosh^2 K) + {1 \over 2N}\tr \ln (1-\tau T^*) \cr
=& \ln (2\cosh^2 K) + {1 \over 2N} \sum_\bq \tr \ln \left(1-\tau
T^*(\bq)\right)\quad .}}
But for any matrix $M$, with eigenvalues $\{\lambda_\alpha\}$,
$$\tr \ln M = \sum_\alpha \ln \lambda_\alpha =
\ln \prod_\alpha \lambda_\alpha = \ln {\rm det} M\quad.$$
Converting the sum over $\bq$ in eq.~\elnZsum\ to an integral leads to
\eqn\elnZdet{\eqalign{{\ln Z \over N}&= \ln (2 \cosh^2 K)+\cr
&{1 \over 2} \int {d^2 \bq \over (2\pi)^2} \ln \left\{{\rm det}
\left |\matrix{
1- \tau e^{-i q_x} & -\tau  e^{-i (q_x- {\pi \over 4})} & 0 & -\tau  e^{-i
(q_x+ {\pi \over 4})} \cr
-\tau  e^{-i (q_y+ {\pi \over 4})} &1- \tau e^{-i q_y} & -\tau  e^{-i (q_y-
{\pi \over 4})} & 0 \cr
0 & -\tau  e^{i (q_x- {\pi \over 4})} & 1- \tau e^{i q_x} & -\tau  e^{i (q_x+
{\pi \over 4})} \cr
-\tau  e^{i (q_y+{\pi \over 4})} & 0 & - \tau e^{i (q_y- {\pi \over 4})} & 1-
\tau e^{i q_y} \cr
}\right |
 \right\}\quad.}}
Evaluation of the above determinant is straightforward, and the final
result is
\eqn\elnZt{{\ln Z \over N} = \ln (2 \cosh^2 K) + {1 \over 2} \int
{d^2 \bq \over (2\pi)^2} \ln \left[ (1+\tau^2)^2 - 2\tau (1 - \tau^2)
(\cos q_x + \cos q_y) \right]\quad.}
Taking advantage of trigonometric identities, the result can be
simplified to
\eqn\elnZc{{\ln Z \over N} = \ln 2 + {1 \over 2} \int_{-\pi}^\pi
{d q_x d q_y \over (2\pi)^2} \ln \left[ \cosh^2 (2K) - \sinh (2K)
(\cos q_x + \cos q_y) \right]\quad.}
While it is possible to obtain a closed form expression by performing
the integrals exactly, the final expression involves a hypergeometric
function and is not any more illuminating.

\section{Singular Behavior}

To uncover the singularity in the free energy of the
two dimensional Ising model in eq.~\elnZt, we start with the simpler
expression obtained by the unrestricted sum over random walks
in  eq.~\ephfe, (specializing to $d=2$)
\eqn\elSRW{f_G=\ln (2 \cosh^2 K) - \int {d q_x d q_y \over (2\pi)^2}
\ln \left[ 1 - 2\tau  (\cos q_x + \cos q_y) \right]\quad.}
Apart from the argument of the logarithm, this expression is similar
to the exact result. Is it possible that such similar functional forms
lead to distinct singular behaviors? The singularity results from
the vanishing of the argument of the logarithm at $\tau_c=1/4$. In the
vicinity of this point we make an expansion as in eq.~\eprop,
\eqn\eExpg{A_G(\tau ,\bq)=(1-4\tau)+\tau q^2+\CO(q^4)\approx \tau_c\left( q^2
+4{\delta \tau \over \tau_c} \right)\quad,}
where $\delta \tau =\tau_c-\tau $. The singular part of eq.~\elSRW\ can be
obtained by focusing on the behavior of the integrand as $\bq\to{\bf 0}$,
and replacing the square Brillouin zone for the range of the integral
with a circle of radius $\Lambda\approx 2\pi$,
\eqn\efGsing{\eqalign{f_{{\rm sing.}}
=&-\int_0^\Lambda{2\pi qdq\over 4\pi^2}
\ln\left( q^2+4{\delta \tau \over \tau_c} \right)\cr
=&-{1\over 4\pi}\left[\left( q^2+4{\delta \tau \over \tau_c} \right)
\ln\left( {q^2+4{\delta \tau / \tau_c}\over e} \right)  \right]_0^\Lambda
\quad.
}}
Only the expression evaluated at $q=0$ is singular, and
\eqn\efGsin{f_{{\rm sing.}}=-{1\over\pi}\,\left({\delta \tau \over \tau_c}
\right)\,\ln\left({\delta \tau \over \tau_c}  \right)\quad.}
The resulting heat capacity $C_G\propto {\partial^{2}f_G
/ \partial^{2} t}$, diverges as $1/\delta \tau $. Since eq.~\elSRW\ is
not valid for $\tau>\tau_c$, we cannot obtain the behavior of heat capacity
on the low temperature side.

For the exact result of eq.~\elnZt, the argument of the logarithm is
\eqn\eAsqex{A^*(\tau ,\bq)=(1+\tau^2)^2-2\tau (1-\tau^2)(\cos q_x+\cos q_y
)\quad.}
The minimum value of this expression, for $\bq={\bf 0}$, is
\eqn\eAsqmin{A^*(\tau ,{\bf 0})=(1+\tau^2)^2 - 4 \tau (1 - \tau^2) =
(1 - \tau^2)^2 + 4\tau^2 - 4 \tau (1 - \tau^2) = (1 - \tau^2 - 2 \tau)^2\quad.}
Since this expression (and hence the argument of the logarithm) is
always non--negative, the integral exists for all values of $\tau$. As
required, unlike eq.~\elSRW, the exact result is valid at {\it all\/}
temperatures. There is a singularity when the argument vanishes for
\eqn\esqtc{\tau_c^2 + 2 \tau_c - 1  = 0\qquad\Longrightarrow\qquad
\tau_c = -1 \pm \sqrt{2}\quad.}
The positive solution describes a ferromagnet and leads to a value
of $K_c = \ln (\sqrt{2} + 1)/2$. Setting $\delta \tau  =\tau-\tau_c$, and
expanding
eq.~\eAsqex\ in the vicinity of $\bq\to 0$ gives
\eqn\eAsqapp{\eqalign{A^*(\tau ,\bq)\approx&
\left[ (-2\tau _c-2)\delta \tau  \right]^2+\tau_c(1-\tau_c^2)q^2+\cdots\cr
\approx&2\tau _c^2\left[ q^2+4\left( \delta \tau \over \tau_c
\right)^2\right]\quad.}}
The important difference from eq.~\eExpg\ is that $(\delta \tau /\tau_c)$
appears at quadratic order. Following the steps in eqs.~\efGsing\
and \efGsin, the singular part of the free energy is
\eqn\efSqsing{\eqalign{
\left.{\ln Z \over N}\right|_{\rm sing.}=& {1\over2}\int_0^\Lambda
{2\pi qdq\over 4\pi^2}\ln\left[q^2+4\left({\delta \tau \over \tau_c}
\right)^2\right]\cr
=&{1\over 8\pi}\left[\left( q^2+4\left({\delta \tau \over
\tau_c}\right)^2\right)
\ln\left( {q^2+4({\delta \tau / \tau_c})^2\over e} \right)
\right]_0^\Lambda\cr
=&{1\over\pi}\,\left({\delta \tau \over \tau_c}  \right)^2\,
\ln\left|{\delta \tau \over \tau_c}  \right|\quad.}}
The heat capacity is obtained by taking two derivatives and diverges as
$C(\delta \tau )_{\rm sing.}=A_\pm \ln|\delta \tau |$. The logarithmic
singularity corresponds to the limit $\alpha=0$; the peak is
symmetric and characterized by the amplitude ratio of $A_+/A_-=1$.

The graphical method presented in this section was originally developed
by Kac and Ward \ref
\kac{M. Kac and J. C. Ward, Phys. Rev. {\bf 88}, 1332 (1952).}.
The main ingredient of the derivation is the result that the correct accounting
of the paths can be achieved by including a factor of  $(-1)$ for each
intersection.
(This was originally a conjecture by Feynman \ref
\feyn{R. P. Feynman, {\it Statistical Mechanics} (Addison-Wesley,
Reading, MA, 1972).},
later proved by Sherman \ref
\Sherman{S. Sherman, J. Math. Phys. {\bf 1}, 202 (1960).}.)
The change of sign is reminiscent of the exchange factor between fermions,
and indeed the final result can be obtained by mappings to free fermions \ref
\rSML{T. Schultz, D. Mattis, and E. Lieb, Rev. Mod. Phys. {\bf 36}, 856
(1964).}.

In addition to the partition function, the correlation functions
$<\sigma_{\bf i}\sigma_{\bf j}>$ can also be calculated by summing
over paths \ref
\rItzyD{C. Itzykson and J.-M. Drouffe, {\it Statistical field theory: 1}
(Cambridge
University Press, New York, 1989).}.
Since the combination $q^2+4(\delta \tau /\tau_c)^2$ in eq.~\eAsqapp\
describes the behavior of these random walks, we expect a
correlation length $\xi\sim |\tau_c/\delta \tau |$, i.e. diverging
with an exponent $\nu=1$ on both sides of the phase transition,
with an amplitude ratio of unity. The exponents $\alpha$ and $\nu$
are related by the hyperscaling identity $\alpha=2-2\nu$. The
critical correlations at $\tau_c$ are more subtle and
decay as $<\sigma_{\bf i}\sigma_j>_c\sim 1/|{\bf i-j}|^\eta$
with $\eta=1/4$. Integrating the correlation functions yields the
susceptibility, which diverges as $\chi_\pm\simeq C_\pm
|\delta \tau |^{-\gamma}$, with $\gamma=7/4$ and $C_+/C_-=1$.

\section{The Two Dimensional Spin Glass}

The key to the exact solution of the non--random two dimensional Ising model
presented in the previous section is the reduction of the graphical expansion
to sums over non--interacting (fermionic) random walks.  This reduction
depends on the geometrical properties of paths in $d=2$ and is independent
of the uniformity of the bonds $J_{ij}$.  The same method can be applied to
the random bond problem, reducing the problem to (fermionic) random walks
in a random medium. This is an undirected version of the problem of
DPRM, extensively mentioned in these notes.  Due to randomness, we can
no longer diagonalize the transfer matrix that generates these walks by
Fourier transformation. However, we can still examine such walks numerically
by successive multiplications of the transfer matrix.  Thus all random bond
Ising models in $d=2$ can in principle be {\it solved exactly in polynomial
time\/} in their size $L$.
This is not true for the three dimensional versions.

We shall demonstrate the potentials of such exact methods by developing
an algorithm for calculating the partition function of the $\pm J$ spin glass
\toul.
This is the model introduced in sec.~I, where its high temperature correlation
functions were examined with directed paths. The advantage of the model
(or any other random mixture of $+J$, $-J$, and absent bonds) is that all
computations can be performed in integer form, thus avoiding any floating point
errors. Of  course the main interest in the spin glass problem stems from the
complexity of its low temperature states.
Despite the great deal of work on spin glasses over the past decades
\nref\bind{K. Binder and A. P. Young, Rev. Mod. Phys. {\bf 58}, 801 (1986).}%
\nref\chow{D. Chowdhury, {\it Spin Glasses and Other Frustrated Systems}
(Princeton University Press, Princeton, NJ, 1986).}\nref
\fisc{K. H. Fischer and J. A. Hertz, {\it Spin Glasses}
(Cambridge University Press, Cambridge, 1991).}\refs{\rBeyond,\bind{--}\fisc},
the description of the phase transition and the nature of the glassy state
remain controversial subjects \nref
\braya{A. J. Bray and M. A. Moore, Phys. Rev. Lett. {\bf 58}, 57 (1987).}\nref
\dfish{D. S. Fisher and D. A. Huse, Phys. Rev. B {\bf 38}, 386 (1988).}%
\refs{\braya,\dfish}.
Interactions with infinite-range \ref
\sher{D. Sherrington and S. Kirkpatrick, Phys. Rev. Lett. {\bf 35}, 192
(1975).}\ lead to a solution with broken replica
symmetry \rBeyond.  It is not known, however, to what extent this
mean-field result captures the behavior of short-range interactions
\nref\mcka{S. R. McKay, A. N. Berker, and S. Kirkpatrick, Phys. Rev.
Lett. {\bf 48}, 767 (1982).}\refs{\dfish,\mcka}.
Monte Carlo simulations of spin glass problems are notoriously difficult due
the ease with which the system can get trapped in long--lived metastable
states. It is thus quite valuable to provide some exact information about
the equilibrium low temperature behavior of spin  glasses, even if that
knowledge is limited to two dimensions.

We start with the Edwards-Anderson (EA) Hamiltonian \ref
\edwa{S. F. Edwards and P. W. Anderson, J. Phys. F {\bf 5},  965 (1975). }
\eqn\eHam{H = \sum_{\langle ij \rangle} J_{ij} \sigma_i \sigma_j\quad,}
where the nearest neighbor  quenched random bonds $J_{ij}$
are chosen from the bimodal distribution
\eqn\eBimodal{p(J_{ij}) ={1 \over 2}\delta (J_{ij} - J) + {1 \over 2}\delta
(J_{ij} + J)\quad,}
with $J>0$. On a lattice with periodic boundary conditions (BCs), there are
exactly $2N$ bonds, with $N=L^2$ the total number of spins.
The high temperature expansion for the partition function takes the form
\eqn\eZpoly{Z = 2^N \cosh^{2N}K \sum_{\ell=0}^{2N} A_\ell \tau^\ell\quad,}
where the coefficients $A_l$ are pure integers.  Note that $A_\ell=0$ for
odd values of $\ell$ since closed loops on the square lattice necessarily
traverse an even number of bonds.

We can use the diagrammatic method introduced in the previous sections to
transform the problem of summing the high temperature series into
one of evaluating a local random walk. Every step proceeds exactly as
before up to eq.~\erecW. However, the $4N \times 4N$ transfer matrix in
eq.~\eTMs\ has to be modified to take care of the randomness in bonds.
Each element of the matrix that connects sites $i$ and $j$ has to be
multiplied with the reduced bond variable $s_{ij} = s_{ji} = J_{ij}/J$,
equal to $+1$ for ferromagnetic bonds and $-1$ for antiferromagnetic
bonds.  There is an additional complication in dealing with finite lattice
size.
For simplicity we shall use periodic boundary conditions on an
$L\times L$ lattice.  We must then take proper account of diagrams
which wrap around the lattice. The correct result, based on the
combinatorics of closed loops on periodic lattices \ref
\rpott{R. B. Potts and J. C. Ward, Progr. Theoret. Phys. (Kyoto)
{\bf 13}, 38 (1955).}\
is $Z = (-Z_1 + Z_2 + Z_3 + Z_4)/2$ with
\eqn\eeZlambdaDet{Z_\lambda = 2^N \cosh^{2N}K
\sqrt{\det \left[1 - T^*_\lambda \tau\right]}\quad.}
Here, $T^*_1$ is the original $4N \times 4N$ transfer matrix,
while $T^*_2$, $T^*_3$, (and $T^*_4$) are obtained respectively
by changing the sign of a horizontal, vertical, (or both) column of bonds.
The  linear combination $(-Z_1 + Z_2 + Z_3 + Z_4)/2$
ensures that all diagrams, including those which loop the entire lattice,
are weighted correctly in the final expression for the partition function.

We have implemented \ref
\saul{L. Saul and M. Kardar, Phys. Rev. E {\bf 48}, R3221 (1993);
Nucl. Phys. B, in press (1994).}\
this algorithm on the computer as follows. Given a
set of bonds $\{J_{ij}\}$, we first construct the $4N \times 4N$
matrices $T^*_\lambda $ and compute the traces
$\tr({T^*_\lambda}^\ell)$ for $\ell\leq N$. This step of the algorithm
is the most computationally intensive.
The coefficients of the series expansions for $\ln Z_\lambda$ are
related to the traces by
\eqn\eLnZinf{{\frac{\ln Z}{N}} = \ln [2 \cosh^2 K] - {\frac{1}{2N}}
\sum_{\ell=0}^\infty {\frac{1}{\ell}}\tr(T^{*\,\ell}) \tau^\ell\quad.}
Next, we compute the high temperature series for $Z$.  This is done by
exponentiating the series for $\ln Z_\lambda$, followed by taking the linear
combination that incorporates periodic boundary conditions.
The high temperature expansion for $Z$ is a polynomial in $\tau$
with integer coefficients; the last term, of order $2N$, is derived from the
graph that traverses every bond on the square lattice.  These $2N$ coefficients
have an end-to-end symmetry that enables one to compute them from the first
$N$ powers of the transfer matrix.  Finally, we expand powers of
$\cosh K$ and $\tanh K$ and rewrite $Z$ as a polynomial
in $e^{-\beta J}$; the end result $Z = \sum_{E} g(E)e^{-\beta E}$
yields the density of states.  For an Ising model with $\pm J$ bonds,
we can perform all these operations using only integer arithmetic.

The algorithm has several desirable features.  First, it returns the
partition function~$Z$ as an exact integer result.  In this way, it
not only avoids the statistical uncertainties inherent in Monte Carlo
simulation; it also avoids the floating point errors that creep into
numerically ``exact'' calculations of $Z$ in large systems at low
temperatures.  The algorithm thus provides us with an efficient and
reliable way to investigate energies and entropies at low
temperatures.  This is particularly important in a system that
exhibits a phase transition at $T=0$, such as the $\pm J$ spin glass.
We can also calculate other quantities, such as the roots of the
partition function in the complex plane, or the number of low-level
excitations, that are otherwise inaccessible. Unfortunately, the
necessity of handling large integers (of order $2^N$) complicates what
would otherwise be a rather straightforward algorithm to implement.

A second advantage of the algorithm is that it executes in polynomial
time.  We estimate the algorithm's performance as follows. Computing
the traces requires $O(N^3)$ arithmetic operations on integers of
order $2^N$, while in general, the power series manipulations take
much less time.  We therefore expect the computation time to scale as
$\tau \sim N^\delta$, with $3 < \delta < 4$.  This stands in contrast
to the numerical column to column transfer matrix method of Morgenstern and
Binder \ref
\morg{I. Morgenstern and K. Binder, Phys. Rev. B {\bf 22}, 288 (1980).}\
which has time and memory requirements that grow
exponentially with system size.  We obtained most of our results on
dedicated DEC 3100 workstations.  The largest lattice
that we examined had $N=36 \times 36$ spins.  Finally, we mention that
the computation of the traces can be broken down into $O(N)$
independent computations, so that a faster, parallel implementation of
the algorithm (on a supercomputer or spread across several
workstations) should be possible.

\section{Results for the Two Dimensional Spin Glass}

We examined the $\pm J$ spin glass on lattices of size $L=4$ to
$L=36$.  Several realizations of
randomness were studied: 8000 for $L=4$, 6, 8; 2000 for $L=10$, 12, 14;
800 for $L= 16$, 18; 80 for $L=20$, 22, 24; and 4 for $L=32$, 36. We
performed quenched averages by assigning an equal probability to each
random sample: $\overline\theta = (1/S)\sum_{s} \theta_s$.  To reduce
the amount of statistical error, we only considered lattices in which
exactly half the plaquettes were frustrated \morg.  We also found it
quite revealing to compare our results with those
for the fully frustrated Ising model \nref
\vill{J. Villain, J. Phys. C {\bf 10}, 1717 (1977).}\nref
\andr{G. Andre, R. Bidaux, J. P. Carten, R. Conte, and L. de Seze,
J. Phys. (Paris) {\bf 40}, 479 (1979).}\refs{\vill,\andr},
as both models undergo phase transitions at $T=0$.
The typical output of our algorithm is a set of integers $g(E)$ for
the number of states.  Using the density of states we can perform calculations
in either the
microcanonical or canonical ensemble. In the limit of infinite size,
of course, the two ensembles should yield identical results.

We used the algorithm first to study the thermodynamic properties of
the ground state, fitting the data for the ground-state energy and
entropy to the finite-size form $f_L= f_\infty + a/L^2$. Extrapolating to
infinite lattice size, we estimate $\overline{e_0}/J = -1.404 \pm 0.002$
and $\overline{s_0} = 0.075 \pm 0.002$. These results are consistent
with previous MC \nref
\wang{J.-S. Wang and R. H. Swendsen, Phys. Rev. B {\bf 38}, 4840
(1988).}\nref
\berg{B. A. Berg and T. Celik, Phys. Rev. Lett. {\bf 69}, 2292 (1992).}%
\refs{\wang,\berg}\
and column to column  transfer matrix \ref
\cheu{H.-F. Cheung and W. L. McMillan, J. Phys. C. {\bf 16}, 7027
(1983).}\
estimates. We also used the algorithm to study the number of low-level
excitations.  On a lattice with periodic BCs, the lowest excited state
has an energy $4J$ above the ground state.  The quantity $e^{\Delta S}
= g(E_0+4J)/g(E_0)$ measures the degeneracy ratio of these excited states.  We
find
that $\overline{\Delta S}_{SG}$ grows faster than $\ln N$ indicating
that the low-lying excitations of the $\pm J$ spin glass involve spin
flips on large length scales.

The abundance of low-lying excitations affects the low-temperature
behavior of the heat capacity.  In a finite system with energy gap
$4J$, the heat capacity vanishes as $C \sim \beta^2 e^{-4\beta J}$.
As pointed out by Wang and Swendsen \wang, this behavior can
break down in the thermodynamic limit.  The 1D Ising model with
periodic BCs shows how this can happen: the energy gap is $4J$, but
the heat capacity of an infinite system vanishes as $C_{1D} \sim
\beta^2 e^{-2\beta J}$. The anomalous exponent reflects the fact that
the number of lowest excited states grows as $N^2$.  From Monte Carlo and
column to column transfer matrix studies, Wang and Swendsen \wang\ conclude
that $C_{SG}\sim\beta^2 e^{-2 \beta J}$ for the 2D $\pm J$
spin glass as well.  Our results find a disagreement in slope
between $\Delta S_{1D}$ and $\overline{\Delta S}_{SG}$ versus $N$,
leading us to suggest a different form for
$C_{SG}$.  As motivation, we appeal to another exactly soluble model
with a phase transition at $T=0$: the fully frustrated (FF) Ising
model on a square lattice \vill.  On a periodic lattice, the
lowest excited states of the FF model have energy $4J$ above the
ground state.  The large number of low-lying excitations, however,
causes the heat capacity to vanish as $C_{FF} \sim \beta^3 e^{-4\beta
J}$.  Note the extra power of temperature.  Our data compare much
better to $\Delta S_{FF}$ than to $\Delta S_{1D}$, suggesting a
similar behavior may describe the $\pm J$ spin glass, e.g.\ $C_{SG}
\sim \beta^{2+\rho} e^{-4\beta J}$ with $\rho \neq 0$.  As we shall
see below, there are other reasons to favor this form.

One way to investigate phase transitions is to examine the
roots of the partition function $Z$ in the complex plane.  This was
first done by Fisher \ref
\fish{M. E. Fisher, {\it Lectures in Theoretical Physics}
(University of Colorado Press, Boulder, 1964), Vol. 7c.}\
in a study of the 2D Ising model
with uniform $+J$ bonds.  Fisher calculated the distribution of roots
of the partition function in the complex $z=e^{-2\beta J}$ plane.  In
the limit of infinite lattice size, he showed that the roots condense
onto two circles centered at $z=\pm 1$, and related the singular
behavior in the free energy to the distribution of roots in the
vicinity of the positive real axis.  Since a system of finite size
does not exhibit non-analytic behavior, it is clear that the roots of
the partition function can only close in on the positive real axis in
the thermodynamic limit.
The zeros of partition functions are thus subject to finite-size
scaling \ref
\itzy{C. Itzykson, R. B. Pearson, and J. B. Zuber,  Nuclear Physics
{\bf B220}, 415 (1983).}.
At a finite-temperature phase transition,  the complex zero $T(L)$ closest
to the positive real axis obeys $|T(L) - T_c|\sim L^{-y_t}$; likewise,
the correlation length diverges as $\xi \sim
(T-T_c)^{-\nu}$ with $\nu = 1/y_t$.  On the other hand, at a $T=0$
phase transition, such as occurs in the 1D Ising model,
one finds $|z(L)| \sim L^{-y_z}$ with $\xi
\sim z^{-1/y_z}$, where $z(L)$ is the smallest root in the
complex $z=e^{-2\beta J}$ plane.
For example, the partition function of a 1D Ising chain with
periodic boundary conditions in eq.~\eZclosed,
\eqn\eZidIs{Z =2^L \cosh^L(\beta J)\left[1 + \tanh^L(\beta J)\right]\quad.}
has its smallest root at $\tanh(\beta J) =
e^{\pm i\pi /L}$, or $z(L) = \pm i \tan(\pi/2L)$.
As $L \rightarrow \infty$, the magnitude of the root
scales as $z(L) \sim L^{-1}$, consistent with the fact that
the correlation length in the 1D Ising model diverges as
$\xi \sim e^{2\beta J}$.

In the $\pm J$ spin glass, we observed that, for most realizations of
randomness, the smallest root $z(L)$ falls on the imaginary axis.
One might expect that the probability distribution for the magnitude
of this root assumes a scale-invariant form as $L \rightarrow \infty$.
We were unable to verify this hypothesis due to insufficient data on
large lattices.  Instead, we examined the statistics of $u(L)$, where
$u = z^2 = e^{-4\beta J}$.  On a square lattice with periodic
boundary conditions, the partition function for a  $\pm J$ spin
glass is polynomial in $e^{-4\beta J}$.  We therefore
looked at the scaling of roots in the complex $u = e^{-4\beta J}$
plane.  The results could be fitted to $|u(L)|\sim L^{-2.2}$ with
$y_u = 2.2 \pm 0.1$; this suggests
to us that the correlation length in the $\pm J$ spin glass diverges
as $\xi \sim e^{2\beta J}$.  Additional powers of temperature and/or
finite-size effects might explain the slight deviation from $y_u = 2$.
Note that this behavior for the correlation length is consistent with the
hyperscaling relation $\ln Z_{\rm singular}\propto \xi^{-2}$, and our claim
that, up to
powers of temperature, the heat capacity diverges as $C \sim
e^{-4\beta J}$.  This result disagrees with previous
studies \nref\mcmib{ W. L. McMillan, J. Phys. C. {\bf 17}, 3179
(1984).}\refs{\wang,\cheu,\mcmib}\
that report $\xi \sim T^{-\nu}$, with $\nu \approx 2.6 - 2.8$.
An analytical approach, closely related to averaging the fermionic path
integrals \ref
\rJug{G. Jug, Phys. Rev. Lett. {\bf 53}, 9 (1984); and private
communications.},
does support exponential correlations, but with a gap of $4J$.

A great deal of information on spin glasses has been obtained by
examining `defects' (droplets) in finite systems. The cost of a defect
of size $L$ is related to the difference in free energies with
periodic and anti-periodic BCs.  At $T=0$, this reduces to the
difference in energy between the ground states.  On an $L\times L$
lattice, the defect energy measures the effective block
coupling \nref\mcmia{W. L. McMillan, Phys. Rev. B {\bf 28}, 5216 (1983).}%
\nref\brayb{A. J. Bray and M. A. Moore, in {\it Heidelberg Colloquium
on Glassy Dynamics}, edited by J. L. van Hemmen and I. Morgenstern
(Springer-Verlag, Heidelberg, 1986).}\refs{\mcmia,\braya,\dfish,\mcka,\brayb}\
$J^\prime$ on length scale
$L$.  Let $p(L)$ be the fraction of $L \times L$ blocks for which
$J^\prime \neq 0$.  Scaling arguments \brayb\ suggest that $p(L)
\sim L^{-\eta}$, where $\eta$ is the critical exponent that
characterizes the power law decay of correlations
$\overline{<\!\sigma_0 \sigma_L\!>^2}$ at $T=0$.  Plotting $p(L)$
versus $L$, we find $\eta=0.22 \pm 0.06$ in
agreement with previous results \refs{\brayb,\wang}.  Besides the
defect energy, we also looked at the defect entropy $\delta S_{L}$,
i.e.  the difference in zero-temperature entropies with periodic and
anti-periodic BCs.  The data could be fitted to
$\overline{\delta S_{L}^{2}} \sim L^{2y_S}$ with $y_S = 0.49\pm 0.02$.
This is curiously close to the result $\delta S_L \sim L^{1/2}$,
expected if entropy changes due to reversing the different bonds along
the boundary are statistically independent.
The defect entropy in the fully
frustrated Ising model approaches a constant value with $1/L^2$ corrections.
It is straightforward, moreover, to show that in the 1D Ising model
the defect entropy scales as
$\delta S_{L} \sim \ln L$.  Both these behaviors are markedly
different from the spin glass.  We do not know any obvious relation between
the finite-size scaling of the defect entropy and other quantities at $T=0$.
More details of the algorithm and results can be found in ref.~\saul.

\listrefs\bye